\documentclass{raa}       

\usepackage{subcaption}
\usepackage{float}
\usepackage{graphicx,times} 
\usepackage{hyperref}
\usepackage{amsmath,amsfonts,amssymb}
\hypersetup{colorlinks=true,urlcolor=magenta,filecolor=magenta,linktoc=page,citecolor=blue,linkcolor=blue}
\usepackage{color}
\newcommand{\kms}{km s$^{-1}$}
\newcommand{\msun}{M$_{\odot}$}
\usepackage{natbib}
\bibpunct{(}{)}{;}{a}{}{,}

\begin{document}
	
\title{Extragalactic HI  survey with FAST : First look of the pilot survey results}
	\volnopage{Vol.0 (20xx) No.0, 000--000}       
	\setcounter{page}{1}  
	
	\author{Jiangang Kang\inst{1,2,3*}
		\and Ming Zhu\inst{1,2*}
		\and Mei Ai\inst{1,2}
		\and Haiyang Yu\inst{1,2,3},
		\and Chun Sun\inst{1,2}}
	\setcounter{page}{1} 
	\institute{National Astronomical Observatories, Chinese Academy of Sciences, Beijing 100101, China; {\rm $^*$corresponding author:kjg@nao.cas.cn}\\
		\and
		CAS Key Laboratory of FAST,National Astronomical Observatories, Chinese Academy of Sciences, Beijing 100101, China, {\rm $^*$corresponding author: mz@nao.cas.cn} \\
		\and
		School of Astronomy and Space Science,University of Chinese Academy of Sciences, Beijing 100049, China \\ 
	}
	
	\abstract{  As first data release of a pilot extragalactic HI survey with  Five-hundred-meter Aperture Spherical radio Telescope (FAST),we extracted  544 extragalaxies from three-dimensional(3D) spectral data to perform interactive searching and computing, yielding global parameters for these detections, extending redshift ranges of HI 21cm line up to z = 0.04 ,which covers part of the sky region in right ascension(R.A. or $\alpha$) and declination(Dec or $\delta$) range $00^{\rm h} 47^{\rm m}< \rm R.A.(J2000)<23^{\rm h}22^{\rm m}$ and $+24^{\circ}<\rm Dec.(J2000) <+43^{\circ}$ . The S/N of 544 HI detections are greater than 5  flagged with code 1 to 4 based on  baseline qualities or RFI contamination. Besides, we find 16 of which without any counterparts in the existing galaxy catalogs. The catalog can give a  guidence for the future HI observation with FAST.
		\keywords {HI survey: data release --- galaxy:
			extragalaxies: HI 21cm line}
	}

	\authorrunning{J.G Kang et al.}            
	\titlerunning{A  FAST HI survey pilot  result} 
	\maketitle
	
	\section{Introduction}           
	\label{sect:intro}
	 Neutral atomic hydrogen(HI) is one of  key tracer for studying the evolution of galaxy and matter distribution in the Universe. The hydrogen content in a galaxy is usually presented in
	three states: neutral molecular ($\rm H_2$), neutral atomic(HI) and
	ionized (HII). HI is usually
	the major phase observed in the outer regions of galaxy discs
	\citep{2004IAUS..220..369W,2005A&A...433L...1B,2011IAUS..277...59H,2018MNRAS.478.1611K}.
	HI emissions are also detected in filaments, plumes and/or bridges,
	tracing the gravitational interactions with neighbouring
	galaxies\citep{2016MNRAS.459.1827P,2003MNRAS.339.1203K,Meyer_2004}.
	For many HI rich galaxies, the HI discs of spirals are usually much more extended than the stellar discs, leading to them highly susceptible to external forces such as tidal
	interactions, gas accretion, and ram pressure stripping, while
	providing fuel for star formation
	\citep{1996hst..prop.6639M,2005MNRAS.362..609B,2017IAUS..321..220B,2012MNRAS.422.1835S,2020}. Therefore,
	the amount and extent of HI in galaxies varies significantly with
	environment \citep{2009,2013,2014,2017}.
	
	Over the past decades, a number of HI surveys have being conducted to detect HI  gas in the
	local Universe\citep{2020}, including the the  Parkes
	All-Sky Survey
	(HIPASS;\citep{2000ASPC..217...50S,2001MNRAS.322..486B,2004yCat.8073....0M}) in the southern hemisphere,
	and the Northern HIPASS extension (NHICAT;\cite{2006}), and
	the Effelsberg Bonn HI Survey (EBHIS;\cite{2011}) in
	north hemisphere. The Arecibo Legacy Fast ALFA Survey
	(ALFALFA;\cite{Giovanelli_2005}) provides improved spatial resolution
	and sensitivity over 7000 $\rm deg^2$.  These surveys 
	have played a
	key role in mapping the HI distribution, revealing the gas properties of nearby
	galaxies and providing valuable census of the
	cosmic HI content at low redshift
	\citep{2021,2018A&A...609A..17J,2020}.  See \citep{2020} for a review of the HI surveys.
	
	As the current largest single dish radio telescope in the world, one
	of the key scientific goals of FAST is to perform a blind
	extragalactic HI sky surveys over a cosmologcially signifcant volum to exploit its superior sensitivity and
	angular resolution\citep{NAN_2011}. The 
	FAST HI survey will enlarge the survey area to a solid angle of 23000 $deg^2$
	with the declination range of $-14.3^{\circ}< \rm Dec < +66.7^{\circ}$
	and up to a redshift of 0.35 to carry out a census of HI galaxies in
	the northern sky\citep{Li_2018,jiang2019commissioning}. 
	As one of the long term scientific program, the extra-galactic HI survey is
	scheduled to be carried out simultaneously with serveral other
	projects(Galactic HI survey,Pulsar search,Fast Radio Burst search) in
	the Commensal Radio Astronomy Fast Survey (CRAFTS) with 19-beam
	receivers in 1.05-1.45 GHz frequency range(L band) since
	2020\citep{Li_2018,jiang2019commissioning}. One of the focus
    of FAST HI survey is to investigate the HI Mass Function (HIMF)
	in lower redshifts and its dependence on galaxy group and cluster
	environment. The HIMF plays a key role in revealing the galactic
	evolution history as a function of redshifts and explaining several
	crises between observations and simulations at small scales in frame of
	$\Lambda$ Cold Dark Matter($\Lambda$CDM)
	model\citep{Obreschkow_2009,Duffy_2012,Weinberg_2015,Jones_2018}. For
	instance, the slope at the faint end of the HIMF
	\citep{2005ApJ...621..215S,Zwaan_2005,Jones_2018} is still a puzzle,
	various HI surveys measure an abundance of low-mass systems that is
	far less than the number of dark matter sub-halos inferred from
	numerical simulations \citep{Briggs_1997,2000ASPC..218..223R,2001MNRAS.322..486B,Giovanelli_2005,NAN_2011,2019SCPMA..6259506Z,1999ApJ...522...82K}. With
	the superior sensitivity and large sky coverage, FAST should be able
	to detect thousands of low mass galaxies with HI masses less than
	$10^{8}$M$_{\sun}$, and therefore will be able to increase the
	abundance of low mass HI galaxies and thus provide a viable solution
	to the ``missing satellites''
	problem\citep{1999ApJ...522...82K,Moore_1999,Strigari_2007,Giovanelli_2005,Weinberg_2015}.Besides,
	according to the recent observations, some of the Milky Way(MW)
	subhalos predicted by $\Lambda$CDM model are significantly more
	massive than those resulted from stellar kinematics in its satellites,
	suggesting  that the MW is missing a portion of
	sub-halos, which is known as, the "too big to fail"(TBTF) problem
	\citep{Boylan_Kolchin_2011,Boylan_Kolchin_2012}. This phenomenon has
	been confirmed in M31 environment\citep{Tollerud_2014} as well as in a
	population of field dwarfs \citep{Ferrero_2012,2015A&A...574A.113P},
	which indicates that this paradox has nothing  with environmental
	effects. The effective way to investigate the
	connection between observations and $\Lambda$CDM predictions is
	offered by the rotational velocity
	function\citep{Giovanelli_2015,Peebles_2001,Klypin_2015}. The dark
	matter halo mass function corresponds to the rotational velocity
	function of galaxies. The rotational velocity of a galaxy can be
	inferred from its velocity width of HI profile as the HI gas can
	extend further than any other directly observable component. A large
	number of galaxies are required to derive the HI velocity function and
	to tackle the TBTF problem
	\citep{Baldry_2008,Zwaan_2010,Papastergis_2011,Cattaneo_2014,Giovanelli_2015,2018ApJ...862...48A,Papastergis_2013,Zhang_2020}.

	In order to test our ability to discriminate cosmic signals from
	radio-frequency interference and to optimize data-taking and
	calibration procedures, we have carried out a pilot HI survey. This
	paper present the first data release from this pilot HI survey, in a
	catalog containing  544 HI extragalaxies from the sky region in
	$+24^{\circ}< \rm \delta<+43^{\circ}$ . Among them, 302 HI detections
	are also presented in th ALFALFA catalog over the same sky area at the
	range $ \rm +24< \rm \delta < +36^{\circ}$\citep{Haynes_2018}, which enable us
	to make full comparaison for which measured by the FAST and ALFALFA. The
	FAST HI pilot survey was designed to cover a vast variety of cosmic
	environments, which includes several nearby high density regions like the
	Virgo superclusters in nearby Universe\citep{Saintonge_2008}.
	
	The rest of the paper is organized as following. In section  2,we briefly
	introduce the key FAST survey parameters and describe sky area covered, give an overview of the FAST observation and data processes and present a search of 544 detections. In section 3, we present the statistical properties of  these detections and  list 16 sources that no optical counterparts to meet them and discuss the implications of these  galaxies. Finally in section 4 we summarise this
	work.  We assume $ H_0$ (Hubble Constant) =70
	kms$^{-1}$Mpc$^{-1}$,$\Omega_m$ (Density parameter of matter) = 0.3
	and $\Omega_{\Lambda}$(Density parameter of dark energy) = 0.7 and we
	use nature unit for the light speed  throughout the paper.
	
	\section{\rm The Data }
	\subsection{\rm Sky Coverage }
	
\begin{figure}[ht]
		\centering
		\includegraphics[width=0.7\textwidth]{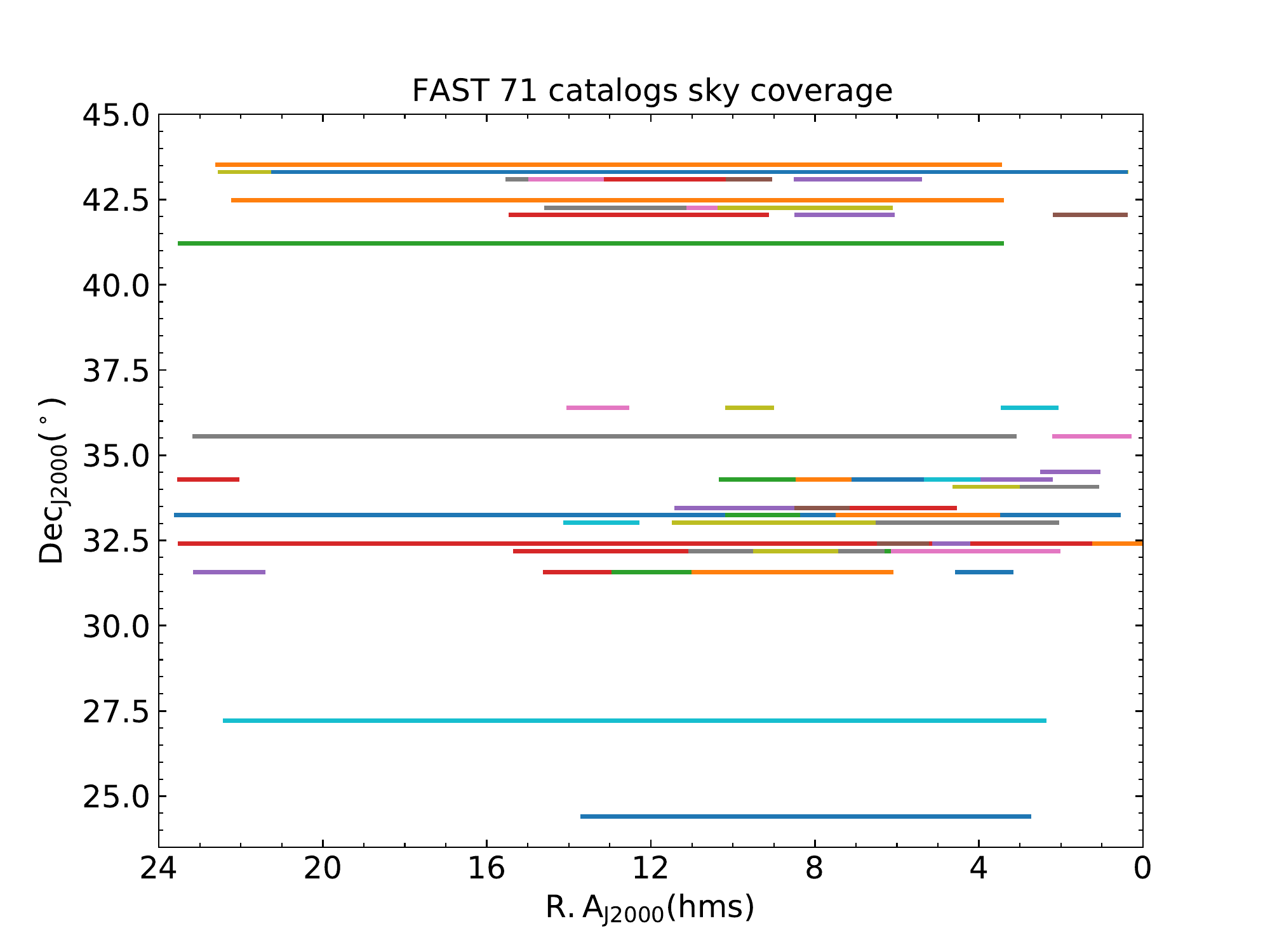}
		\caption{ The 71 datacubes sky coverage for this catalog  in $\rm  + 24^{\circ}<Dec< +43^{\circ}$ declination strip,the each Dec line width about 20-30 arcmins.\label{skycover}}
	\end{figure}

	Figure.\ref{skycover} present the footprint of 71 discontinuous
	datacubes in this data release of the pilot HI survey. 
	They are distributed in the range of $\rm +24^{\circ} < \delta < +
	43^{\circ}$ and 
	$00^{\rm h}47^{\rm m} <R.A.<   + 24^{\rm h}00^{\rm m}$.  These
	specific regions were surveyed as their zenith angles which requires  FAST telescope parameters like  sensitivity or resolution under the optimized status. By drift scanning sky areas over a larger
	R.A. range, and to test the performance of the whole FAST equipment in
	a varity of conditions during the pilot survey
	\citep{NAN_2011,jiang2019commissioning,Zhang_2020}. In order to obtain
	some early science achievements, high density region of clusters in the northern sky  are targeted as a foremost  to discover meaning physics. 
	
	\subsection{\rm FAST Observations and Data Reduction}
	The data reported here was taken during the period of May 2020 to Aug
	2021, and the pilot HI survey was carried out as a time-filler project
	when there are no other programs in the FAST observing queue. Although
	such type of project can not be planned for specified targets, we can
	still do blind search for HI 21cm emission. Our observational set-up was
	fixing the sky declination, and do drift scans during the project
	time. Each scan lasts for about 1-8 hours. We also use some
	commissioning time to do calibration observations.  The FAST HI survey
	is carried out with the drift scan mode using FAST's focal-plane
	19-beam receiver system, which is set in a hexagonal array and works
	in dual polarization mode, with a frequency range from 1050 MHz to
	1450 MHz. The 19-beam receiver was rotated by 23.4 degree so that the
	beam tracks are equally spaced in Declination with $1.14'$
	spacing\citep{Li_2018}. For the backend, we choose the Spec(W)
	spectrometer which has 65536 channels covering the bandwidth of 500
	MHz for each polarization and beam, with a velocity spacing of 1.67
		km/s and a spectral resolution, after Hanning smoothing,  about 5 km/s
	which is sufficient for resolving fine spectral structure and
	obtaining accurate column densities and velocity fields.
	
	The raw data output from each drift scan observation were reduced
	using a python based data redution pipeline HIFAST which was developed
	by Jing et al. (in preparation). This pipeline consists of the
	following steps:(a) identify the calibratoin signal and calibrate the
	unit of the spectrum into kelvin. The calibration signal was from a
	standard 10K noise diode which is injected every 32 seconds. The
	detailed procedure was described in \citep{10.1093/mnras/stab754};(b) convert the
	telescope pointing parameters to the precise position of R.A. and Dec
	for each spectrum ;(c) subtract the baseline for each spectrum using
	the asymmetrically reweighted penalized least squares algorithm
	\citep{2015baek};(d) calibrate the flux unit from $T_A$ Kelvin to Jy
	using the conversion factor 16 K/Jy \citep{jiang2019commissioning};(e)
	correct the Earth's rotation and convert the redshift velocity from
	local standard of rest to the helicenter velocoty in Equatorial
	coordinate system;(e) finally,create the three-dimensional (3D) datacubes consist of
	R.A. and Dec, and Velocity in FITS data format. Once a data file is
	processed by the pipeline, the drift scan data were gridded to form
	3D data cubes with $1'$ spacing. To limit the size
	of the datacube, the velocity coverage of the datacube for the pilot
	survey is limitted to the range from $\rm -2000 $ to 19,000 \kms.
	
	Finally, we perform interactive search for 21cm signals from a galaxy in 71
	datacubes as shown in Figure 1. The datacubes were displayed with the
	Starlink\footnote{http://starlink.eao.hawaii.edu/starlink} software
	package and the HI emission sources were identified by human eyes and
	extracted manually. The final list of the HI catalog containing 544 sources 
	is presented in Table 1.
	
	\subsection{\rm Optical counterparts of the HI detections}
	For each HI detected source, we search for its optical counterparts using
	the NED database. Fields of $4' \times 4'$ around each HI detection
	were inspected. The counterparts were usually found within a circle of
	1.5 arcmin radius centering at the HI source position, and the offset bewteen optical
		velocity and HI radio velocity is less than 300 km/s and the two redshifts is almost equal with error less than 0.001. Once optical sources are found within the search cone, we choose those sources with the type "G" standing
		for a galaxy in NED.  Most of the best matched counterparts were found
	to be from various existing catalogs such as the UGC catalog and
	NGC/IC catalog \footnote{http://www.ngcicproject.org/}, WISE
	catalog \footnote{https://irsa.ipac.caltech.edu/frontpage/},
	2MASS\footnote{http://egg.astro.cornell.edu/alfalfa/data/index.php}
	catalog ,KUG\footnote{http://dbc.nao.ac.jp/cjads.html} catalog,
	MRK\footnote{https://heasarc.gsfc.nasa.gov/W3Browse/rosat/markarian.html}
	catalog, SDSS\footnote{http://skyserver.sdss.org}
	catalog,MGC\footnote{https://heasarc.gsfc.nasa.gov/W3Browse/rosat/markarian.html}
	catalog, PGC\footnote{http://cseligman.com/text/atlas/pgcmisc.htm} and
	GALEX\footnote{http://www.galex.caltech.edu/researcher/data.html}
	catalog,and
	CGCG\footnote{https://heasarc.gsfc.nasa.gov/W3Browse/galaxy-catalog/uzc.html}.
	We also found that 302 FAST detections have been detected by ALFALFA and
	reported in the Arecibo General Catalog
	(AGC) \footnote{http://egg.astro.cornell.edu/alfalfa/data/index.php}.
	
	There are a few objects whose counterparts have not been attributed to
	the catalogs listed above, but they match equatorial coordinates and velocity of
	the FAST detected HI sources properly.  These galaxies were found in the
		VizieR Online Data Catalog. Notes to individual galaxies are as listed in Table \ref{tab:no_catalogs}:
	\begin{table*}[ht]
		\vspace{2mm}
		\begin{center}
			\caption{ HI detections and its optical counterparts without specical catalogs\label{tab:no_catalogs}}
			\resizebox{\textwidth}{1.5cm}{
				\setlength{\tabcolsep}{0.2cm}{
					\begin{tabular}{cccccccccc}
						\hline\hline
						FGC ID& Optical ID  & $\alpha_{\rm J2000}$(HI) &$\delta_{\rm J2000}$(HI)& $\alpha_{ \rm J2000}$(Opt.)& $\delta_{\rm J2000}$(Opt.)& position offset &  cz (FAST HI) & cz (counterpart HI)& velocity offset\\
						&     &  &    & &  & (arcmins)&\kms & \kms    &  \kms   \\
						(1) &  (2) &  (3) & (4)  & (5) &(6)&(7) &(8) &(9)&(10)     \\
						\hline
						83	&[BKB95] 0242+4236B	&2 : 45 : 56.40	&42 : 48 : 18.00&02h45m56s&+42d48.6m05s& 0.309 &  	5039.3& 5192.1& 152.8	\\
						331	&	LSBCF508-03	&	13 : 18 : 5.76	&24 : 44 : 2.40	&	13h18m07.58s&  +24d 44m 31.20s &  1.203  & 	2725.3  & 2793.5	& 68.2\\
						358	&	SBS1415+437	&14 : 17 : 3.36	&	43 : 29 : 42.00	&	14h17m01.408s	& +43d30m05.47s  & 1.13  & 637	&616.1	& 20.9	\\
						376	&[TSK2008]1251	&15 : 15 : 5.04	&	42 : 4 : 48.00	&	15h 15m 3.12s	& +42d 4m 22.80s & 0.837&2388.8 & 2549.1& 160.3	\\
						383	&	KKR 65	&	22 : 2 : 6.72	&42 : 8 : 56.40	&	22h 2m 6.00s	& +42d 8m 56.40s &  0.143 &4356.5  & 4426.3 &69.8	\\
						388	&	LCSBS2687P&22 : 40 : 16.32	&34 : 38 : 31.20	&	22h 40m 14.79s	&  +34d 38m 41.71s  & 0.976 & 8404.1	& 8295.2 &108.9\\
						\hline	
			\end{tabular}}}
		\end{center}
	\end{table*}
	\begin{itemize}	
		\item Col. 1 :   FAST galaxies catalog identification (FGC ID)  .
		\item Col. 2 : Optical counterpart identification (Optical ID) number for these catalog detections
		\item Col. 3 and  Col. 4: The R.A. and Dec of FAST galaxies.
		
		\item Col. 5 and  Col. 6: The R.A. and Dec of  Optical counterpart.
		
		\item Col.7: the position offset of FAST and optical counterparts .
		\item Col.7 and  Col 8 : FAST source HI velocity and optical counterparts velocity.
		\item Col.9 : velocity offset bewteen FAST and counterparts.
	\end{itemize} 
	
\begin{figure}[ht]
		\centering
		\includegraphics[width=0.8\textwidth]{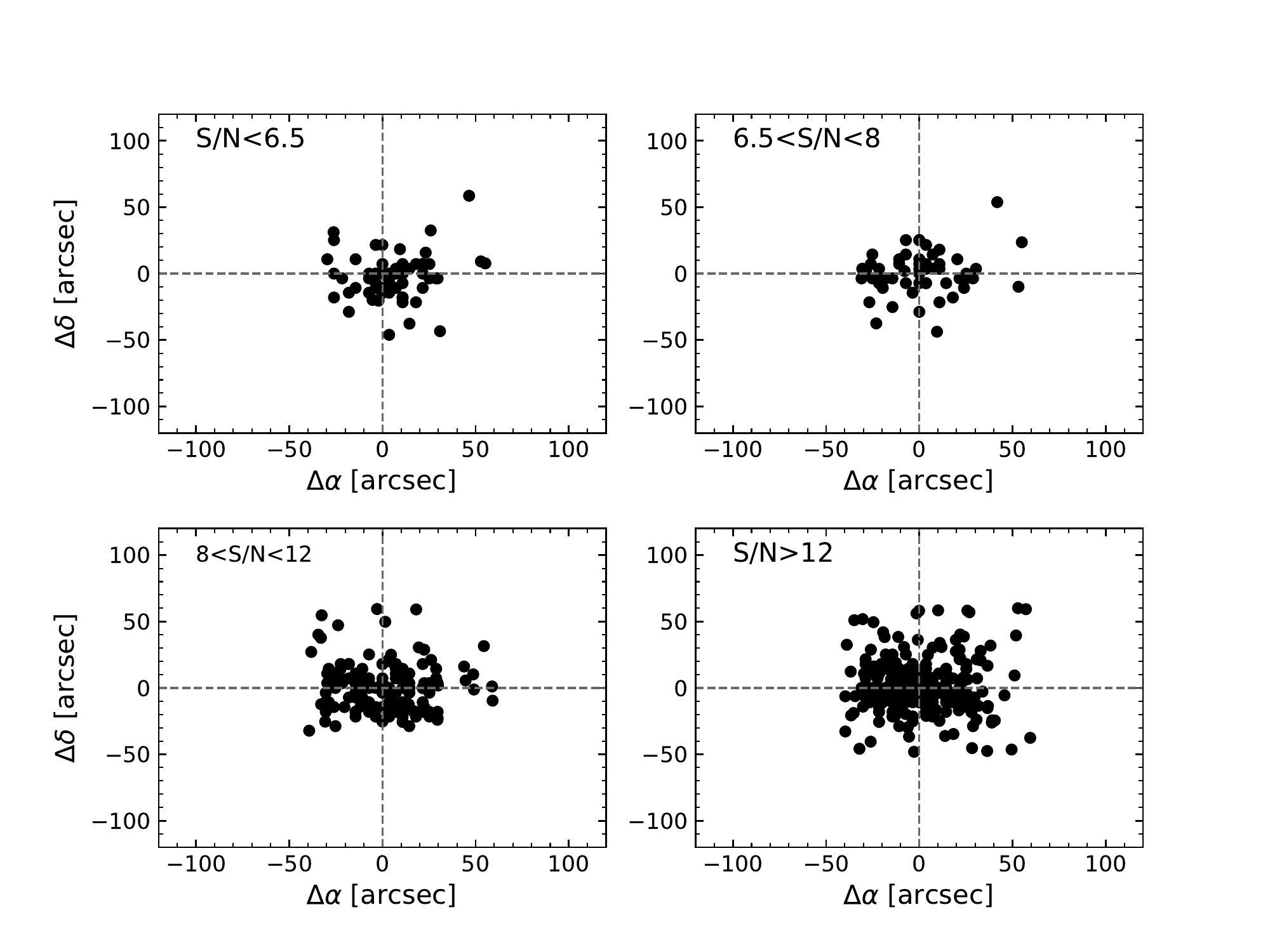}
		\caption{Pointing offsets between measured HI coordinates and the coordinates of the corresponding  527 optical counterparts in four different $S/N$ bins and for which, the mean pointing error in (R.A, Dec) coordinates in unit of arcsecond(\arcsec)  are  (19.4,17.8), (20.5,16.2), (21.6,18.1)  and (20.8,20.1),respectively for four plots from top-left to bottom-right. \label{opt_comp}}
	\end{figure}

	With the drift scan mode, most of the FAST observations are sampled at
	a fixed azimuth angle, and the zenith angle varies for different
	sources. To investigate possible dependence of the overall technical
	performance of the FAST telescope on pointing directions, we check the
	positional accuracy of the sources by comparing the HI detected
	position with the optical counterpart's position.  The foremost key
	limitation factor for the positional accuracy of FAST sources is the
	resolution of the FAST beams, which are in an approximately circle shape,
	with half power full widths of $2.95'$.  The second key parameter
	affecting the quality of the FAST position accuracy is the S/N of the
	HI emission. High-S/N sources locate in more accurate centroids than
	those low-S/N ones.  Positional accuracy can also be affected by other
	factors, including the systematic pointing offsets of telescope,
	asymmetry in the HI profiles and centers of the optical and HI
	emission, statistical errors in the HI sources and mismatch of optical
	counterparts. Figure.\ref{opt_comp} shows the position offsets between
	these HI sources and optical counteparts in the detected 527 sources,
	represented by black points. Each point stands for the offset of
	optical source's RA and DEC coordinates from the center of a
	source coordinates in four different S/N panels.  The average offset is about 
	$20''$. Such accuracy is excellent considering the 
	$2.9^{\prime}$ size of FAST beam.  In the panel of S/N
	objects (S/N \textgreater 12), the mean positions offset for R.A and Dec are 20.8
	and 20.1 arcseconds, respectively. As for velocity offset bewteen copunterpart  and HI emission for 544 sources are within 300 \kms with the largest value 218.6 km/s and average is 94.7 \kms

	\section{\rm Results}\label{results}
	
	\subsection{\rm Catalog Presentation}
	
	Here, we list 544 detections together with their optical counterparts in Table
	\ref{tab:HI_sources}. The paper version lists 50 representives objects
	and the rest table are available online. All detections are obtained
	from 71 different data cubes with radio frequency interference(RFI)
	and background level abroadly removed.  Table \ref{tab:optical} show
	50 of 527 sources and their optical counterparts with optical velocity, and the remaining HI sources and counterparts are
	available online.  There are 16 sources that have been not matched optical
	counterparts so far as listed in table \ref{tab:no_optical}.
	The contents of the different columns in the table 1 and 3 respectively are:
	\begin{itemize}
		\item Col. 1: FAST galaxies catalog identification (FGC ID).

		\item Col. 2: Optical counterpart identification (Optical ID) number for these catalog detections.
		
		\item Cols. 3 \& 4: Centroid position of each HI source in R.A. and Dec, J(2000). The positional average accuracy is about 20\arcsec and the accuracy of the HI positions depends on source emission intensity. 
		
		\item Cols. 5 \&6: J(2000) Dec and R.A. of the optical counterpart of the HI detection.  Counterpart objects are examined based on spatial offset, morphology, galaxy type, magnitude, and velocity or redshift. If a HI source detection's optical counterpart can not be found in the optial survey catalogs, the corresponding optical ID is kept empty.
		
		\item Col. 7: Heliocentric redshift velocity of the HI sources, $cz_{\odot}$, taken at  midpoint between the channels where the flux density reaches maximum value of the two flux peaks or one,  Units are \kms.
		
		\item Col. 8: Velocity width of the spectrum line profile, $\rm W50$, measured at the 50\% level of the flux peaks, as in the description of Col. 6. 
		
		\item Col. 9: Integrated flux density of the source, $F_c$, in Jansky \kms (Jy \kms). This is measured on the integrated full spectrum line after removing the baseline and background noise. Note that the uncertainty in the total flux calibration is about 10\%, mostly due to the fact that no flux calibrators were measured during many of the drift scan observations, and we only use the average conversion factor 16 K/Jy to convert from T$_A$ scale to Jy scale. 
		
		\item Col. 10:  Signal-to-noise ratio S/N of a detection for integral flux, the error in integral flux for each source usually is resulted from three parts: rms noise of each channel and the error from baseline fitting rms and integral range in line width W50. The total error is quantitatively estimated as the $\rm \sigma(F_c) = 4(S/N)_p^{-1}(F_{p}F_c\delta v)^{1/2} $\citep{2004AJ....128...16K}, the S/N for integral flux can be estimated as the format:
		\begin{equation}
		\rm S/N=\frac{F_{c}}{\sigma(F_c)}
		\end{equation}
		where $\rm F_c$ is the integral flux density in Jy \kms,where $\rm (S/N)_p = F_p/\sigma_{\rm rms} $ is peak to noise ratio, $\delta v$  is 5 km/s of velocity resolution.  The detections are classified into four categories 
		based on their S/N and  baseline qualities with code 1 to 4 :(1) 422 sources
		with signal of S/N \textgreater 6.5; (2) 61 sources with signal of $\rm
		5.1\lesssim S/N\lesssim 6.5$; (3) 28  sources with relatively poor baselines; (4) 33 sources that are partly masked  by strong RFIs.
		
		\item Col. 11: Background noise rms about 5 \kms velocity resolution with rebinning  cubes per 3 channels.
		
		\item Col. 12: Object distance in unit of milli-parsecs (Mpc) from observational position, $D_{\rm Mpc}$. The distance is simply  $cz/H_0$; $cz_\odot$ is the recession velocity measured in the Cosmic Microwave Background reference frame and $H_0$ is the Hubble constant, for which we use a value of 70 \kms Mpc$^{-1}$.
		
		\item Col. 13: Log10 values of HI mass in solar unit of a source. That parameter is obtained using the expression $M_{HI}=2.356\times 10^5 D_{\rm Mpc}^2 F_c$\citep{Giovanelli_2015}. 
		
		\item Col. 14: The corresponding entry number in the Arecibo General Catalog(AGC), a special catalog of extragalactic objects  measured by M.P.H. and R.G.\citep{Haynes_2018}.
		
		\item Col.15: Object code, sources with $ \rm (S/N)_p \geq 6.5 $ are labeled code 1, sources with $5\rm \leq    (S/N)_p  < 6.5$   are recorded as code 2 , and sources with relatively poor baselines or that are partly masked  by strong RFIs are flagged with code 3 and 4,  respectively.
		
	\end{itemize}
	\begin{table*}[ht]
		\vspace{2mm}
			
			\caption{ 50 HI  detections with its optical counterparts\label{tab:HI_sources}}
			\resizebox{\textwidth}{7cm}{
				\setlength{\tabcolsep}{0.1cm}{
					\begin{tabular}{ccccccccccccccc}
						\hline\hline
						FGC ID&Optical ID &$\alpha_{\rm J2000}$(HI) &$\delta_{\rm J2000}$(HI)& $\alpha_{ \rm J2000}$(Opt.)& $\delta_{\rm J2000}$(Opt.)& cz$_\odot$& W50&$\rm F_{\rm int}$&S/N&$\sigma_{\rm rms}$&Dist&$\rm logM_{\rm HI}$& AGC ID & Code\\
						&     &      &      &     &    & (\kms)  & (\kms) & (Jy \kms) &  &(mJy) & (Mpc) & ($\rm M_{\odot}$)&  \\
						(1) &  (2) &  (3) & (4)  & (5) & (6)& (7)            &           (8)  & (9)            &(10) & (11) & (12 )& (13) & (14) &(15)\\
						\hline
						1	&	UGC12898	&	0 : 0 : 34.56	&	33 : 35 : 38.40	&	0h 0m 37.44s	&	+33d 36m 3.60s	&	4638.3	&	168.1	&	3.0	&	15.6	&	2	&	66.3	&	9.5	&		&	1	\\
						2	&	UGC12920	&	0 : 2 : 23.04	&	27 : 12 : 10.80	&	0h 2m 23.04s	&	+27d 12m 39.60s	&	7319.4	&	276.5	&	3.0	&	18.9	&	1.5	&	104.6	&	10	&	12920	&	1	\\
						3	&	UGC00063	&	0 : 7 : 50.40	&	35 : 57 : 39.60	&	0h 7m 50.88s	&	+35d 57m 57.60s	&	450.0	&	73.3	&	3.6	&	18.3	&	2.8	&	6.4	&	7.4	&	63	&	3	\\
						4	&	UGC00069	&	0 : 8 : 9.84	&	27 : 31 : 33.60	&	0h 8m 11.04s	&	+27d 31m 40.80s	&	4580.3	&	177.2	&	7.5	&	40.8	&	1.5	&	65.4	&	9.6	&	69	&	1	\\
						5	&	KUG0008+355	&	0 : 10 : 42.24	&	35 : 50 : 56.40	&	0h 10m 41.76s	&	+35d 50m 56.40s	&	6058.6	&	127.9	&	2.5	&	8.2	&	2.8	&	86.6	&	9.4	&	100068	&	4	\\
						6	&	NGC0021	&	0 : 10 : 48.00	&	33 : 20 : 24.00	&	0h 10m 46.80s	&	+33d 21m 10.80s	&	4865.4	&	349.4	&	1.0	&	7.7	&	2	&	69.5	&	9	&	100	&	1	\\
						7	&	UGC00117	&	0 : 12 : 55.68	&	33 : 21 : 39.60	&	0h 12m 54.48s	&	+33d 21m 39.60s	&	4609.3	&	207.1	&	5.1	&	23.1	&	2	&	65.9	&	9.7	&	117	&	1	\\
						8	&	UGC00128	&	0 : 13 : 51.60	&	35 : 59 : 16.80	&	0h 13m 50.88s	&	+35d 59m 38.40s	&	4469.2	&	216.7	&	10.8	&	28	&	2.8	&	63.9	&	9.8	&	128	&	1	\\
						9	&	WISEAJ001713.37+271454.2	&	0 : 17 : 11.52	&	27 : 15 : 10.80	&	0h 17m 13.20s	&	+27d 14m 52.80s	&	3686.6	&	150.8	&	1.9	&	11.5	&	1.5	&	52.7	&	8.9	&	101815	&	1	\\
						10	&		&	0 : 17 : 15.36	&	42 : 7 : 55.20	&		&		&	4844.4	&	215.7	&	4.6	&	16	&	2.7	&	69.2	&	9.7	&		&	1	\\
						11	&	UGC00221	&	0 : 23 : 11.04	&	27 : 26 : 13.20	&	0h 23m 10.80s	&	+27d 25m 55.20s	&	3860.5	&	160.3	&	2.0	&	12.7	&	1.5	&	55.2	&	9.1	&	221	&	1	\\
						12	&	WISEAJ002638.16+360018.7	&	0 : 26 : 36.48	&	35 : 59 : 52.80	&	0h 26m 38.16s	&	+36d 0m 18.00s	&	7179.3	&	143.9	&	1.3	&	5.7	&	2.8	&	102.6	&	9.2	&	104539	&	3	\\
						13	&		&	0 : 28 : 32.40	&	33 : 15 : 50.40	&	0h 28m 34.32s	&	+33d 16m 19.20s	&	4570.7	&	214.2	&	4.0	&	17	&	2	&	65.3	&	9.6	&	284	&	1	\\
						14	&	UGC00288	&	0 : 29 : 5.28	&	43 : 26 : 24.00	&	0h 29m 3.60s	&	+43d 25m 55.20s	&	197.1	&	49.1	&	2.5	&	15.7	&	3	&	3.8	&	7.3	&		&	3	\\
						15	&	UGC00303	&	0 : 30 : 33.36	&	42 : 6 : 46.80	&	0h 30m 31.68s	&	+42d 6m 36.00s	&	5424.1	&	36.2	&	0.7	&	5.2	&	2.7	&	77.5	&	9	&		&	2	\\
						16	&	MRK0953	&	0 : 37 : 10.56	&	35 : 53 : 49.20	&	0h 37m 12.24s	&	+35d 54m 10.80s	&	4369.4	&	69.4	&	2.0	&	7.7	&	2.9	&	62.4	&	8.9	&	100346	&	1	\\
						17	&	UGC00376	&	0 : 37 : 54.72	&	32 : 41 : 13.20	&	0h 37m 54.00s	&	+32d 41m 20.40s	&	4734.9	&	205.7	&	8.4	&	30.7	&	1.9	&	67.6	&	10	&	376	&	1	\\
						18	&	UGC00384	&	0 : 38 : 20.64	&	32 : 38 : 42.00	&	0h 38m 22.08s	&	+32d 38m 16.80s	&	4615.7	&	186.7	&	7.2	&	26.3	&	2.4	&	65.9	&	9.9	&	384	&	1	\\
						19	&	UGC00394	&	0 : 38 : 43.92	&	42 : 0 : 18.00	&	0h 38m 43.44s	&	+41d 59m 49.20s	&	5393.5	&	60.8	&	0.8	&	5.4	&	2.7	&	77.1	&	9.1	&		&	1	\\
						20	&	NGC0226	&	0 : 42 : 53.76	&	32 : 35 : 2.40	&	0h 42m 54.00s	&	+32d 34m 51.60s	&	4715.5	&	137.3	&	7.7	&	33.9	&	1.9	&	67.4	&	9.9	&	459	&	1	\\
						21	&	IC0046	&	0 : 42 : 59.04	&	27 : 14 : 42.00	&	0h 42m 58.08s	&	+27d 15m 10.80s	&	5126.2	&	221.9	&	2.9	&	13.7	&	1.5	&	73.2	&	9.3	&	100482	&	1	\\
						22	&	MRK0346	&	0 : 44 : 54.96	&	27 : 26 : 52.80	&	0h 44m 56.88s	&	+27d 26m 60.00s	&	5058.6	&	144.4	&	1.5	&	10.2	&	1.5	&	72.3	&	9.5	&	100499	&	1	\\
						23	&	UGC00484	&	0 : 46 : 57.36	&	32 : 40 : 1.20	&	0h 46m 55.92s	&	+32d 40m 30.00s	&	4795.0	&	408.2	&	15.1	&	40.7	&	1.9	&	68.5	&	10.2	&	484	&	1	\\
						24	&	KPG017	&	0 : 47 : 1.20	&	32 : 41 : 31.20	&	0h 47m 3.12s	&	+32d 41m 2.40s	&	4953.9	&	79.3	&	2.7	&	18.1	&	2.4	&	70.8	&	9.5	&	1730	&	1	\\
						25	&	UGC00549	&	0 : 54 : 42.72	&	36 : 45 : 36.00	&	0h 54m 42.00s	&	+36d 45m 54.00s	&	6045.7	&	68.2	&	0.6	&	5.2	&	3.1	&	86.4	&	9	&		&	4	\\
						26	&	UGC00602	&	0 : 58 : 21.36	&	36 : 44 : 16.80	&	0h 58m 23.28s	&	+36d 43m 48.00s	&	6126.2	&	212.5	&	2.3	&	9.4	&	2.9	&	87.5	&	9.6	&		&	4	\\
						27	&	MCG +07-03-013	&	1 : 00 : 54.96	&	43 : 41 : 6.00	&	01h00m59.50s	&	+43d40m26.0s	&	5000.6	&	200.1	&	2.2	&	8.5	&	3	&	71.4	&	9.4	&		&	1	\\
						28	&	KUG 0059+356	&	1 : 2 : 38.40	&	35 : 53 : 38.40	&	1h 2m 39.60s	&	+35d 53m 49.20s	&	2184.3	&	84.1	&	1.4	&	6.5	&	2.9	&	31.6	&	8.4	&	115328	&	3	\\
						29	&	KUG 0108+356	&	1 : 11 : 34.56	&	35 : 52 : 8.46	&	1h 11m 34.80s	&	+35d 53m 20.40s	&	9446.6	&	129.0	&	1.5	&	6.3	&	2.9	&	135	&	9.7	&	115609	&	4	\\
						30	&	KUG 0109+357	&	1 : 11 : 54.72	&	36 : 2 : 9.60	&	1h 11m 52.80s	&	+36d 1m 48.00s	&	9505.0	&	99.2	&	1.8	&	52.2	&	2.9	&	135.8	&	10	&	115610	&	3	\\
						31	&		&	1 : 14 : 45.84	&	27 : 10 : 51.60	&		&		&	3570.7	&	100.0	&	5.0	&	22.1	&	1.5	&	51	&	8.8	&		&	1	\\
						32	&	CGCG502-074	&	1 : 24 : 9.84	&	32 : 45 : 28.80	&	1h 24m 10.80s	&	+32d 45m 57.60s	&	5871.7	&	147.4	&	2.3	&	10.8	&	2.2	&	83.9	&	9.6	&	110302	&	1	\\
						33	&	NGC0523	&	1 : 25 : 22.08	&	34 : 1 : 8.40	&	1h 25m 20.64s	&	+34d 1m 30.00s	&	4902.3	&	459.7	&	1.8	&	11.7	&	2.1	&	70	&	9.3	&	979	&	1	\\
						34	&	WISEAJ012622.06+323806.9	&	1 : 26 : 20.40	&	32 : 38 : 24.00	&	1h 26m 21.84s	&	+32d 38m 9.60s	&	4079.8	&	63.8	&	1.0	&	6.2	&	2.4	&	58.3	&	8.9	&	115203	&	1	\\
						35	&	NGC0573	&	1 : 30 : 49.44	&	41 : 15 : 21.60	&	1h 30m 49.44s	&	+41d 15m 25.20s	&	2775.2	&	120.7	&	2.4	&	12.3	&	2.5	&	39.7	&	8.9	&		&	3	\\
						36	&	WISEAJ013141.69+340858.2	&	1 : 31 : 42.48	&	34 : 8 : 56.40	&	1h 31m 41.76s	&	+34d 8m 60.00s	&	7095.5	&	59.8	&	0.8	&	6.1	&	2.1	&	101.4	&	9.3	&	115595	&	1	\\
						37	&	WISEAJ013154.04+271945.5	&	1 : 31 : 52.56	&	27 : 19 : 51.60	&	1h 31m 54.00s	&	+27d 19m 44.40s	&	3768.8	&	55.9	&	0.4	&	5.3	&	1.5	&	53.8	&	8.6	&	112516	&	2	\\
						38	&	CGCG522-006	&	1 : 47 : 42.96	&	35 : 0 : 52.92	&	1h 47m 43.68s	&	+35d 1m 22.80s	&	5475.6	&	91.7	&	3.7	&	18.2	&	2.8	&	78.2	&	9.7	&	110530	&	1	\\
						39	&	NGC0672	&	1 : 47 : 54.48	&	27 : 25 : 51.60	&	1h 47m 54.48s	&	+27d 25m 58.80s	&	406.0	&	195.4	&	120.4	&	455.8	&	1.5	&	5.8	&	8.5	&	1256	&	1	\\
						40	&	CGCG482-017NED01	&	1 : 48 : 35.04	&	27 : 33 : 18.00	&	1h 48m 35.28s	&	+27d 32m 52.80s	&	10473.9	&	131.5	&	2.4	&	15.3	&	1.5	&	149.6	&	9.8	&	110534	&	1	\\
						41	&	IC1731	&	1 : 50 : 12.48	&	27 : 11 : 52.80	&	1h 50m 12.24s	&	+27d 11m 45.60s	&	3536.9	&	174.8	&	6.7	&	26.9	&	1.5	&	50.5	&	9	&	1291	&	1	\\
						42	&	UGC01347	&	1 : 52 : 45.36	&	36 : 36 : 43.20	&	1h 52m 45.84s	&	+36d 37m 8.40s	&	5467.6	&	138.3	&	3.4	&	12.8	&	2.9	&	78.1	&	9.7	&		&	1	\\
						43	&	UGC01355	&	1 : 53 : 37.92	&	43 : 58 : 26.40	&	1h 53m 36.24s	&	+43d 57m 57.60s	&	6076.3	&	282.9	&	2.4	&	7.5	&	3	&	86.8	&	9.6	&		&	1	\\
						44	&		&	1 : 56 : 49.68	&	34 : 8 : 27.60	&		&		&	4739.3	&	385.4	&	2.7	&	18.4	&	2.1	&	67.7	&	9.5	&		&	1	\\
						45	&	WISEA J015708.84+354852.0	&	1 : 57 : 9.60	&	35 : 48 : 54.00	&	1h 57m 7.92s	&	+35d 48m 25.20s	&	4441.9	&	181.7	&	2.7	&	8.8	&	2.9	&	63.5	&	9.3	&	115626	&	1	\\
						46	&	UGC01422	&	1 : 57 : 8.16	&	32 : 47 : 20.40	&	1h 57m 6.72s	&	+32d 47m 20.40s	&	4367.7	&	267.3	&	0.4	&	5.7	&	1.8	&	62.4	&	8.6	&	1422	&	2	\\
						47	&	NGC0753	&	1 : 57 : 41.04	&	35 : 54 : 57.60	&	1h 57m 42.24s	&	+35d 54m 57.60s	&	4710.8	&	375.6	&	18.5	&	42.9	&	2.8	&	67.3	&	10	&	1437	&	1	\\
						48	&	KUG  0156+324	&	1 : 59 : 7.92	&	32 : 43 : 30.00	&	1h 59m 9.36s	&	+32d 43m 4.80s	&	4304.9	&	168.9	&	5.1	&	26.6	&	1.8	&	61.5	&	9.7	&	115273	&	1	\\
						49	&	UGC 01472	&	1 : 59 : 57.36	&	34 : 20 : 9.60	&	1h 59m 58.90s	&	+34d 20m 35.00s	&	4725.6	&	131.5	&	1.5	&	9	&	2.1	&	67.5	&	9.2	&	1472	&	1	\\
						50	&		&	2 : 6 : 30.96	&	43 : 51 : 36.00	&		&		&	5197.1	&	187.6	&	2.5	&	8.4	&	3	&	74.2	&	9.5	&		&	1	\\

						\hline	
\end{tabular}
} }

	\end{table*}
	
	In Table \ref{tab:optical}, we show the first 50 sources of 527 counterparts  with basic parameters, for which column (1) to column (6) are FGC ID, optial ID, HI line velocity, optical velocity, apparent magnitude and objects type,respectively. Here HI radio velocity has been transformed to optical velocity in columns (3).  
	\begin{table*}[ht]
		\vspace{2mm}
		\tiny
		\begin{center}
			\vspace{0.1mm}
			\caption{ 50 HI Candidate Detections and optical counterparts \label{tab:optical}}
			\resizebox{\textwidth}{10cm}{
					\begin{tabular}{ccccccccccc}
						\hline\hline
						FGC ID&Optical ID & HI Velocity(optical)&Optical velocity&Magnitude&Galaxy type& \\
						
						&     &       (km s$^{-1}$)  & (km s$^{-1}$) & 	(grb -band) &  \\
						(1) &  (2) &  (3) & (4)  & (5) & (6)&   \\
						\hline
						1	&	UGC12898	&	4711.1	&	4780	&	16.5	&	G	\\
						2	&	UGC12920	&	7502.4	&	7613	&	15.48	&	G	\\
						3	&	UGC00063	&	450.7	&	441	&	15.34	&	G	\\
						4	&	UGC00069	&	4651.4	&	4637	&	14.49	&	G	\\
						5	&	KUG0008+355	&	6183.4	&	6159	&	15.5	&	G	\\
						6	&	NGC0021	&	4945.6	&	4765	&	13.51	&	G	\\
						7	&	UGC00117	&	4681.2	&	4754	&	14.78	&	G	\\
						8	&	UGC00128	&	4536.8	&	4531	&	16.5	&	G	\\
						10	&	WISEAJ001713.37+271454.2	&	3732.5	&	3704	&		&	G	\\
						12	&	UGC00221	&	3910.9	&	3905	&	14.66	&	G	\\
						13	&	WISEAJ002638.16+360018.7	&	7355.3	&	7302	&		&	UvS	\\
						14	&	UGC00284	&	4641.4	&	4732	&	14.71	&	G	\\
						15	&	UGC00288	&	197.3	&	187	&	16	&	G	\\
						16	&	UGC00303	&	5524.0	&	5623	&	16.5	&	G	\\
						17	&	MRK0953	&	4434.0	&	4441	&	15.5	&	G	\\
						18	&	UGC00376	&	4810.8	&	4820	&	16	&	G	\\
						19	&	UGC00384	&	4687.9	&	4702	&	14.43	&	G	\\
						20	&	UGC00394	&	5492.3	&	5596	&	15.1	&	G	\\
						21	&	NGC0226	&	4790.9	&	4830	&	14.31	&	G	\\
						22	&	IC0046	&	5215.3	&	5286	&	14.75	&	G	\\
						23	&	MRK0346	&	5145.3	&	5186	&	16.5	&	G	\\
						24	&	UGC00484	&	4872.9	&	4864	&	13.86	&	G	\\
						26	&	UGC00549	&	6170.0	&	6062	&	15.79	&	G	\\
						27	&	UGC00602	&	6253.9	&	6145	&	14.68	&	G	\\
						29	&	KUG 0059+356	&	2226.4	&	2205	&	16.5	&	G	\\
						30	&	KUG 0108+356	&	9753.7	&	9789	&	16.5	&	G	\\
						31	&	KUG 0109+357	&	9816.0	&	9741	&	17.5	&	G	\\
						33	&	CGCG502-074	&	5988.9	&	6024	&	15.6	&	G	\\
						34	&	NGC0523	&	4983.8	&	4761	&	12.3B	&	G	\\
						35	&	WISEAJ012622.06+323806.9	&	4136.0	&	4135	&		&	UvES	\\
						36	&	NGC0573	&	2801.1	&	2788	&	14.1	&	G	\\
						37	&	WISEAJ013141.69+340858.2	&	7267.4	&	7262	&		&	IrS	\\
						38	&	WISEAJ013154.04+271945.5	&	3816.7	&	3820	&		&	G	\\
						39	&	CGCG522-006	&	5577.4	&	5559	&	14.79	&	G	\\
						40	&	NGC0672	&	406.6	&	429	&	11.47	&	G	\\
						41	&	CGCG482-017NED01	&	10852.8	&	10979	&	15.3	&	G	\\
						42	&	IC1731	&	3579.1	&	3503	&	14	&	G	\\
						43	&	UGC01347	&	5569.1	&	5543	&	13.49	&	G	\\
						44	&	UGC01355	&	6201.9	&	6322	&	13.97	&	G	\\
						46	&	[DF2014] 01	&	4508.6	&	4592	&		&	G	\\
						47	&	UGC01422	&	4432.3	&	4583	&	14.35	&	G	\\
						48	&	NGC0753	&	4785.9	&	4858	&	12.97	&	G	\\
						49	&	KUG  0156+324	&	4367.6	&	4406	&	16.5	&	G	\\
						50	&	UGC 01472	&	4801.2	&	4849	&	16.07	&	G	\\
						52	&	UGC01601	&	5483.9	&	5591	&	15.49	&	G	\\
						53	&	UGC01602	&	5580.8	&	5489	&	16.5	&	G	\\
						54	&	UGC01626	&	5497.2	&	5543	&	14.11	&	G	\\
						55	&	KUG  0206+355	&	4985.5	&	4973	&	16	&	G	\\
						56	&	UGC01729	&	4492.0	&	4445	&	15.12	&	G	\\
						57	&	UGC01738	&	5570.8	&	5686	&	15.48	&	G	\\
						58	&	IC1784	&	5005.4	&	4817	&	14	&	G	\\
						
						\hline	
			\end{tabular}}
		\end{center}
	\end{table*}

	\subsection{\rm General Properties of the Detections}
	We first describe some basic properties of the 544 HI detections by
	FAST.  About 28.9\% of FAST detections have cz \textless 3000
	\kms. The detection rate decreases when the FAST survey scans a sky
	region much larger than the supergalactic plane and extends to the low
	density environment outside the northern part of the Virgo Cluster.
	In this data release we did not include any sources that are in any doubt of RFIs. Many tentative detections need to be confirmed with future observations and are not included in this catalog. Thus the HI detection rate with single pass of drift scan is much lower than that of ALFALFA. We estimate a detection rate of about 0.8 per square degree for the pilot FAST  survey area  of about 681  square degree, which is much lower than the average rate of 5.4 objects per square degree found in
	\citep{Haynes_2018} where the survey area crosses the northern part of
	the Virgo supercluster.
	
	\begin{figure}[ht]
		\centering
		\includegraphics[width=0.7\textwidth]{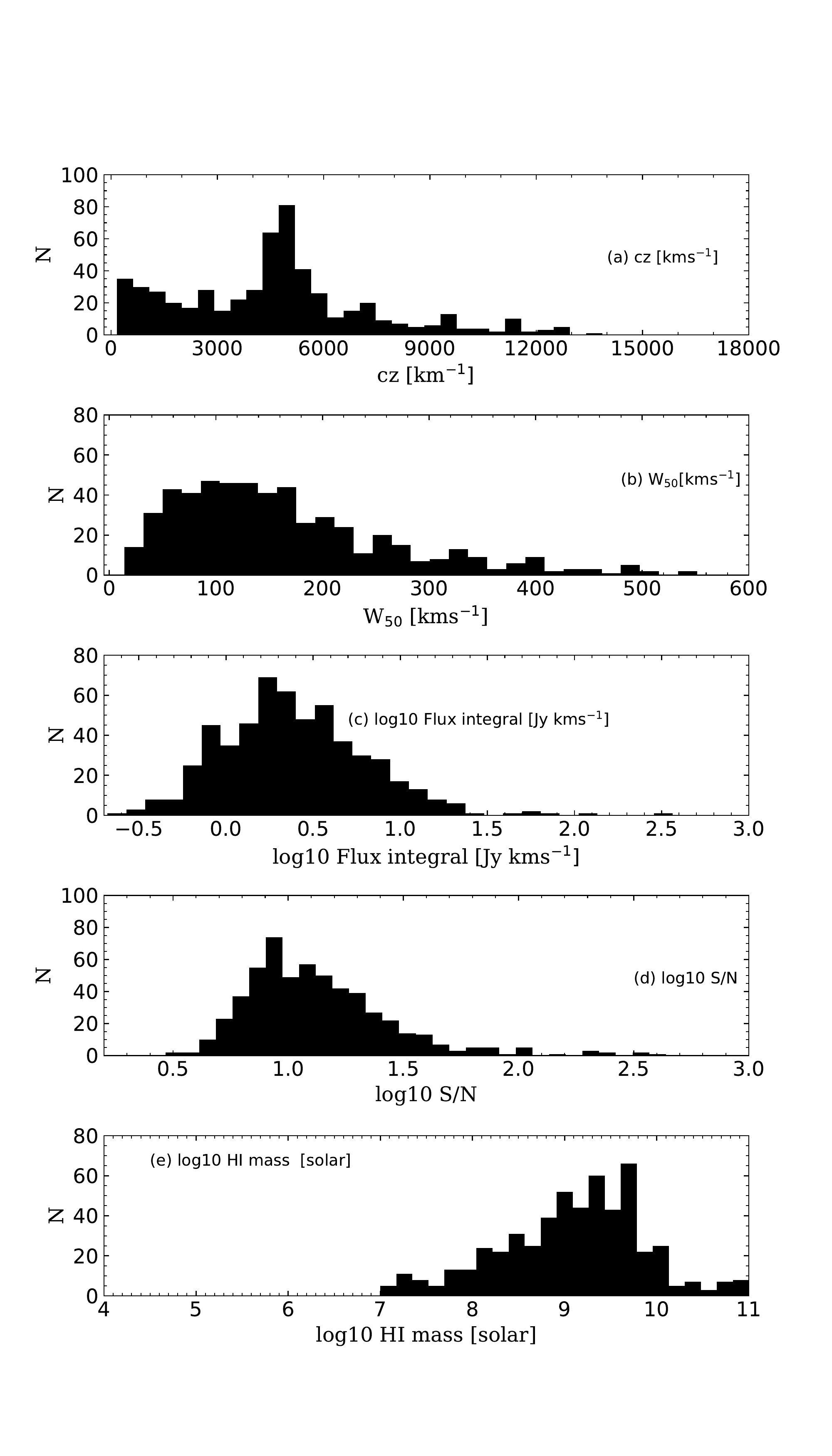}
		\caption{Histograms of the HI detections from top to buttom: heliocentric recessional velocity in \kms,  HI line width at half power ($W50$) in \kms, logarithm of the flux integral in Jy \kms, logarithm of the signal-to-noise ratio, and logarithm of the HI mass in solar units. \label{hists}}
	\end{figure}
	
	Figure \ref{hists} presents the distribution in values of
	heliocentric velocity cz, velocity width W50,S/N, and HI mass $\rm
	M_{HI}$ for the HI detections in our catalog. The
	distances are inferred from redshifts alone with the help of cosmic
	expansion dynamical theory. Due to the fact that many objects with less than 3000 \kms
		are likely to be located in the Virgo Cluster sky region or even the local group region, the model
		yields great uncertainties caused by the peculiar motions. Thus, the real distance could be different from that listed in Table 2. However, it would be easy to scale the data once a more reliable distance is obtained. 
		Plot (e)
		indicates somewhat over abundance in the sources with  HI mass less than $10^8 M_\odot$ in our catalog,  which could be due to the larger uncertainties in the distance measurements for the nearby sources.
	
	The median redshift of the velocity distribution is 4681.33 \kms and
	galaxies in this catalog have velocity widths $W50$ which varies from
	24.9 \kms \ to 539.3 \kms (see Fig.\ref{hists}). Some detected sources with narrow
	spectra turn out to be  nearby dwarf
	galaxie with low HI masses. The upper limit is at 539.3 \kms 
	for the line width, close to the upper-limit of the velocity width
	measured by previous HI samples
	\citep{2004AJ....128...16K,2005yCat.8077....0S,2006AJ....132.1426S}. It
		can also be seen that most of the line widths range from 100 to 200 \kms , the fraction of line width more than 300 \kms is significantly reduced and the largest line width is 539.3 \kms.  The third panel of Figure.\ref{hists}
	shows the integrated flux distribution. The distribution of fluxes
	ranges from 0.21 to 365.8 Jy \kms, and the median of the distribution
	is 2.3 Jy\kms. The signal-to-noise ratios of the galaxies in Table 2 range from 5.1
	to 436.8 with the lowest HI mass down to 7.1 logarithmic solar
	mass. As shown in the last panel of Fig.\ref{hists}, we found 36
	galaxies with $\log_{10}(M_{HI})<8.0$ and the median of the mass
	distribution is $\sim 10^{9.2}$ \msun, while the galaxy with
	the largest HI mass in this catalog is $\sim 10^{10.9}$
	\msun.
	\begin{figure}
		\centering
		\includegraphics[width=0.7\textwidth]{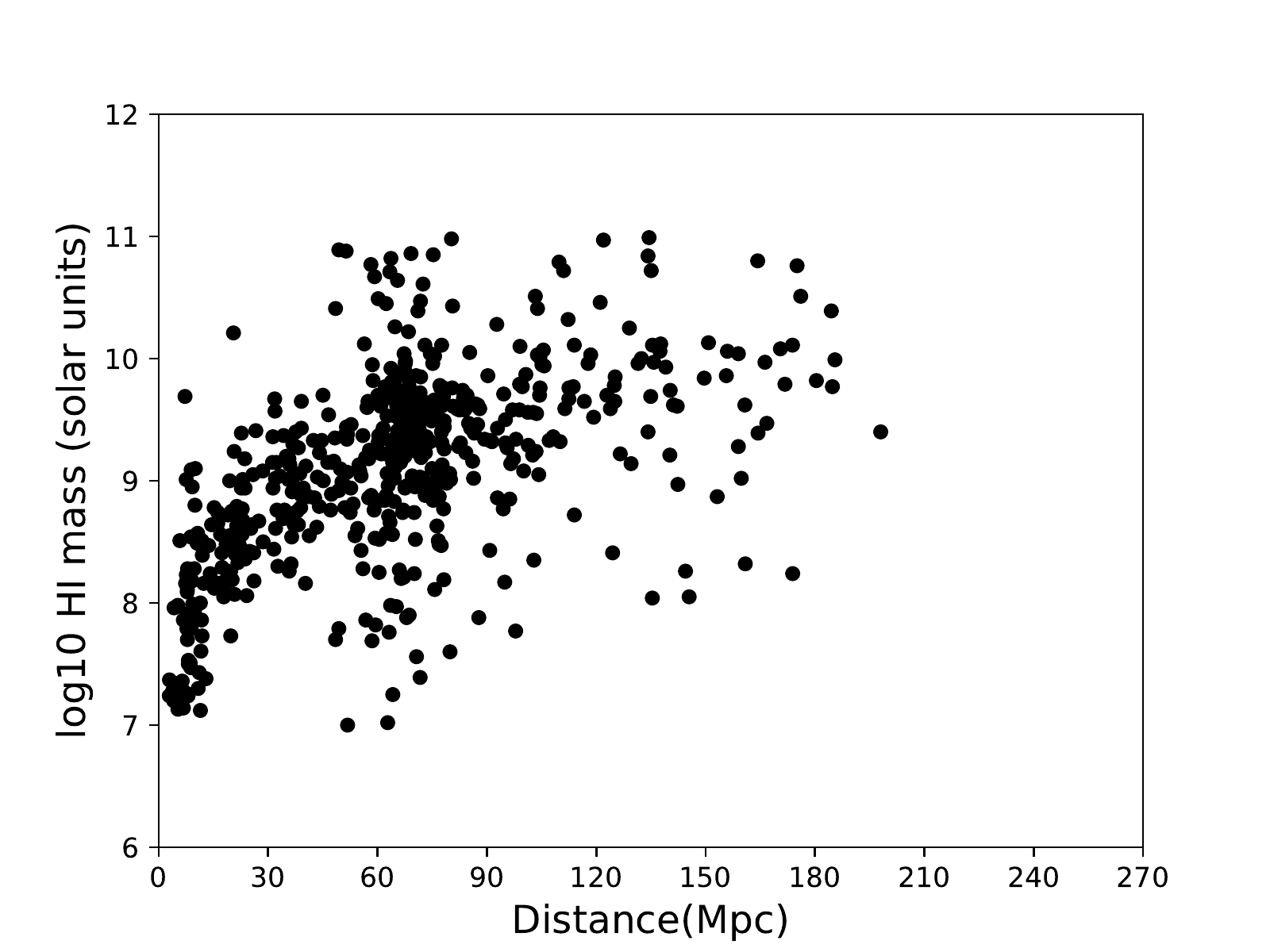}
		\caption{Log distribution of measured HI mass as a function of  recession distance. \label{HImass}}
	\end{figure}
	
	The distribution of the HI mass of the detections as a function of
	heliocentric recession velocity is presented in Figure
	\ref{HImass}.It shows that 72.4 percent of 544 sources were  dectected  within 120 Mpc  and the farthest distance is 185.5 Mpc with HI mass about $\rm 10^{9.9} M_\odot$. There are 489 sources  with HI mass less than  $\rm 10^{10} M_\odot$. 
	\begin{figure}[ht]
		\centering
		\includegraphics[width=0.7\textwidth]{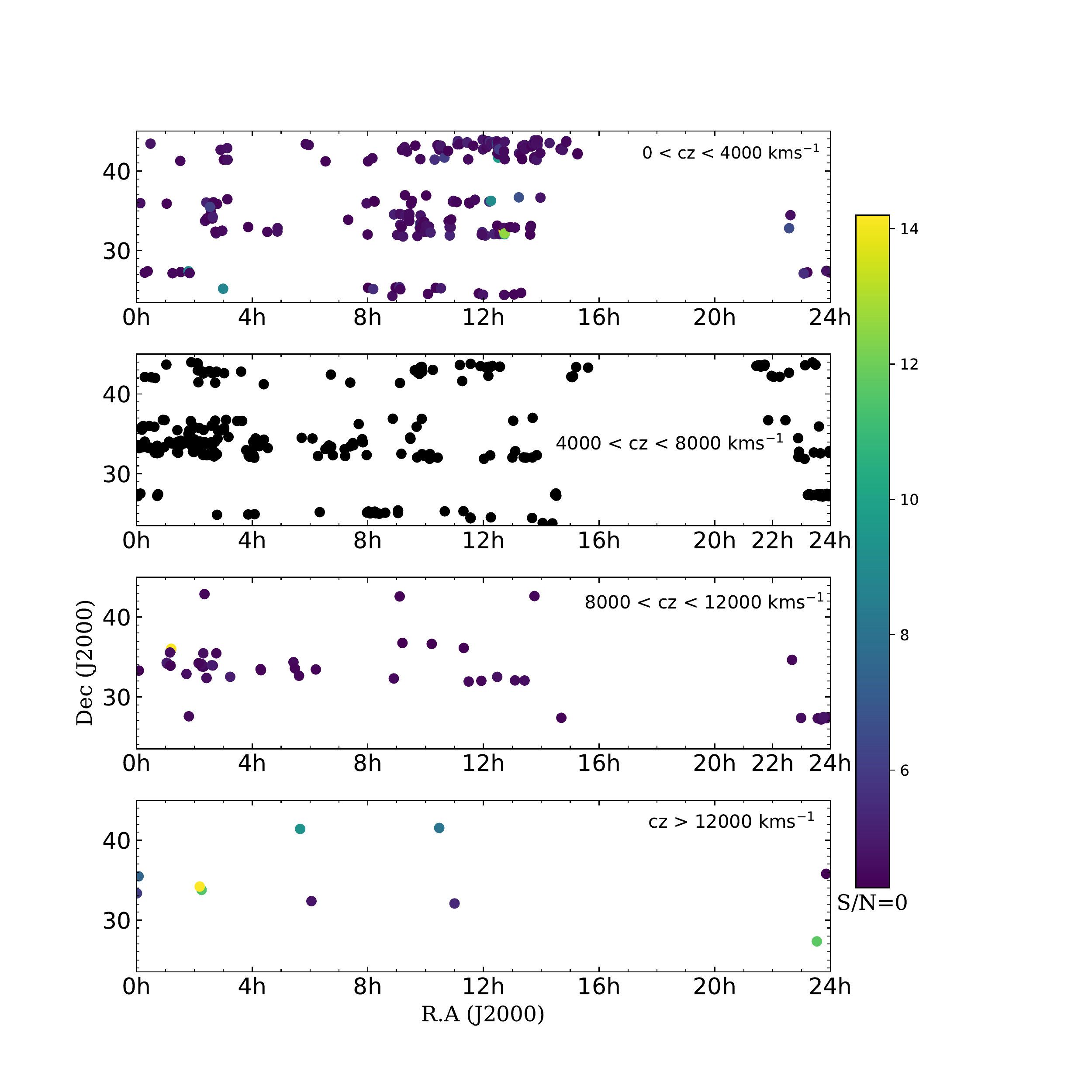}
		\caption{The velocity distribution of all dectections with the ratio of signal to noise in five different velocity $cz$ intervals, presented as color scatters.\label{fig:sn}}
	\end{figure}
	Furthermore, we show the S/N distribution in different velocity bins
	for all detections in Figure .\ref{fig:sn}.  Some galaxies, such as FGC 30,FGC 39, FGC
		87, FGC 209,FGC 314, FGC 318, FGC 321, FGC 327 have very high S/N
		($>100$), indicating that these galaxies have high HI fluxes. Some of them have very high HI masses, such as FGC 30, with a distance of 135.8 Mpc and HI mass of $\rm 10^{10} M_\odot$. In fact, a total of 55 gas rich galaxies  have HI mass greater than  $\rm 10^{10} M_\odot$.
		There are also many dwarf galaxies detected whitin the distance of 25 Mpc with HI mass larger than $\rm 10^{7} M_\odot$. FAST HI survey will detect large sample of these type of objects with its superior sensitivity and efficiency.
	
	\subsection{\rm The properties of the  FAST detections}

	The HI survey completeness can be defined as the fraction of cosmic
	sources of a given integrated flux density within the survey solid
	angle that are detected by a radio telescope\citep{2005b}.  Although the
	sky coverage of the pilot survey is small,  and the sky
	distributions of the datacubes are discontinuous, we can still try to 
	get a sense of the completeness of the FAST survey.
	
	Our results shows that the spectral noise level at 5 km/s resolution is about 2.5 mJy on average, 
	which reaches the  expected sensitivity of FAST for the one-pass drift scan 
	survey.  The detection threshold of a given spectral profile depends
	on both the shape and width of the profile. For wider profiles we can smooth the spectra further 
	and the rms noise can decrease accordingly in term of the rms value of $2.5\times(\rm res/5)^{1/2}$
	mJy, where res is the spectral resolution of the data after smoothing
	in unit of \kms.   The detection threshold is
	\citep{2005b} can be obtained with the formula:
	\begin{equation}
	\rm F_{th} = (S/N )\times (rms) \times W50  
	\end{equation}
	where S/N is a fiducial S/N  for a detection and rms
	is the root mean square background noise at a pixel after smoothing spectrum line, the noise rms in unit of Jy and the width in \kms. 
	
	Assuming a critical parameter of 200 \kms as the
	threshold width, for the one-pass drift scan of FAST survey, the
	empirical relationship between the integrated flux density detection
	threshold ($\rm S_{21,th}$,in Jy \kms and the profile width (W50, in
	\kms) from the pilot observations of FAST survey can be expressed 
	in terms of the S/N ratio as \citep{2005b,2011b}:\\
	\begin{equation}\label{eq:sn_t}
	S_{21,th}=
	\begin{cases} 
	0.11 \rm (S/N) \times (W50/200)^{1/2}, & W50 < 200 \\
	0.11 \rm (S/N) \times (W50/200), & W50 \ge 200
	\end{cases}
	\end{equation}
	\begin{figure}[ht]
		\centering
		\includegraphics[width=0.7\textwidth]{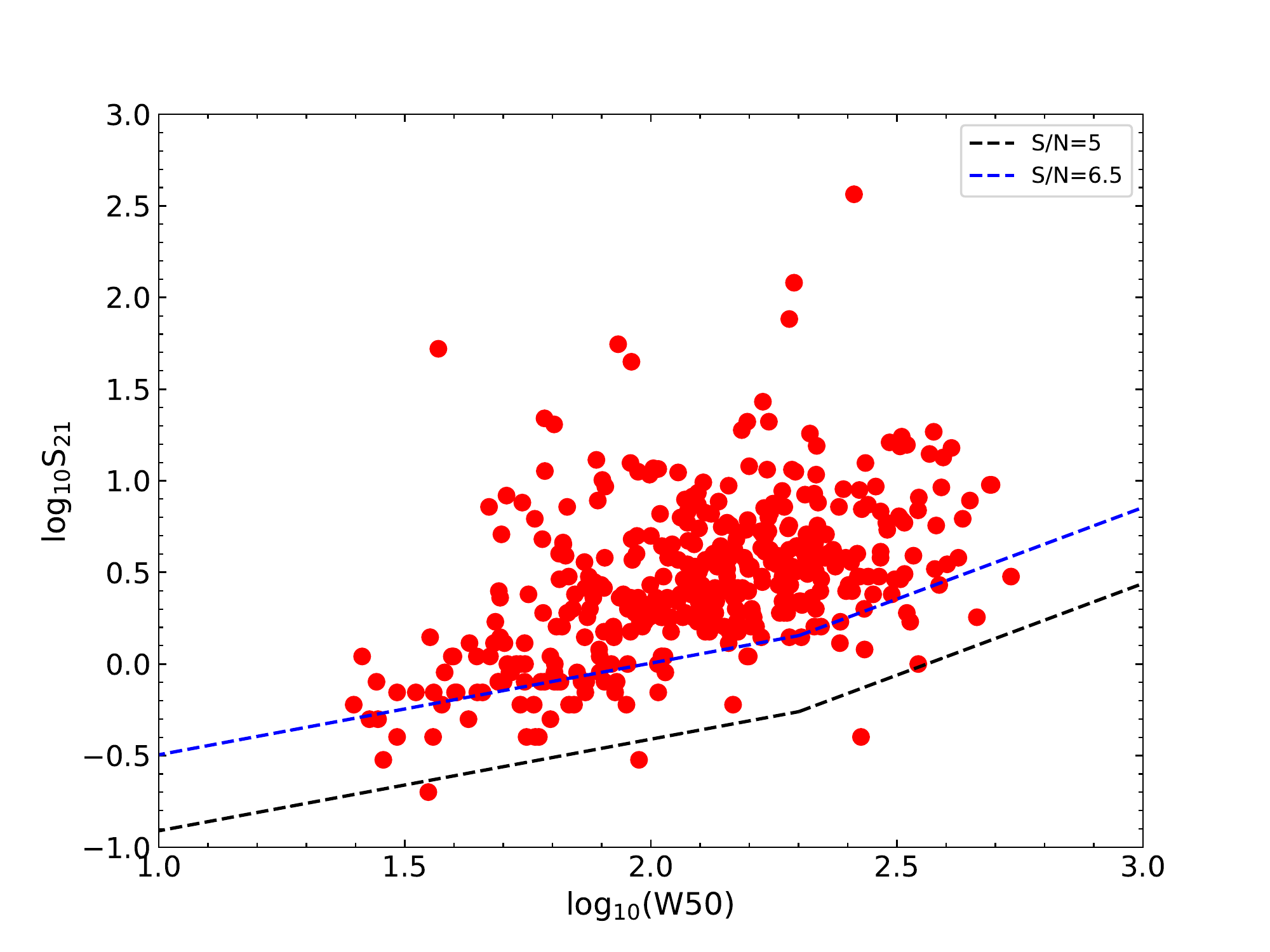}
		\caption{HI flux integral plotted versus velocity width, for the detected sample. The three dashed
			lines correspond to the flux threshold Fth one obtains for a signal-to-noise ratio of, respectively, 5 and 6.5, using Eqs.\ref{eq:sn_t}.\label{flux}}
	\end{figure}
	
	Note that the normalization factor of 0.11 here is different from that in Haynes et al. (2011) because the FAST data have a lower rms (2.5mJy at 5 km/s resolution) than that of ALFALFA (3.5mJy at 10 km/s resolution). It should also be noted that there are about 10\% of uncertainties in the flux calibration for the pilot FAST survey data due to the lack of observation time for calibrator measurements. 
	
	Figure \ref{flux} displays a plot of the HI integrated fluxes of the 
	HI detections versus their line width. This figure
	shows that  the flux sensitivity limit of the survey
	depends on the W50. Lower fluxes  can be detected 
	for smaller line widths in Eqs.\ref{eq:sn_t}. The blue and black dashed
	lines in Fig \ref{flux} show the possible detection thresholds at levels of
	S/N =5 and 6.5,respectively.  The
	median $\rm W50$ for a typical galaxy is 143.56\kms. The 
	flux is positively correlated with the line width. The integrate 
		flux traces the HI gas mass, while the line width traces the galaxy
		rotation curve which is related to  the total mass of the galaxy. These two parameters are naturally correlated with each other because the HI mass is correlated with the total mass of the galaxy\citep{2018ApJ...862...48A,2020ApJ...894...92G}.

	Since the data from the FAST pilot survey were reduced with a HI
	pipeline in-development, the RFI mitigation and standing wave removal
	depend heavily on human interactions. The automated source finding
	codes for the FAST survey are also under developement, and the HI
	detections were identified by human eyes.  Hence, it is possible that
	a few weak HI sources could be missed due to human mistakes or due to
	RFI contamination in the spectra.  As shown in Figure \ref{flux}, most
	of the detected sources have relately high S/N (with S/N $> $ 6.5).
	The sources with $\rm 5 < S/N < 6.5$ are relatively rare, suggesting that the HI sources
	in this S/N range were not completely identified. In the next
	section we will further evaluate the completeness of our FAST
	detetions by comparing with the ALFALFA dataset.
	
	\subsection{\rm Comparison with ALFALFA detections}

\begin{figure}
		\centering
		\includegraphics[width=0.49\textwidth]{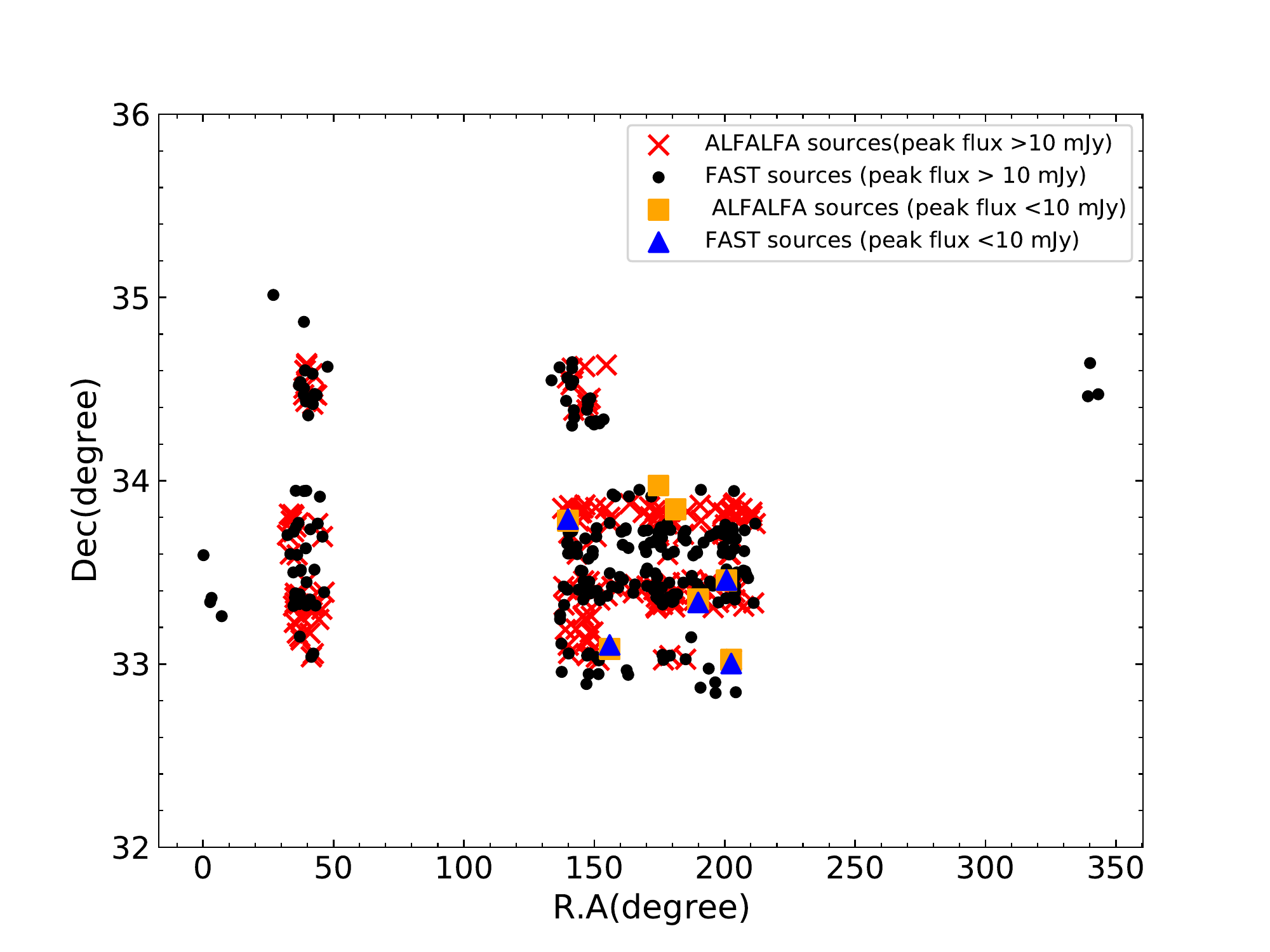}
		\caption{The comparions of ALFALFA and FAST sources for Dec 33$^\circ$, +34.5$^\circ$,red crosses denote ALFALFA sources and black solid dots label FAST source with  the peak flux 10 mJy, orange squares and blue triangles stand for ALFALFA and FAST sources with the peak flux less than 10 mJy, respectively .\label{fig:com_af}.}
	\end{figure}

	The ALFALFA Survey used the seven-horn Arecibo L-band Feed Array (ALFA) to
	blindly survey nearly 7000 $\rm deg^2$ of high Galactic latitude sky
	over 4400 nighttime hours\citep{Haynes_2018}.  Of the 544 FAST
	detected sources, 302 of them are also detected by ALFALFA. .  Many of
	the FAST observed regions suffered serious RFI contamination, thus we
	chose two regions with less RFI effects to make fare comparision with
	the ALFALFA datasets. 
	The left panel of Figure \ref{fig:com_af} compares the sources
	found in two Declinations range at $ 33.02^\circ < \rm Dec <
	33.9^\circ$ with $32.2^\circ <\rm R.A <212^\circ$,and $ 33.3^\circ <
	\rm Dec < 34.6^\circ$ with and $338.5^\circ < \rm R.A < 48.5^\circ$
	and $137.7^\circ < \rm R.A < 157.5^\circ$. In these ranges, FAST
	detect 194 sources, while ALFALFA contains 204 sources satisfying
	common heliocentric velocity range $100 < cz < 13000$ \kms. The ALFALFA
	survey was conducted in two parts at $\rm 07h30m < R.A. < 16h30m $ and $\rm 22h <
	R.A. < 03h $ over $\rm  +0^\circ <Dec  < +36^\circ $.  We found that more than
	90\% of the ALFALFA sources with peak flux more than 10 mJy are also
	detected by FAST.  Of the 7 ALFALFA sources with peak fluxes less than
	10 mJy, 5 of them are detected by FAST. There are about 10 to 20
	percent of ALFALFA sources not detected by FAST due to RFI contamination.
	
	Figure \ref{fig:com_af} show the FAST detection rate is slightly lower than  
	that of ALFALFA for the regions we compared. The detection rate is much lower in the edge of 
		the FAST scan maps due to lower sensitivity in these area. Besides, there is strong RFIs in the frequency range of 1100--1300MHz which could impact the detection rate \citep{2021arXiv211111018Z}.  As a results, the baselines are poor for some FAST sources, and some galaxies' spectra miss a part of the flux due to RFI masking.

	To futher check the telescope performance and our calibrations, we
	compare the FAST detected HI spectra with that of ALFALFA
	survey. Figure.\ref{lines} show ten representative HI spectrum
	profiles overlaid with the corresponding ALFALFAL sources, where the
	x-axis labels the radio velocity in unit of $\rm km/s$ and y-axis
	refers to the flux density in milliJansky (mJy).  The spectra
	measured with FAST telescope matches that of ALFALFA within 10\% of
	uncertainty. The FAST spectral resolution have been smoothed to 10
	km/s to match that of ALFALFA\citep{Haynes_2018} and baseline fittings are
	general good for all sources.

	
	
	\begin{figure*}
		\centering
		\includegraphics[width=0.48\textwidth]{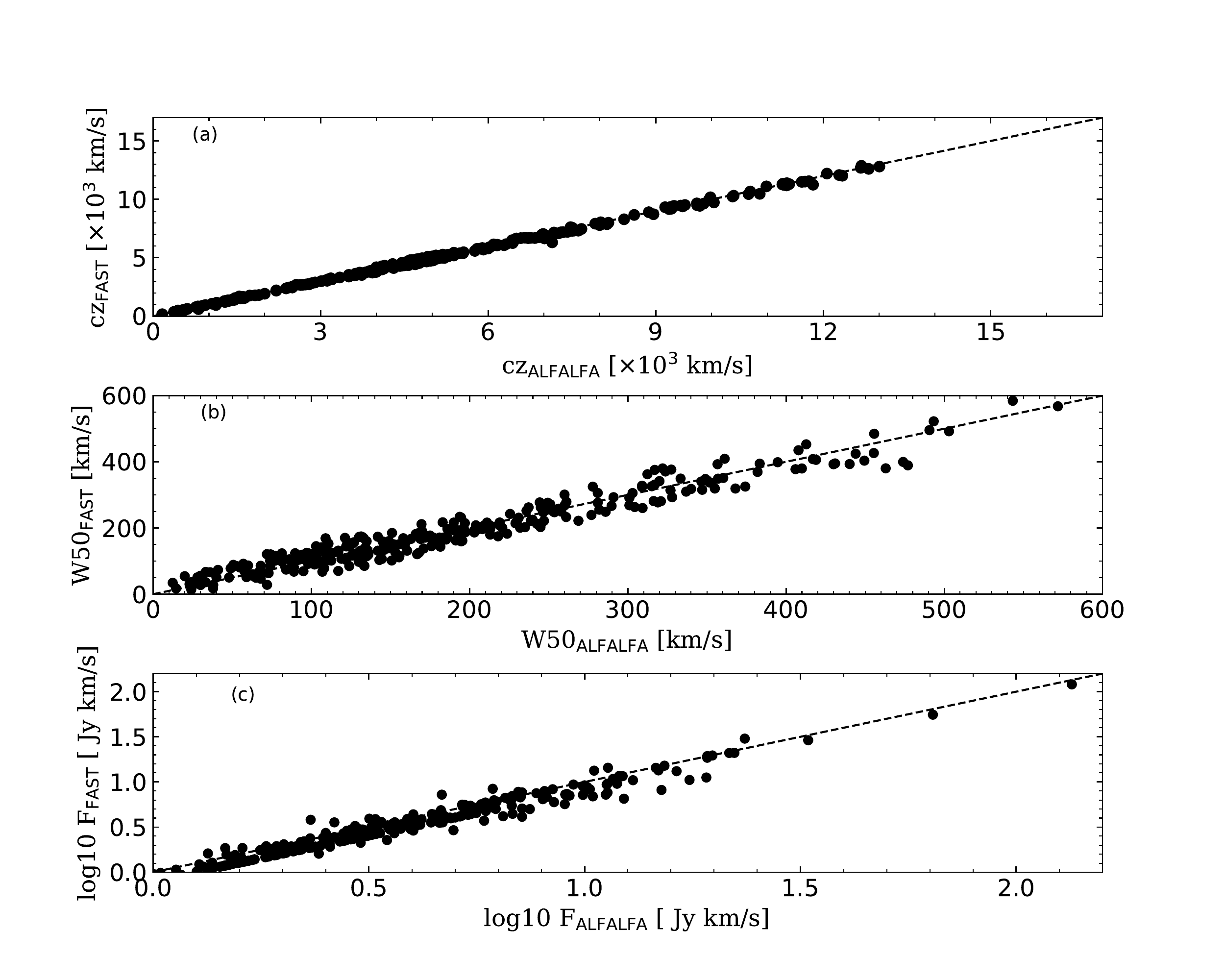}
		\includegraphics[width=0.48\textwidth]{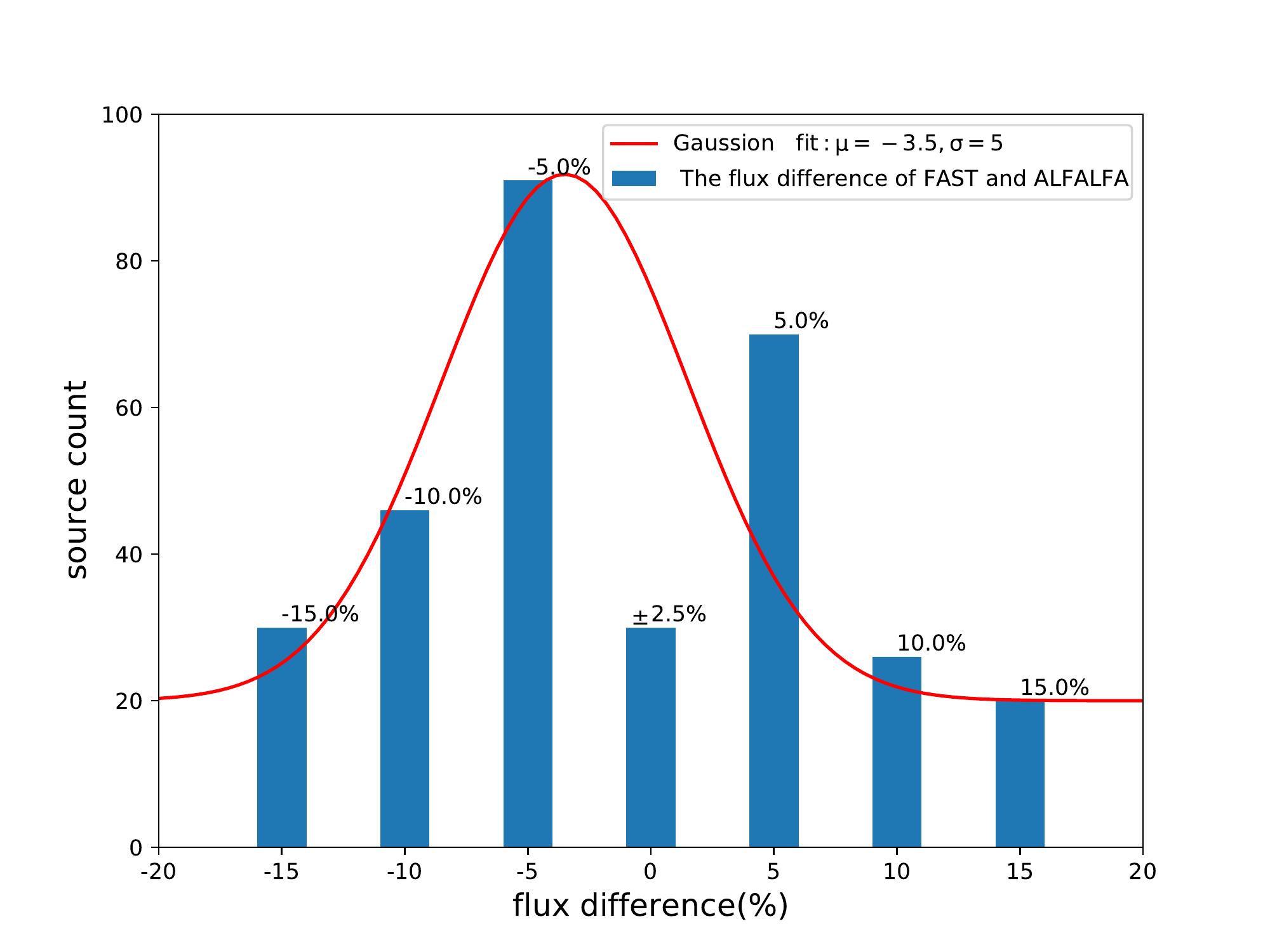}
		\caption{Left:Testing   velocity in \kms,  W50 at half of peak spectrum in \kms, logarithm of the  total flux  in Jy\kms for  302 detections of FAST with its counterparts in ALFALFA catalog\citep{Haynes_2018} from  (a) to (b) panel,repectively: \label{fa_af}. Right panel: Bar char for the flux  difference in percentage for the sources detected by both FAST and ALFALFA.}
	\end{figure*} 
	\begin{equation}\label{eq:error}
	\rm flux  \quad error \quad (\%) = \frac{Fc_{FAST}- Fc_{ALFALFA}}{Fc_{FAST}} 
	\end{equation}
	
	Figure.\ref{fa_af} in left panel compares the ALFALFA measured values of velocity
	$cz$, line widths W50, and integrated fluxes $\rm F_c$ to those
	measured with the FAST HI drift scan observations for the 302
	galaxies in common. The mean offsets are 57.5 km/s,40.2 km/s,and standard deviation are 121.6 km/s, 30.7 km/s for the $\rm cz$ and $\rm W_{50}$ measurements. The FAST integrated fluxes are somewhat smaller than that measured with ALFALFA, which could be due to poor baseline fitting or affected by RFIs and standing waves.  The flux difference estimated with equation \ref{eq:error} for the 302 FAST-ALFALFA detected sources is presented as bar diagram in Figure \ref{fa_af} right panel. The flux differences are mostly with $\pm$15\%. The mean offset $\mu$  and standard deviation $\sigma$ for the gaussian fit to the bar char (Figure \ref{fa_af}) are -3.5\% and  5\%, respectively. From these results,we find no significant difference bewteen the FAST and ALFALFA data sets considering the uncertainties in flux calibration for the pilot FAST HI survey which is about 10\%. 
	\begin{figure*}[ht]
		\includegraphics[width=0.49\textwidth]{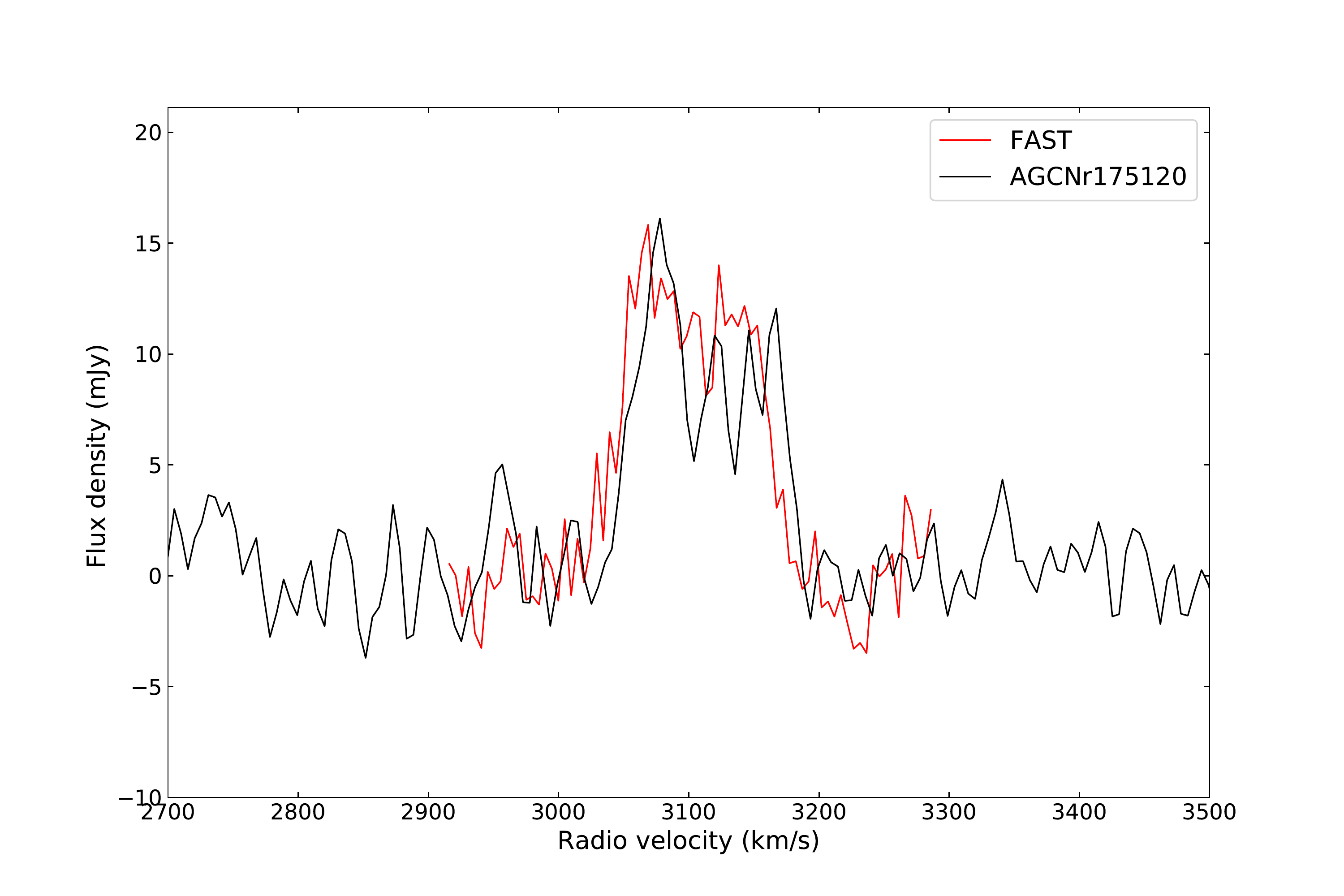}
		\includegraphics[width=0.49\textwidth]{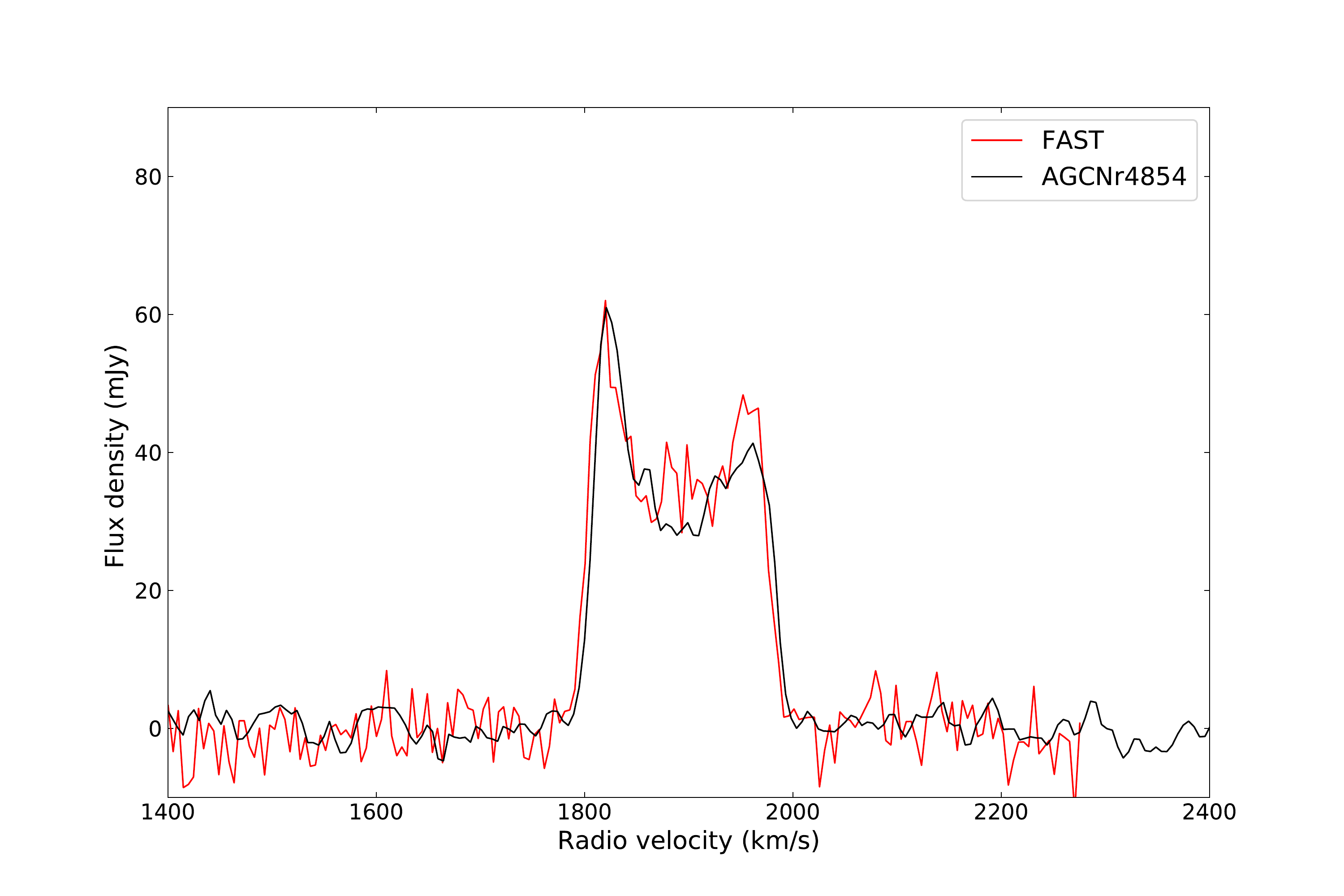}
	\end{figure*}
	\begin{figure*}
		\includegraphics[width=0.49\textwidth]{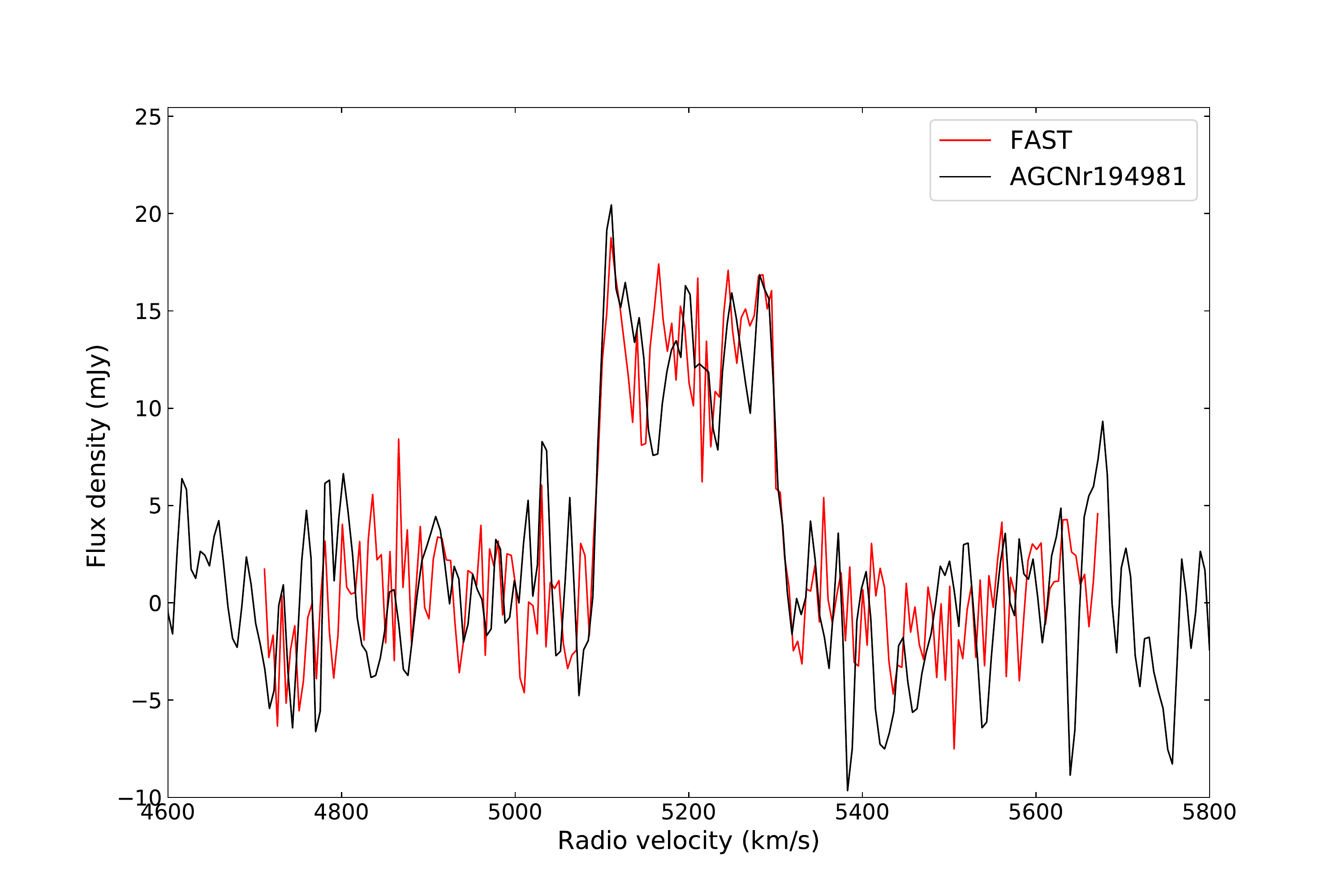}
		\includegraphics[width=0.49\textwidth]{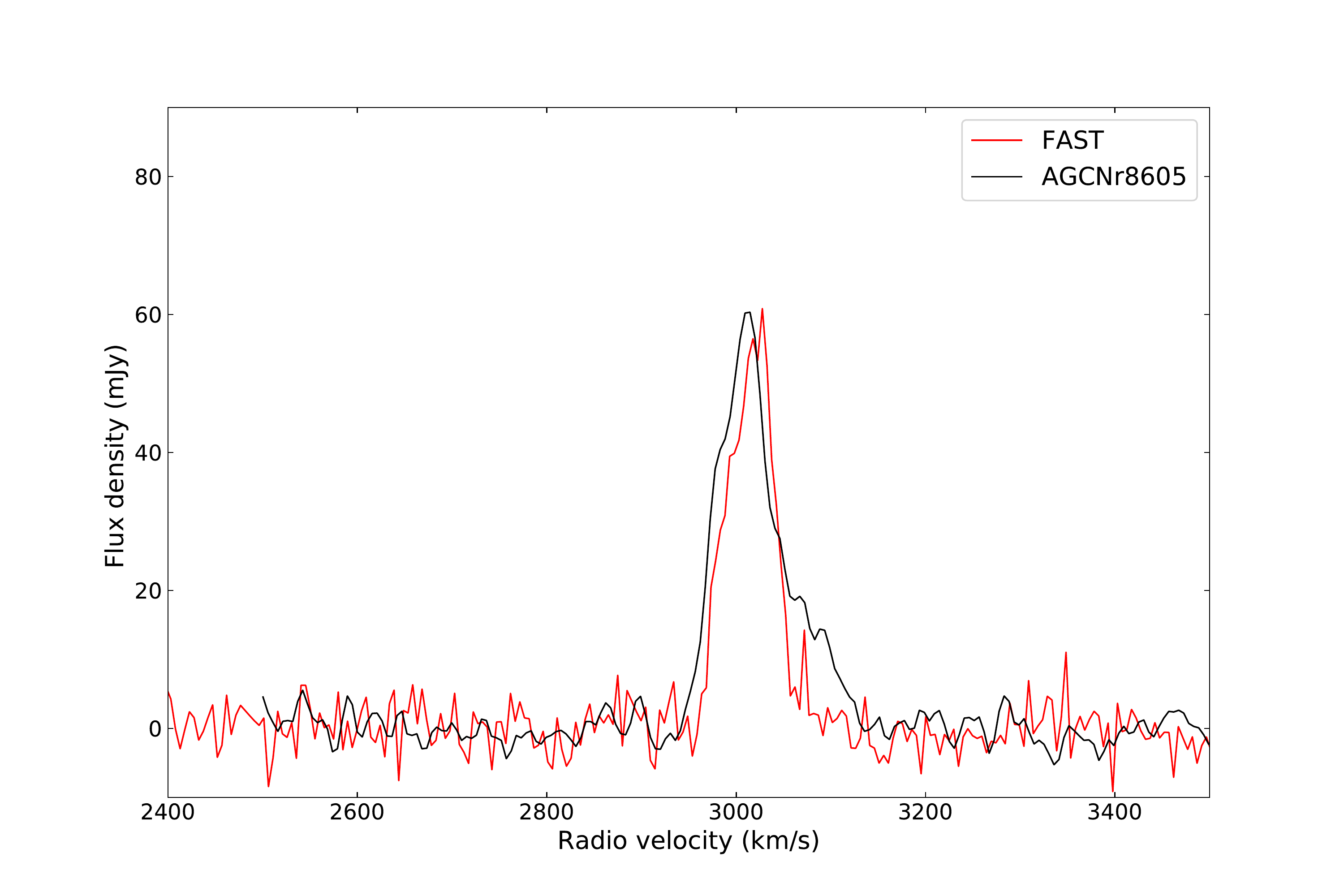}
	\end{figure*}
	\begin{figure*}
		\includegraphics[width=0.49\textwidth]{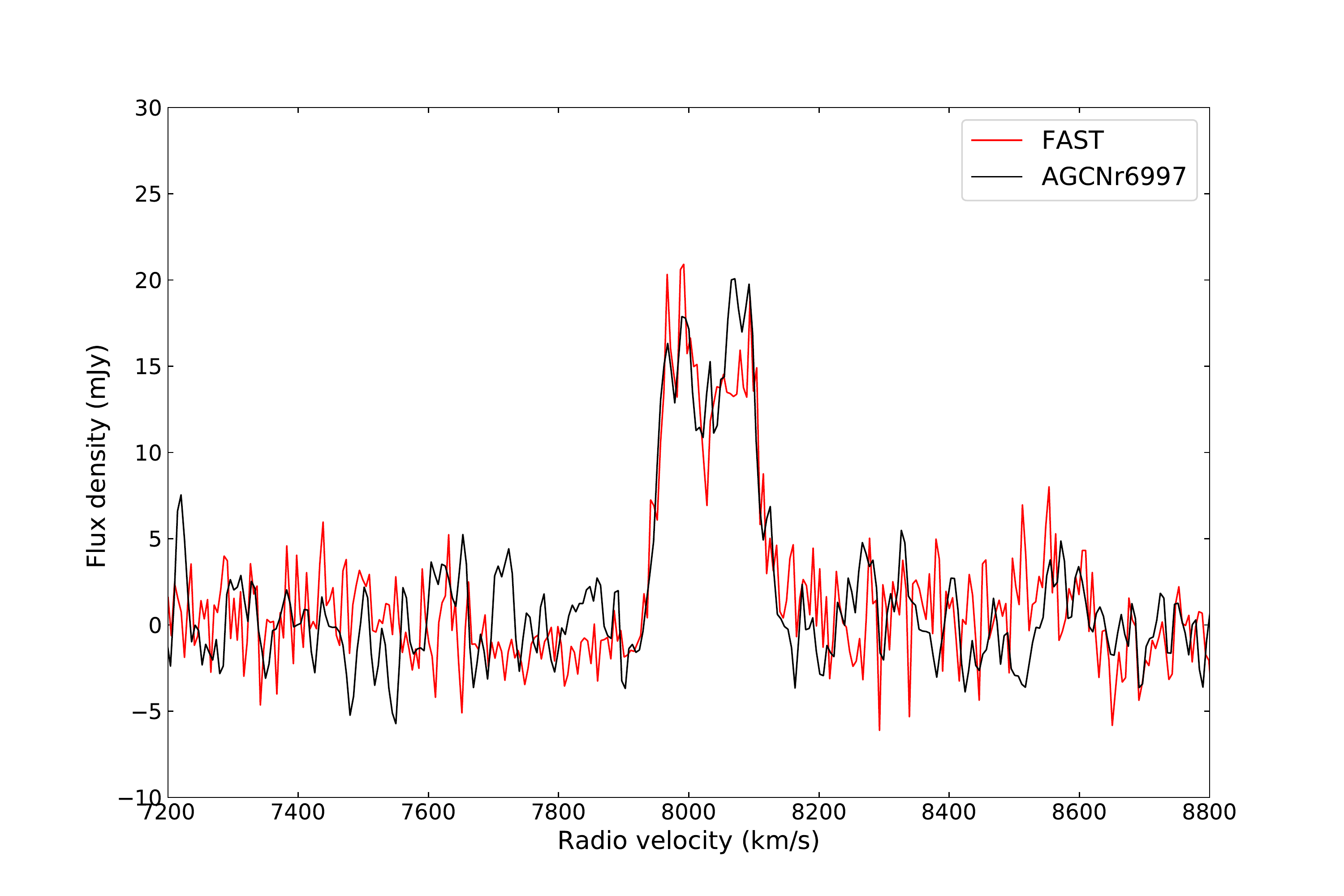}
		\includegraphics[width=0.49\textwidth]{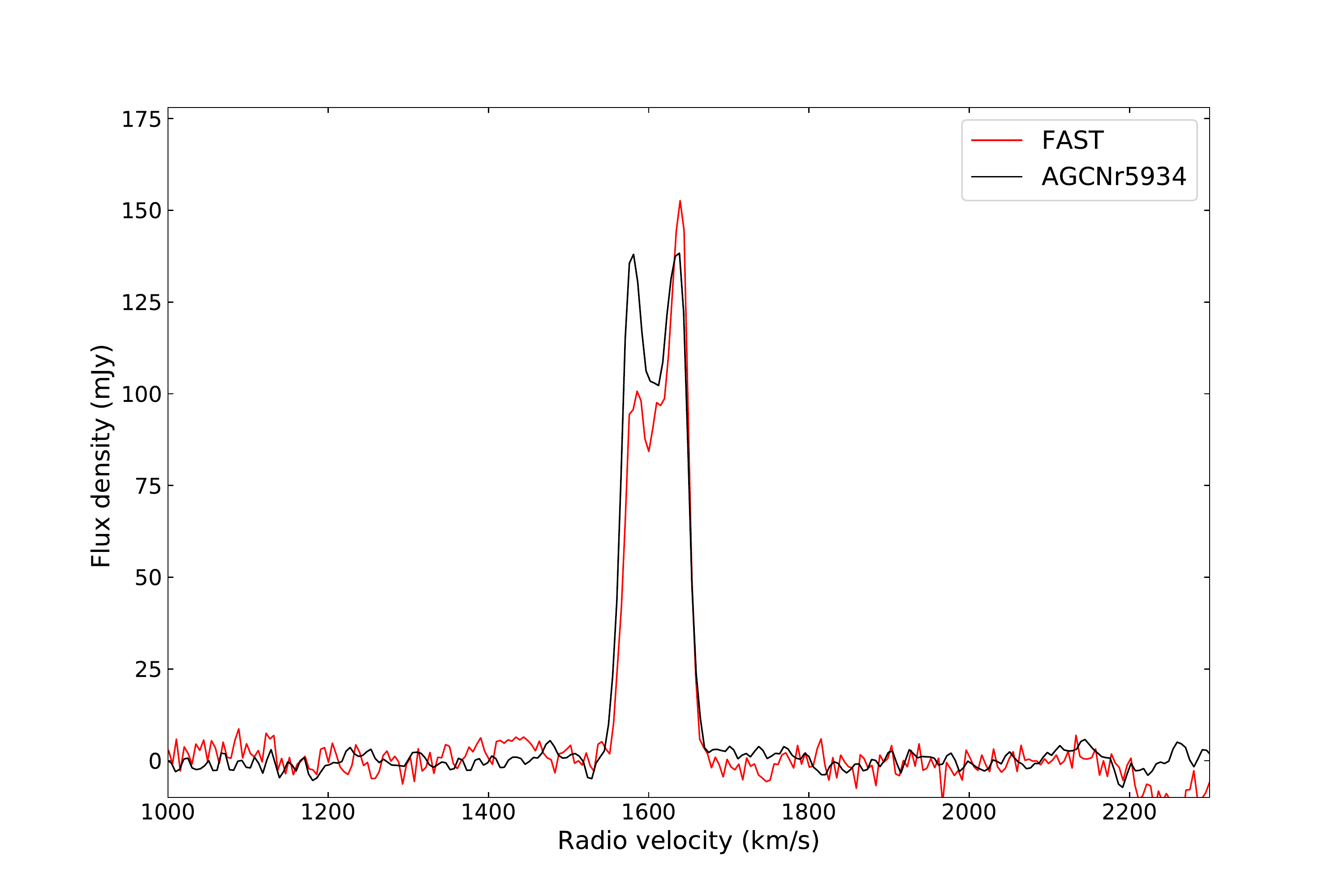}
	\end{figure*}
	\begin{figure*}
		\includegraphics[width=0.49\textwidth]{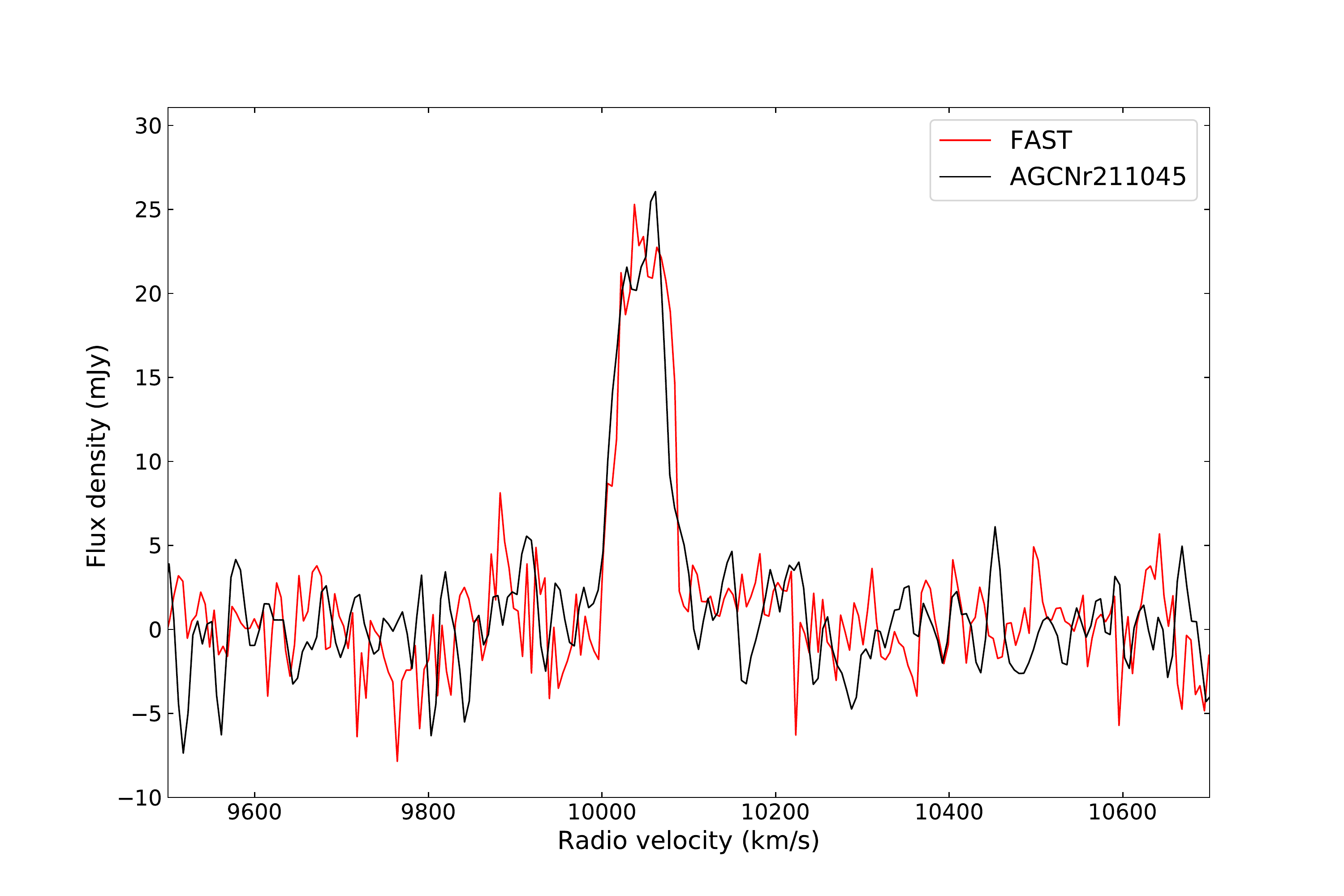}
		\includegraphics[width=0.49\textwidth]{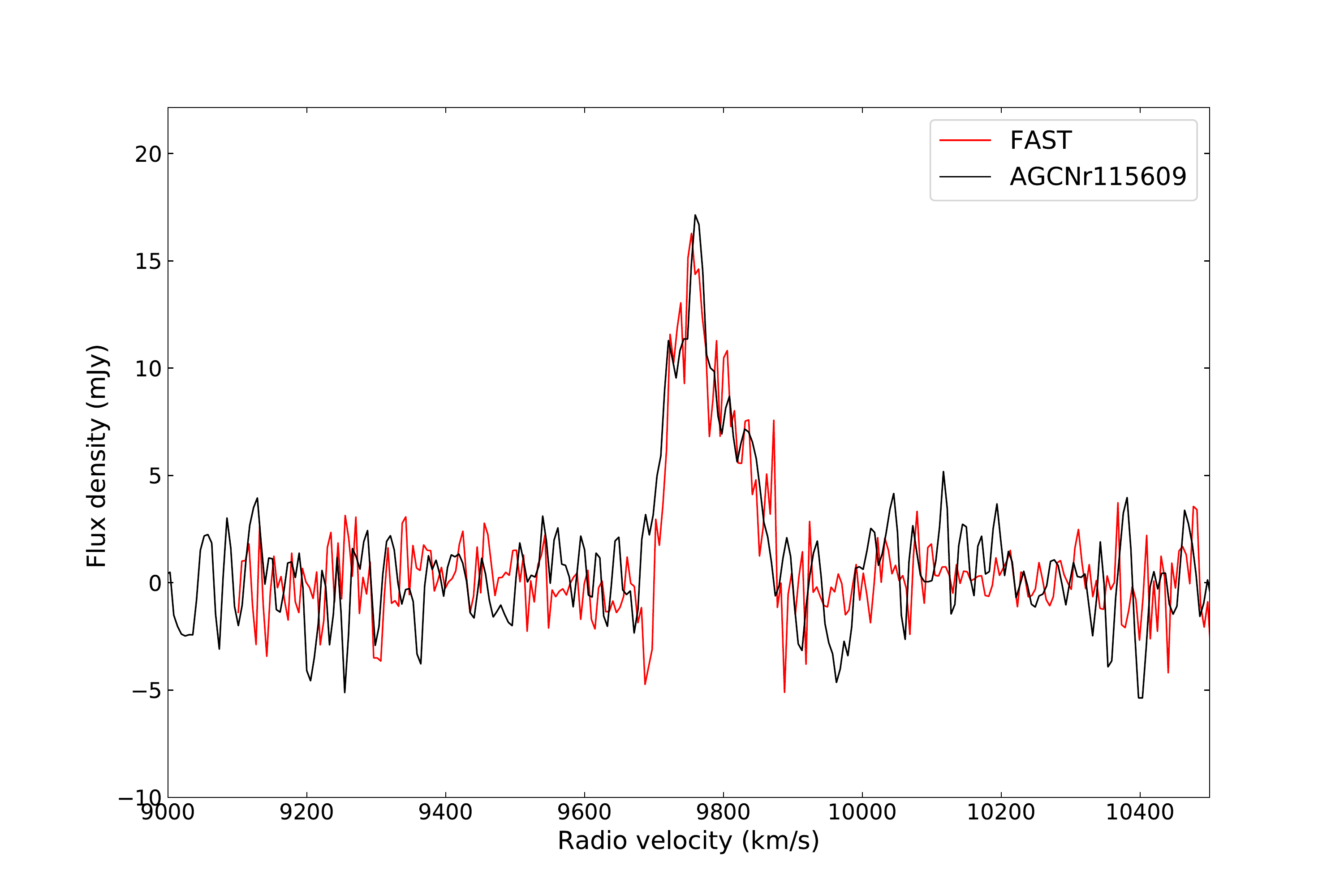}
	\end{figure*}
	\begin{figure*}
		\includegraphics[width=0.49\textwidth]{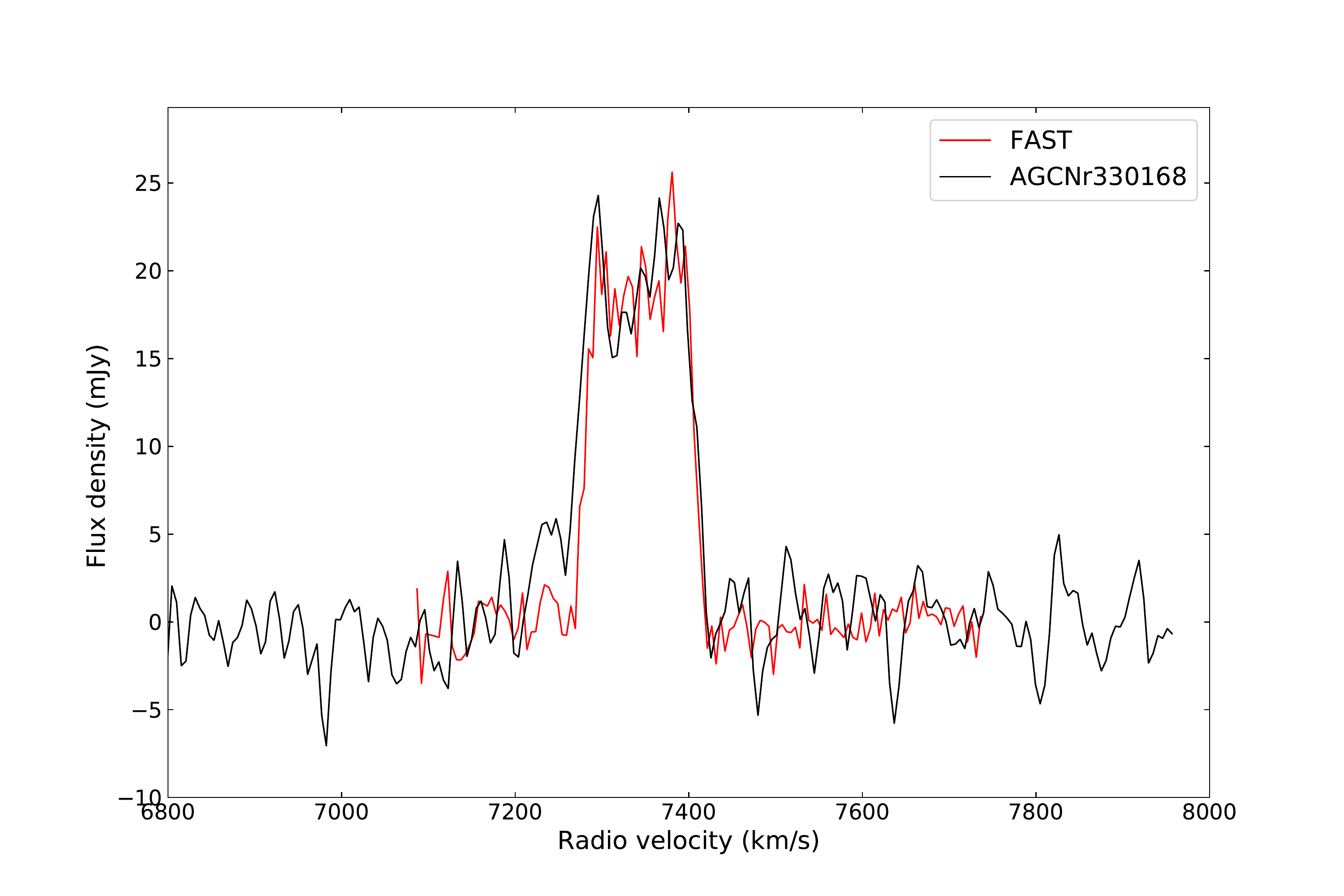}
		\includegraphics[width=0.49\textwidth]{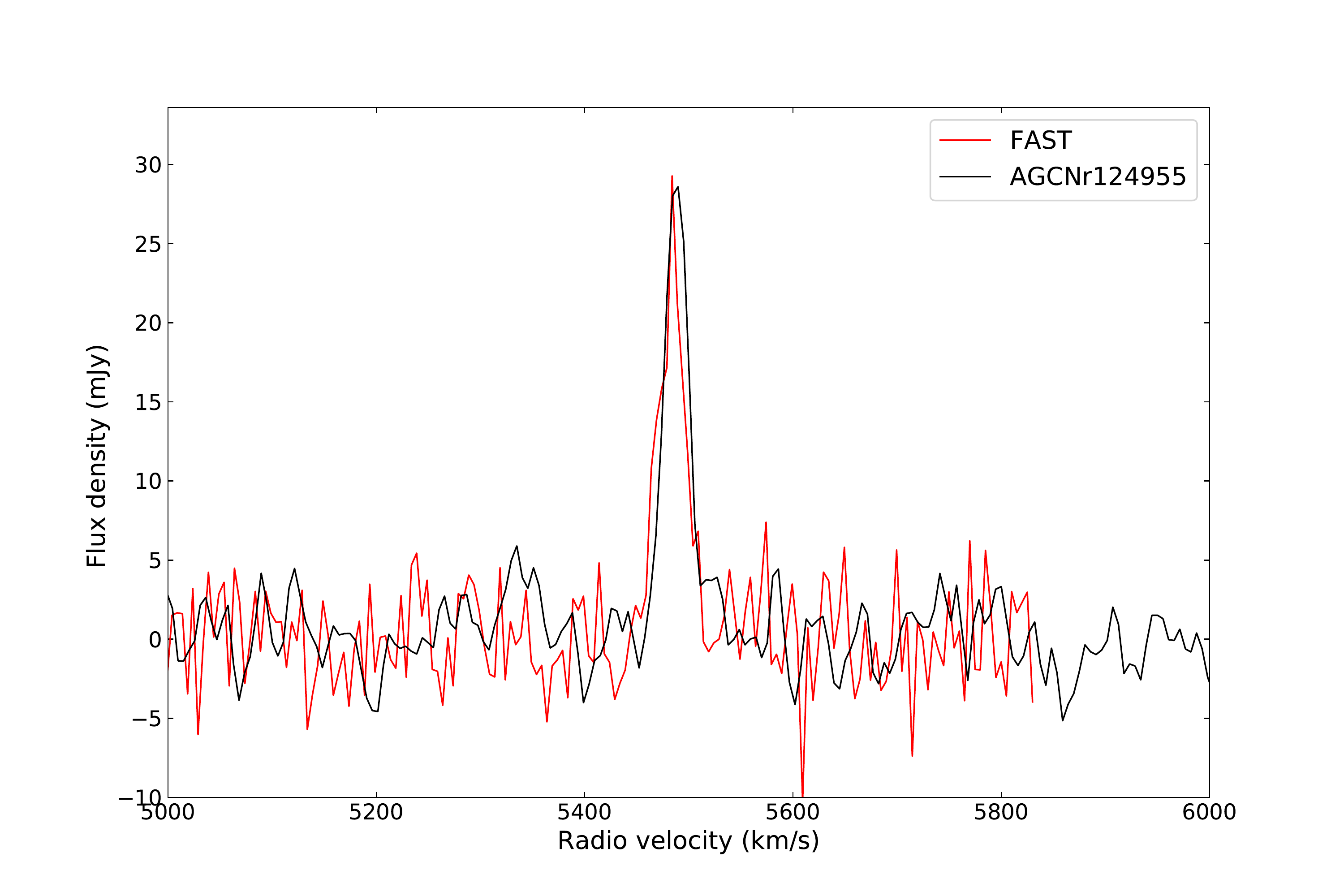}
	\end{figure*}
	\begin{figure*}
		\includegraphics[width=0.49\textwidth]{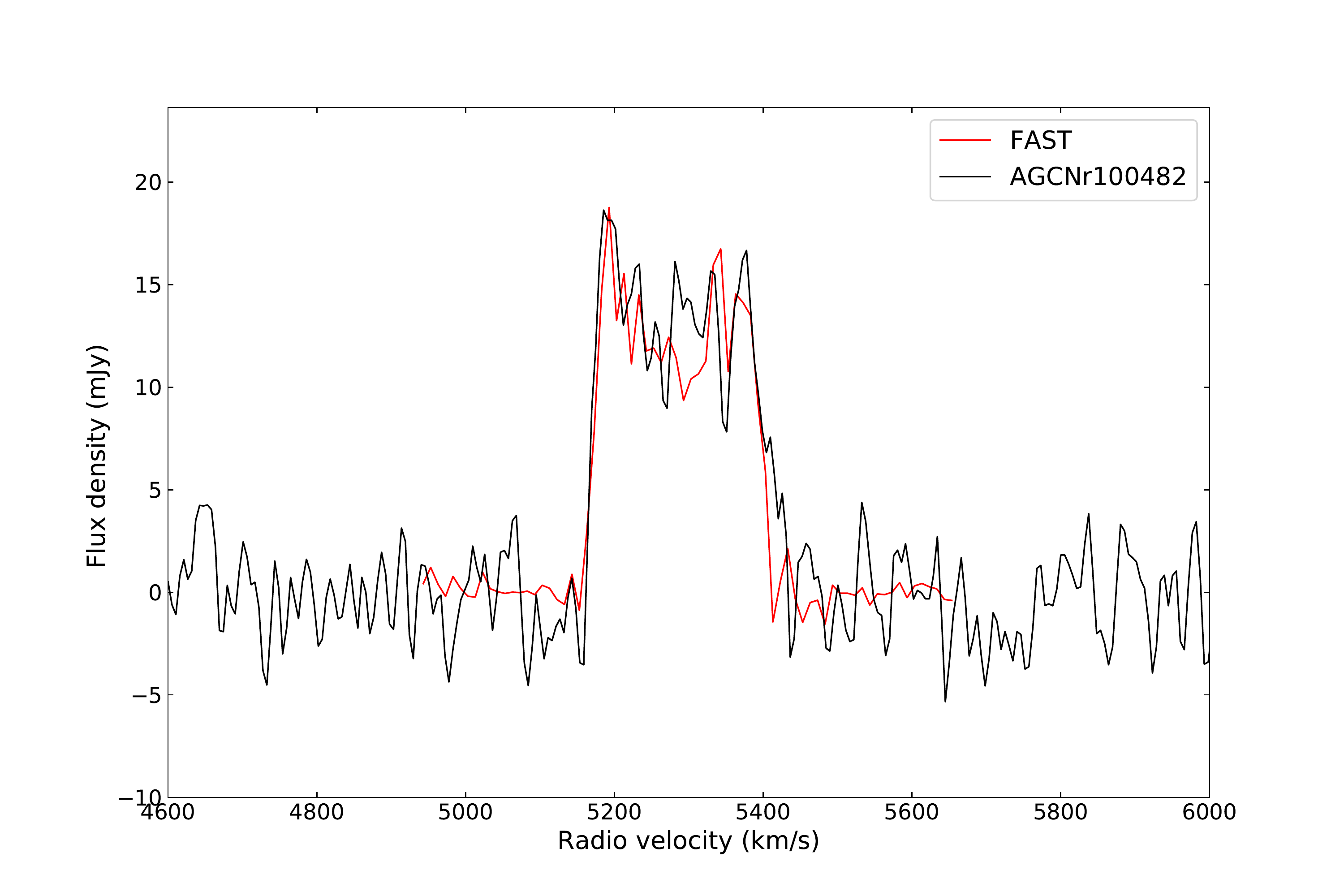}
		\includegraphics[width=0.49\textwidth]{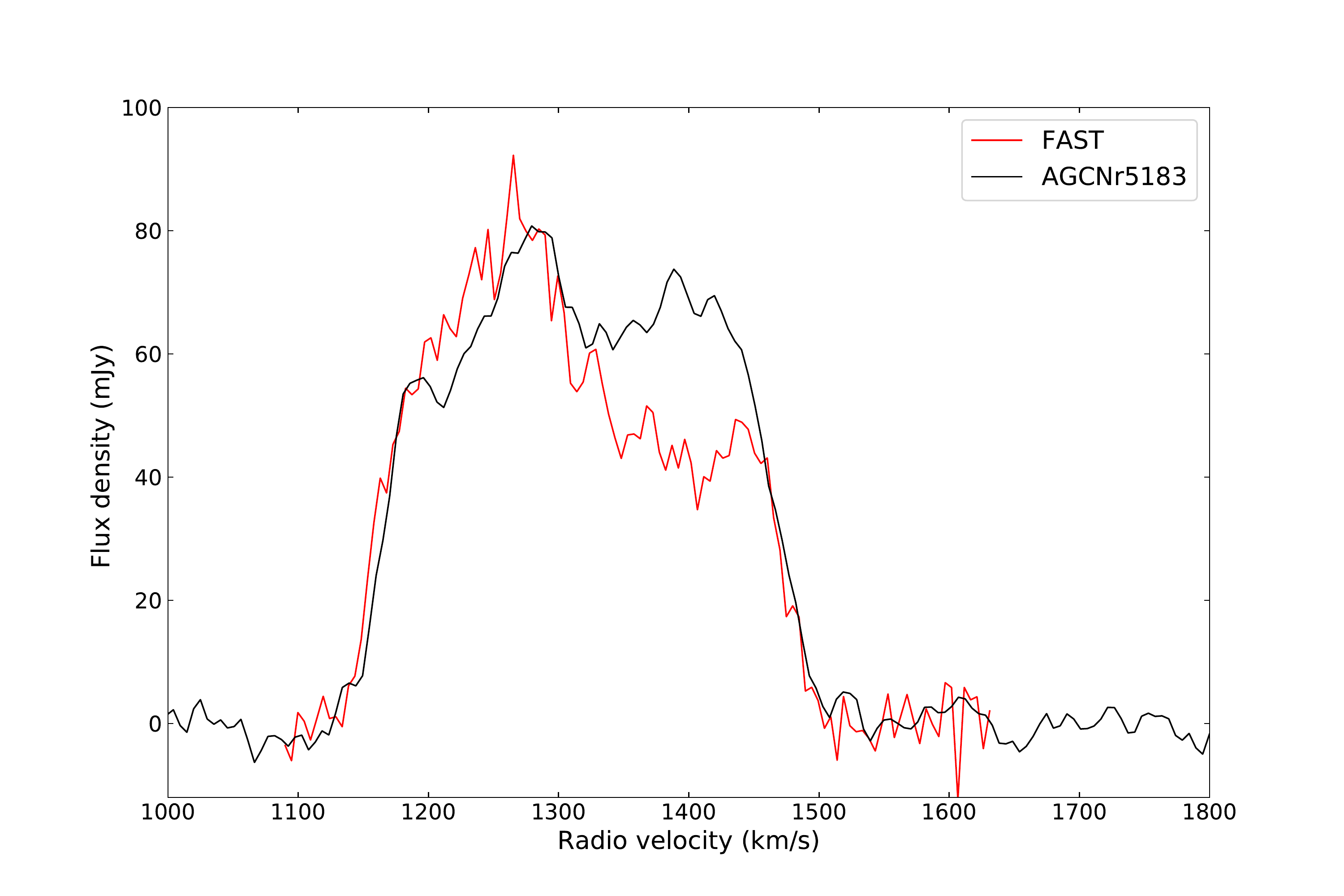}
	\end{figure*}
	\begin{figure*}[ht]
		\includegraphics[width=0.49\textwidth]{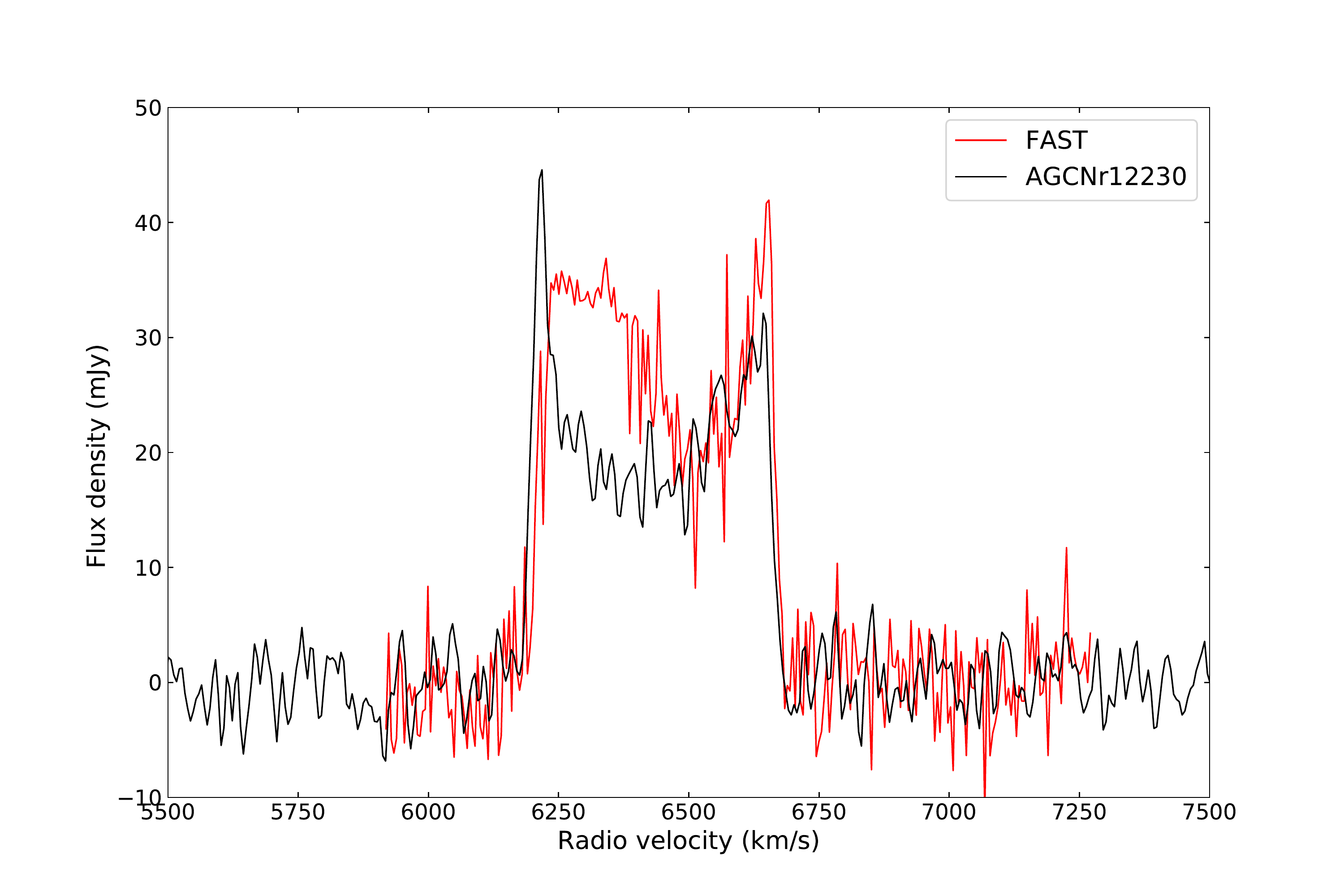}
		\includegraphics[width=0.49\textwidth]{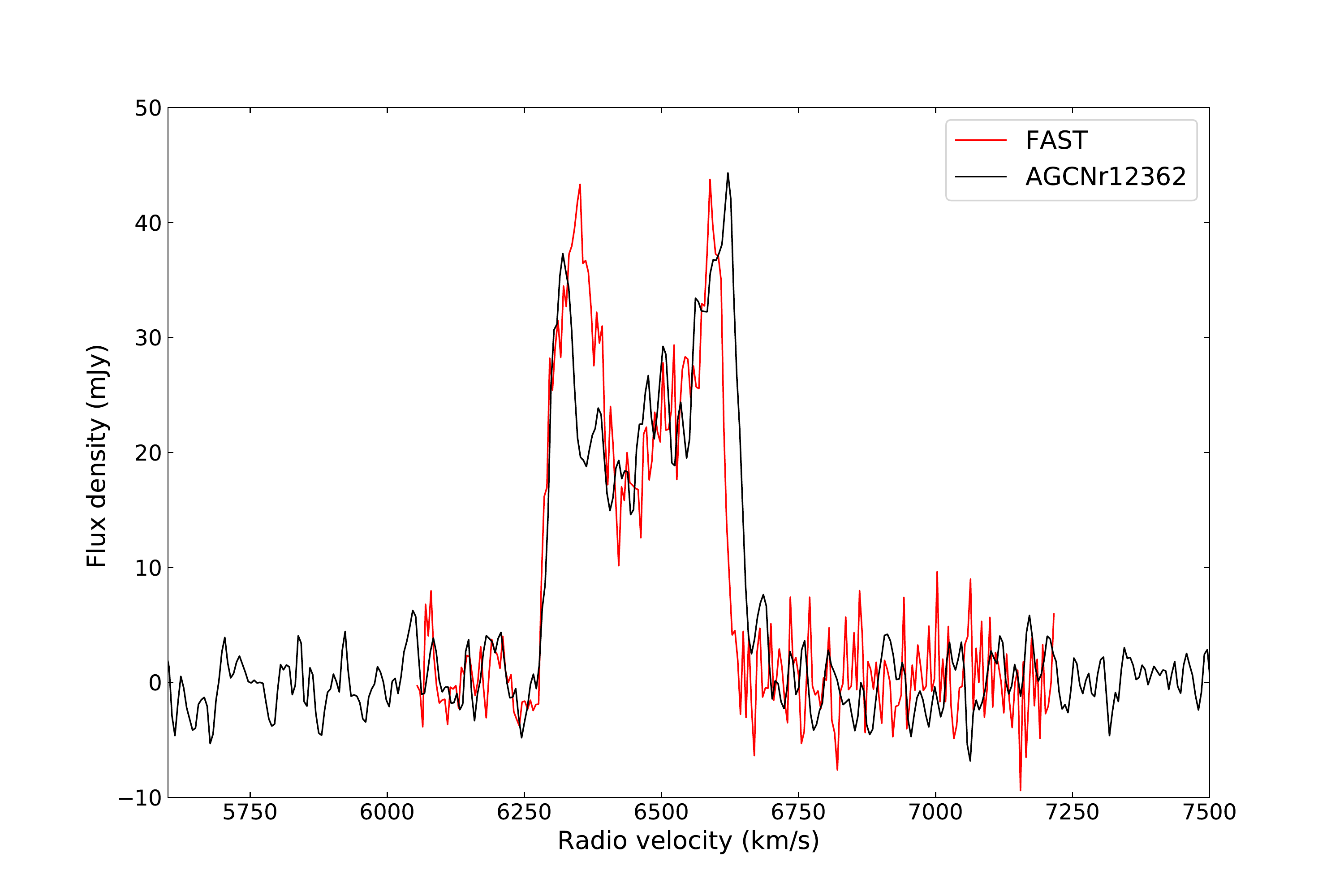}
		\caption{ Ten of typical  HI spectrum lines  of all the sources in unit of $\rm Jy  kms^{-1}$ for source flux at  different radio velocity(km/s) range and overlap corresponding ALFALFA source spectra. \label{lines}}
	\end{figure*}
	
	\subsection{\rm HI sources with no optical counterparts}\label{no counterparts}
	
\begin{figure*}[ht]
		\includegraphics[width=0.15\textwidth]{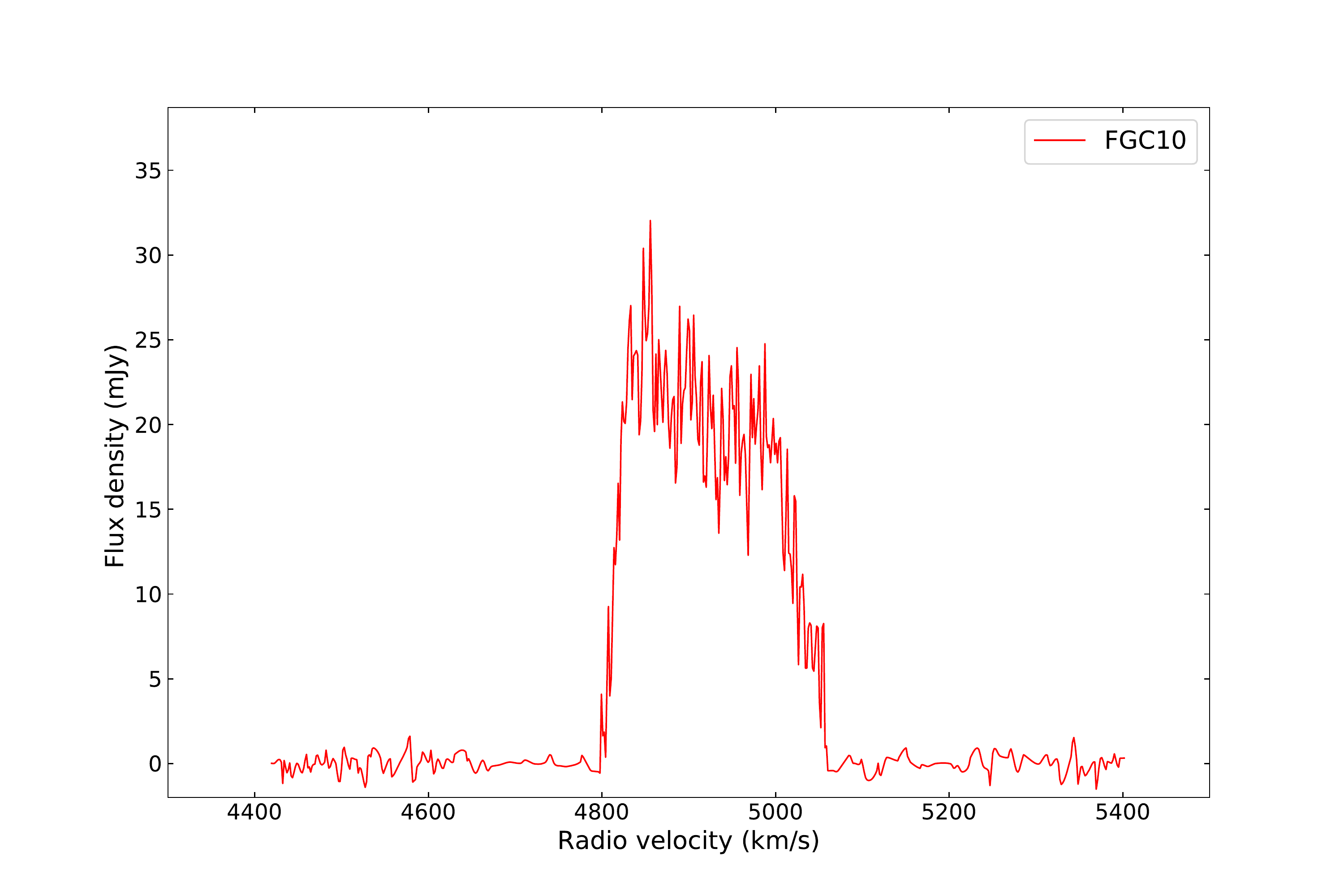}
		\includegraphics[width=0.16\textwidth]{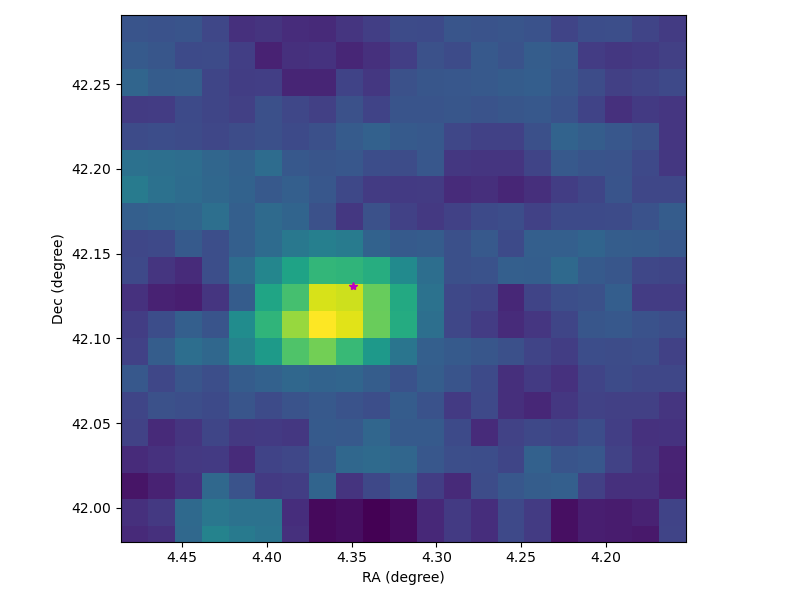}
		\includegraphics[width=0.17\textwidth]{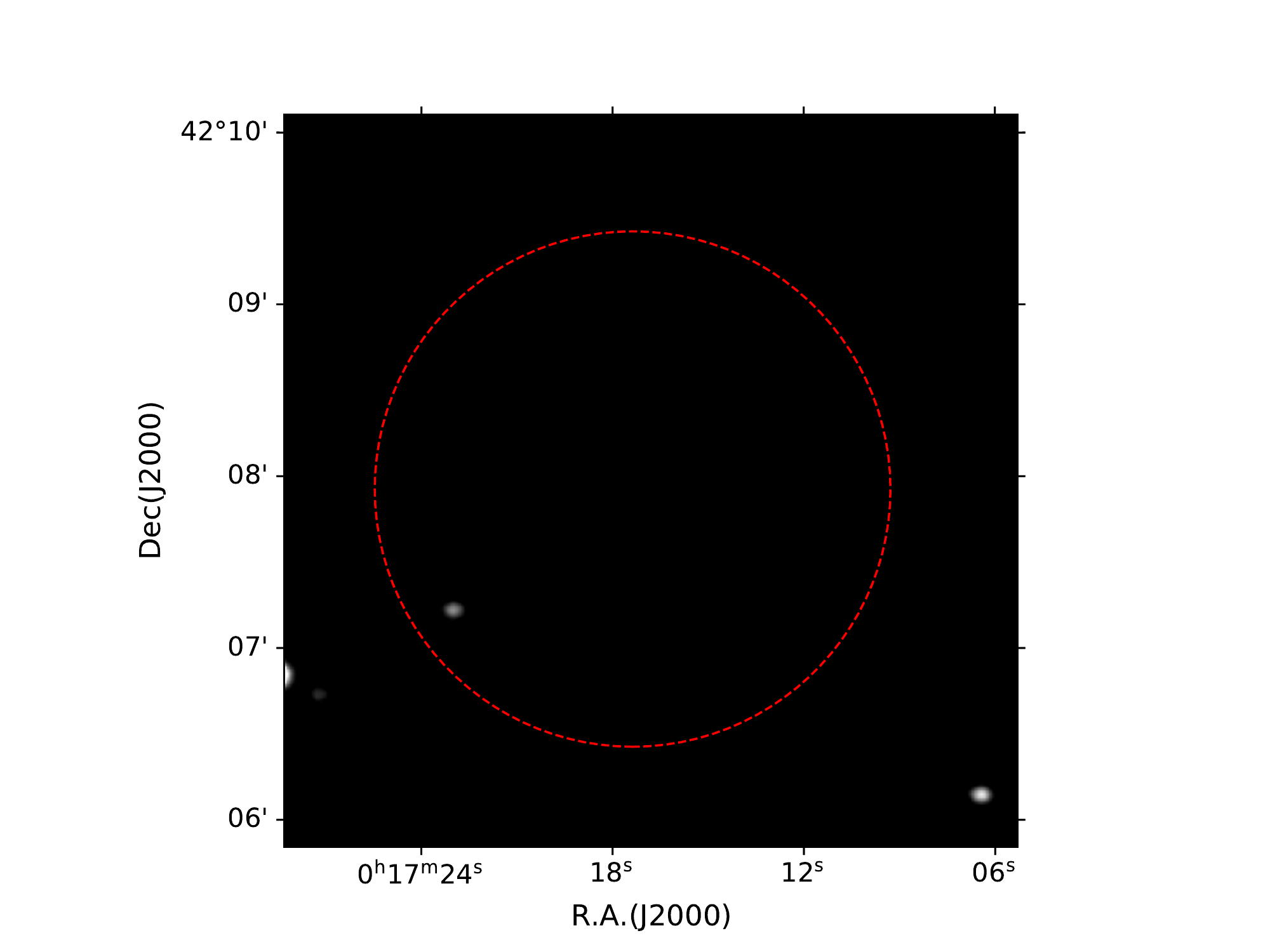}
		\includegraphics[width=0.15\textwidth]{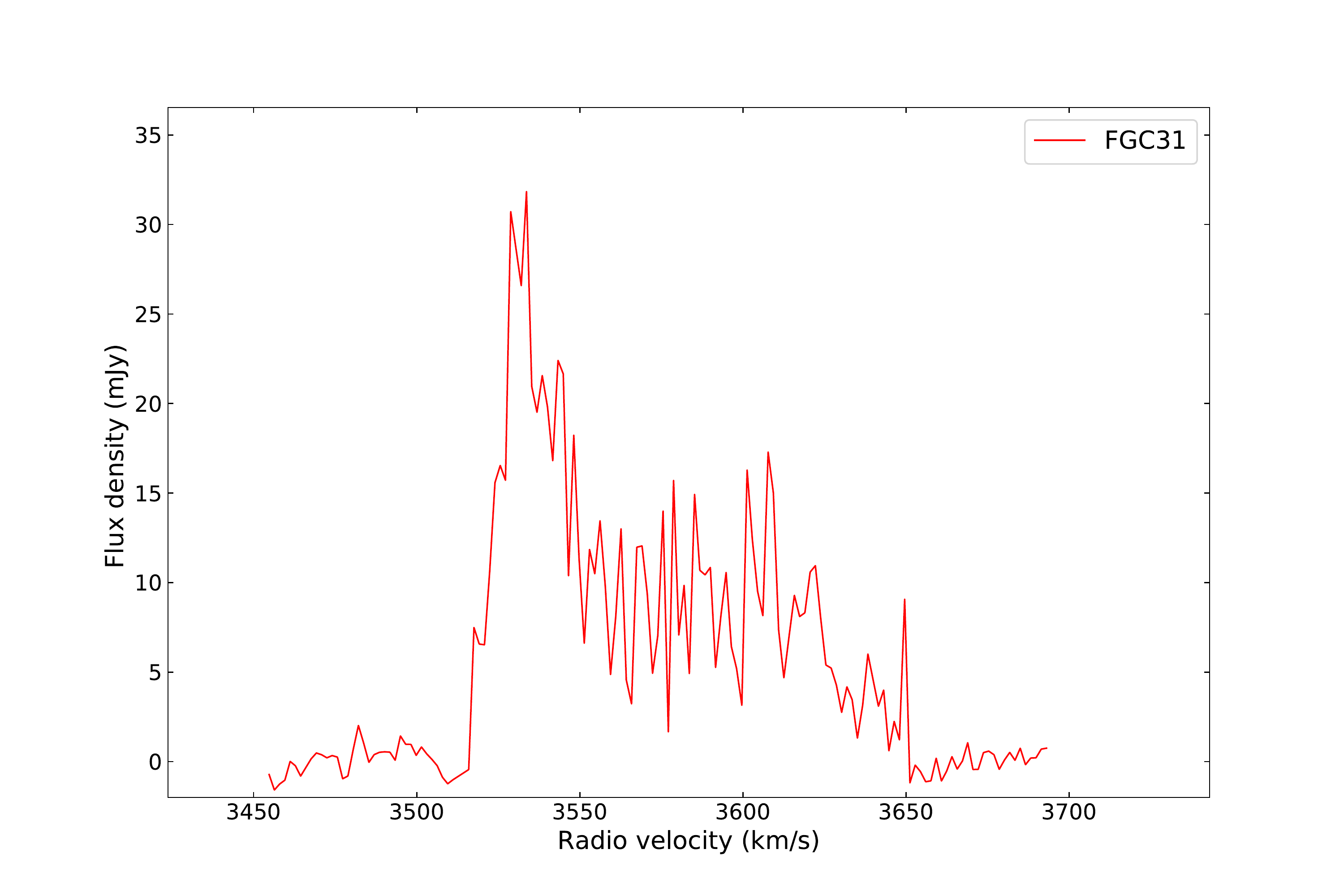}
		\includegraphics[width=0.16\textwidth]{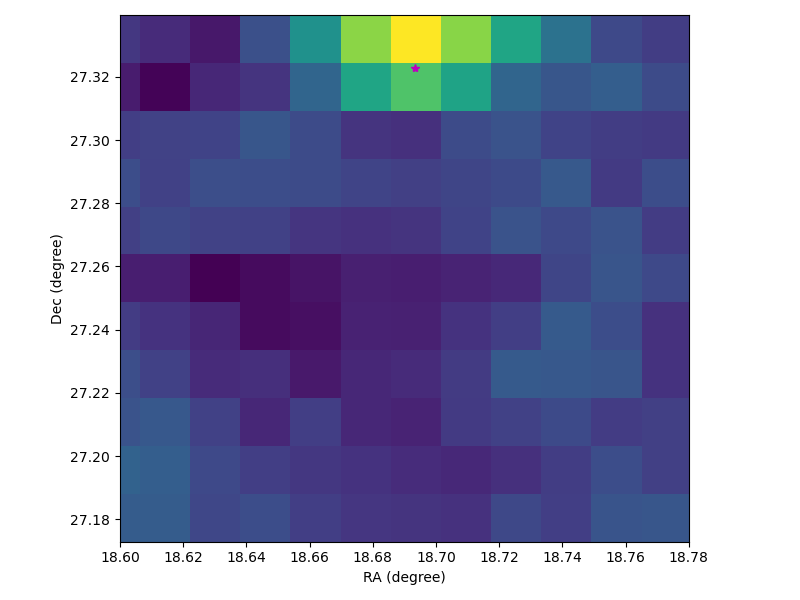}
		\includegraphics[width=0.16\textwidth]{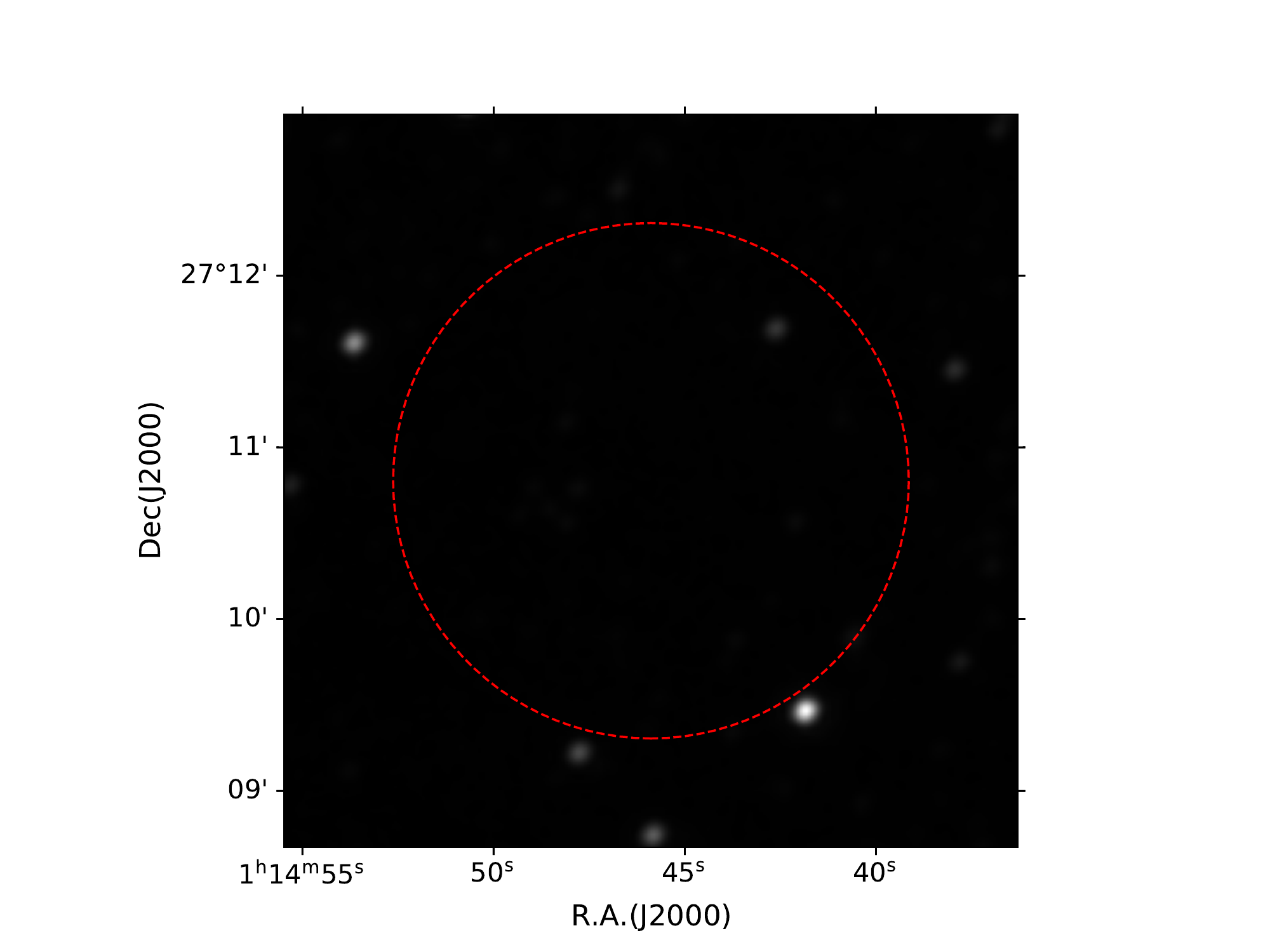}
		\caption{Left and middle plot show  HI source spectrum line, HI intensity mappings,respecitively, and its counterparts image in SDSS grz-band  in right panel,Left and middle plot show  HI source spectrum line, HI intensity mappings,respecitively, and its counterparts image  in unWISE W1/W2 NEO6 band  in right panel \label{map2}}
	\end{figure*}

\begin{figure*}[ht]
		\includegraphics[width=0.15\textwidth]{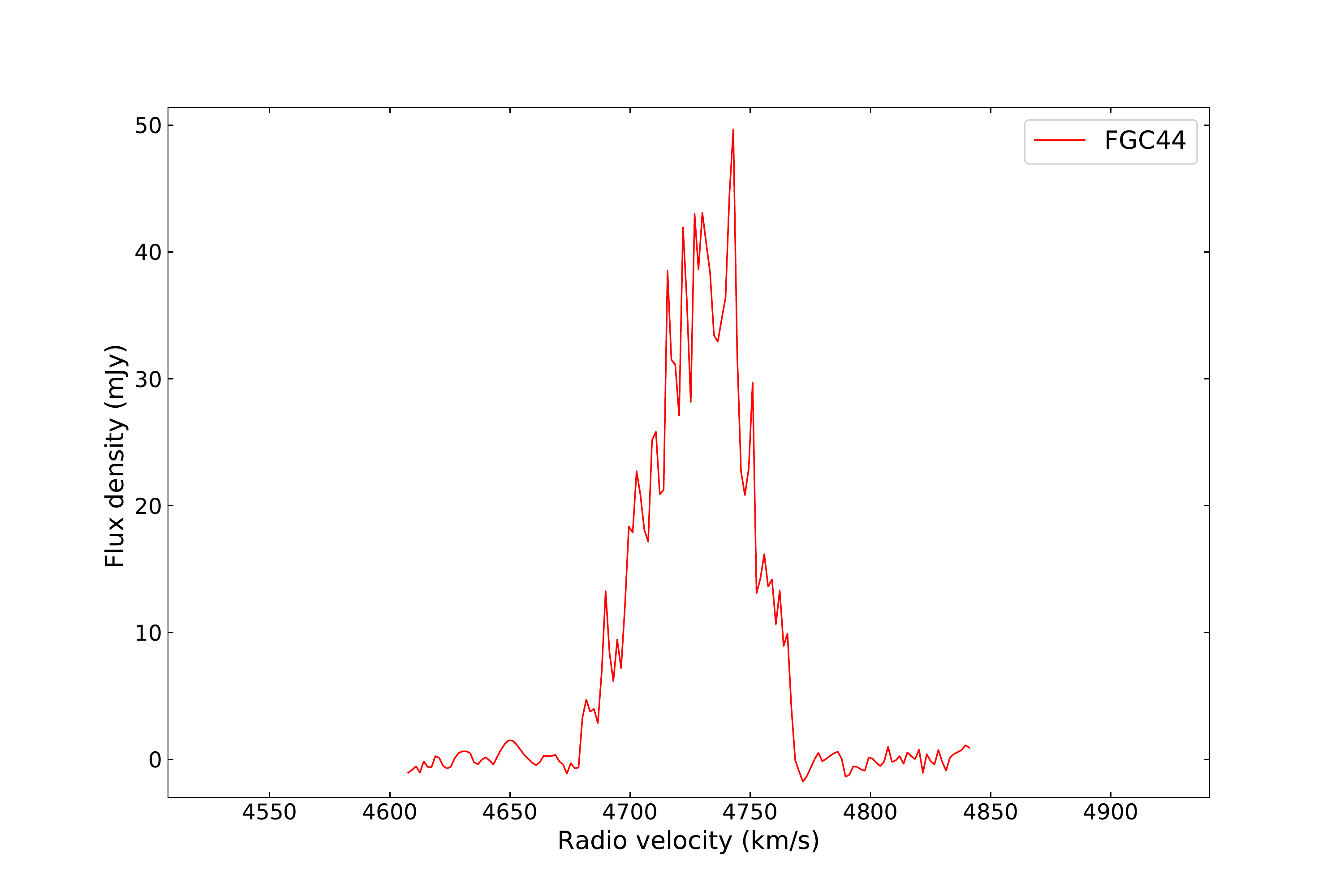}
		\includegraphics[width=0.15\textwidth]{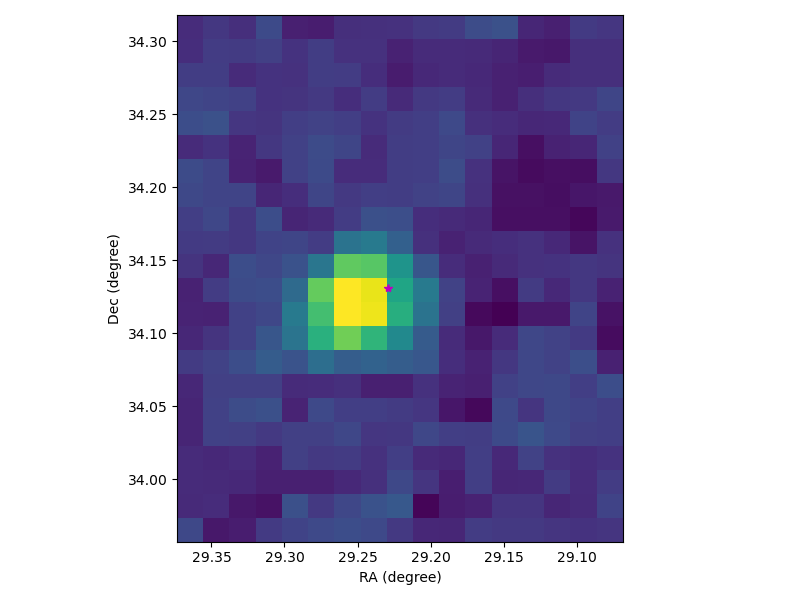}
		\includegraphics[width=0.18\textwidth]{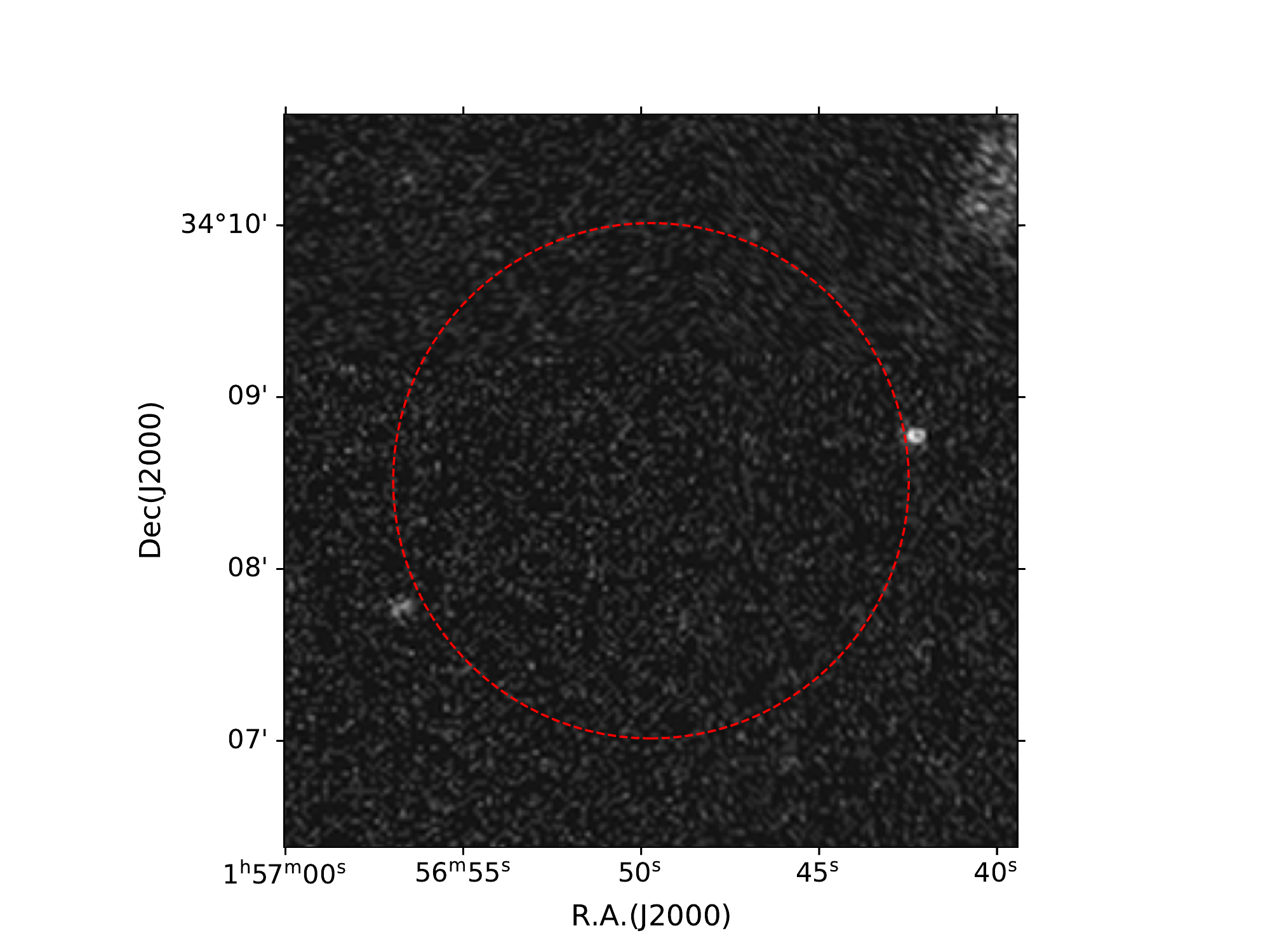}
		\includegraphics[width=0.16\textwidth]{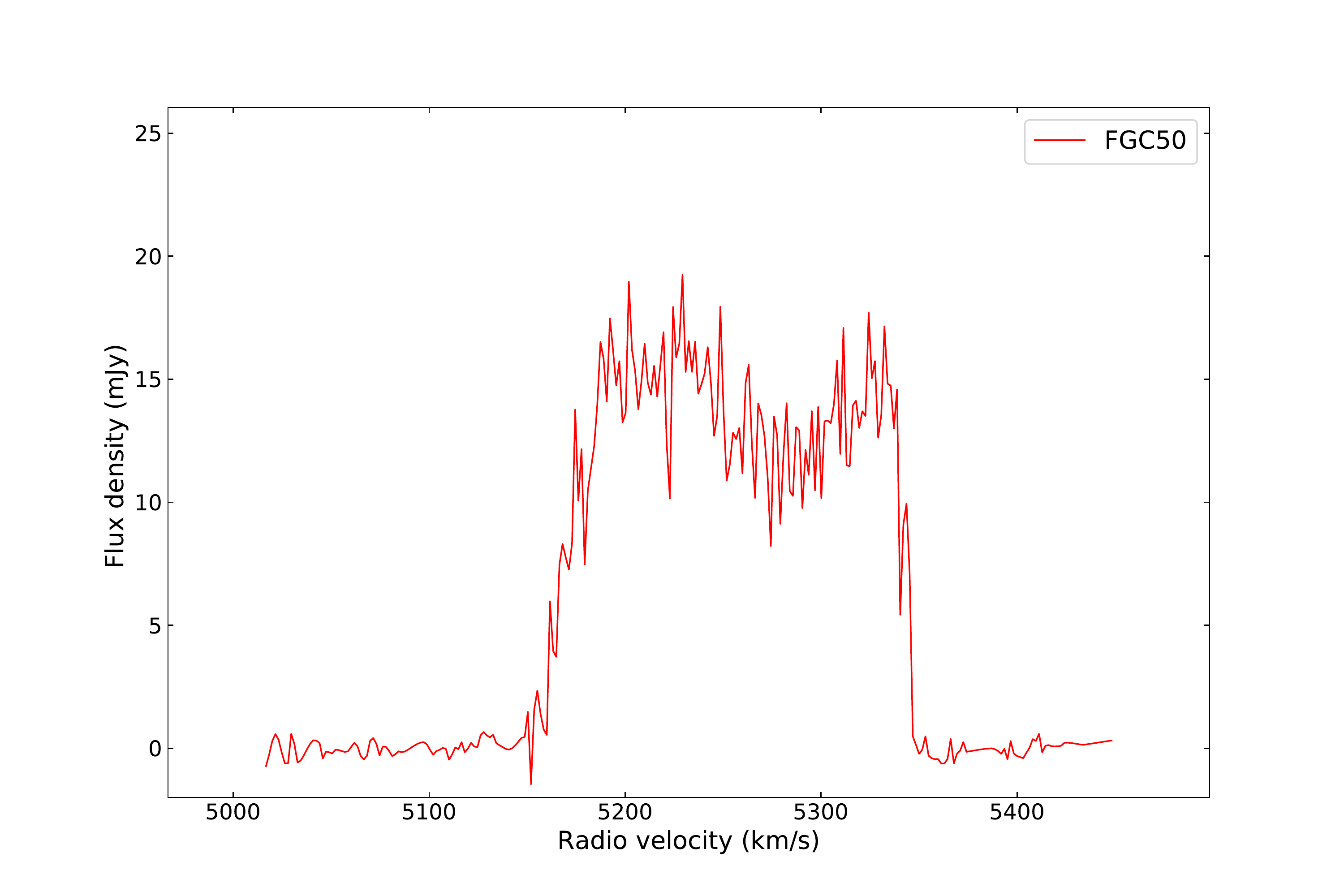}
		\includegraphics[width=0.15\textwidth]{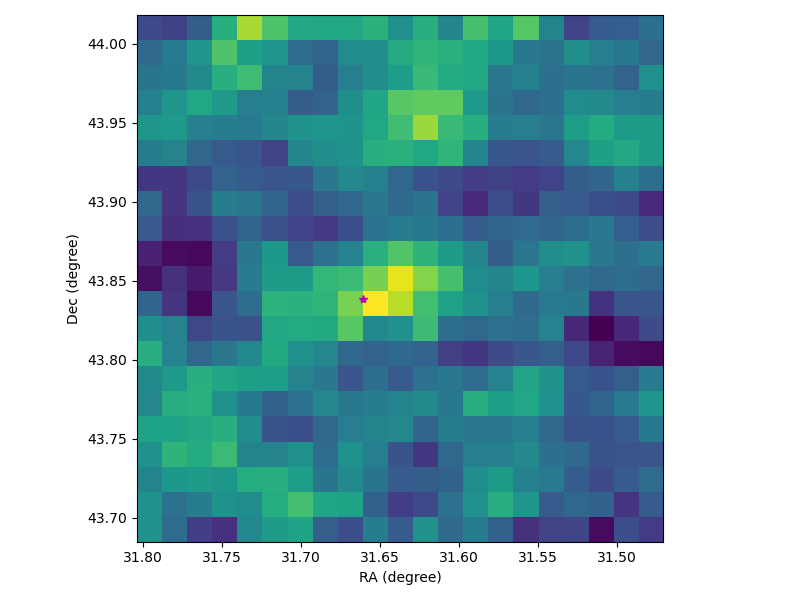}
		\includegraphics[width=0.16\textwidth]{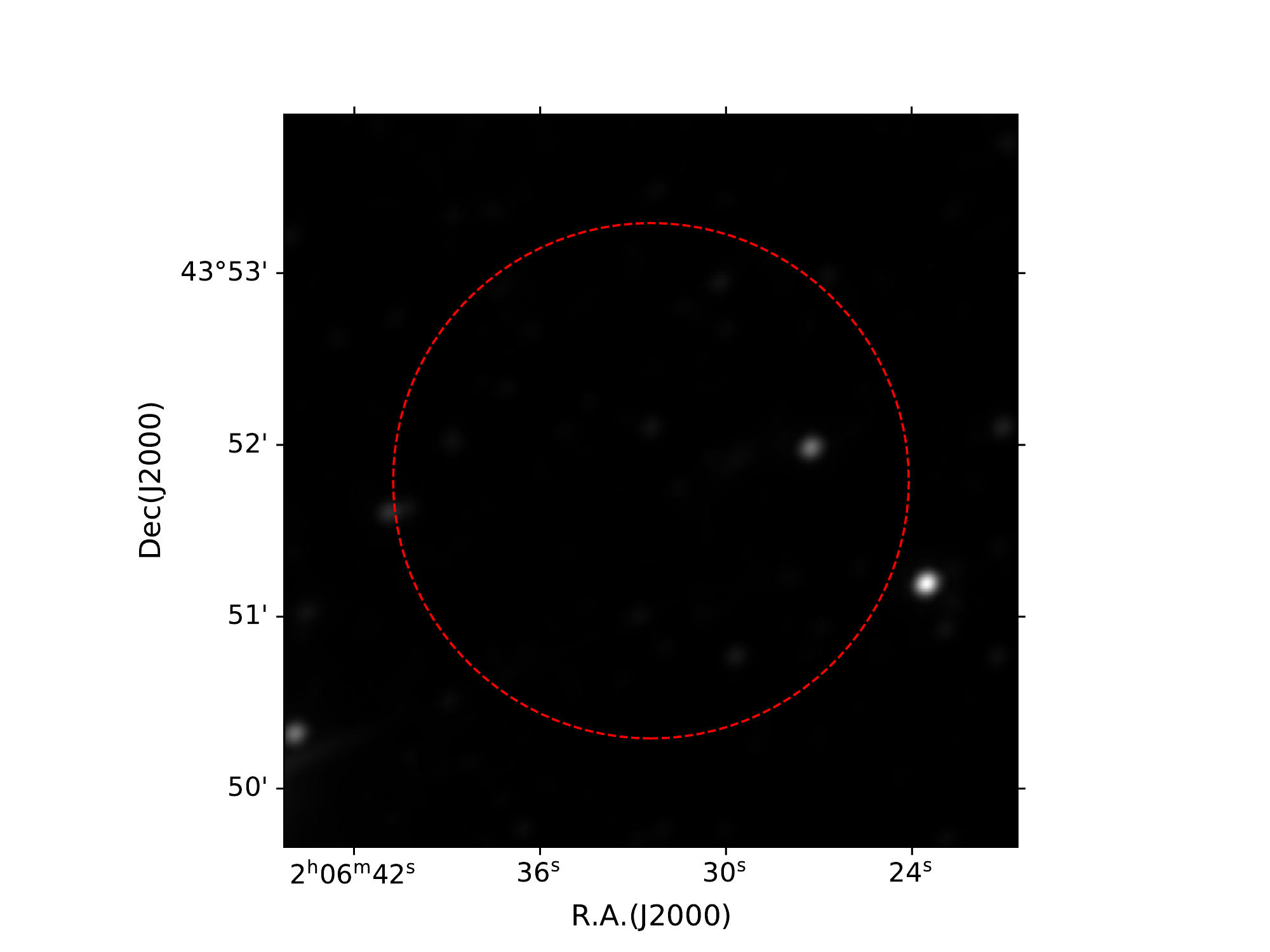}
		\caption{Left and middle plot show  HI source spectrum line, HI intensity mappings,respecitively, and its counterparts image in unWISE W1/W2 NEO6 band  in right panel,Left and middle plot show  a few spectrum lines, HI intensity mappings,respecitively, and its counterparts image in unWISE W1/W2 NEO6 band in right panel.\label{map3}}
	\end{figure*}

\begin{figure*}[ht]
		\includegraphics[width=0.16\textwidth]{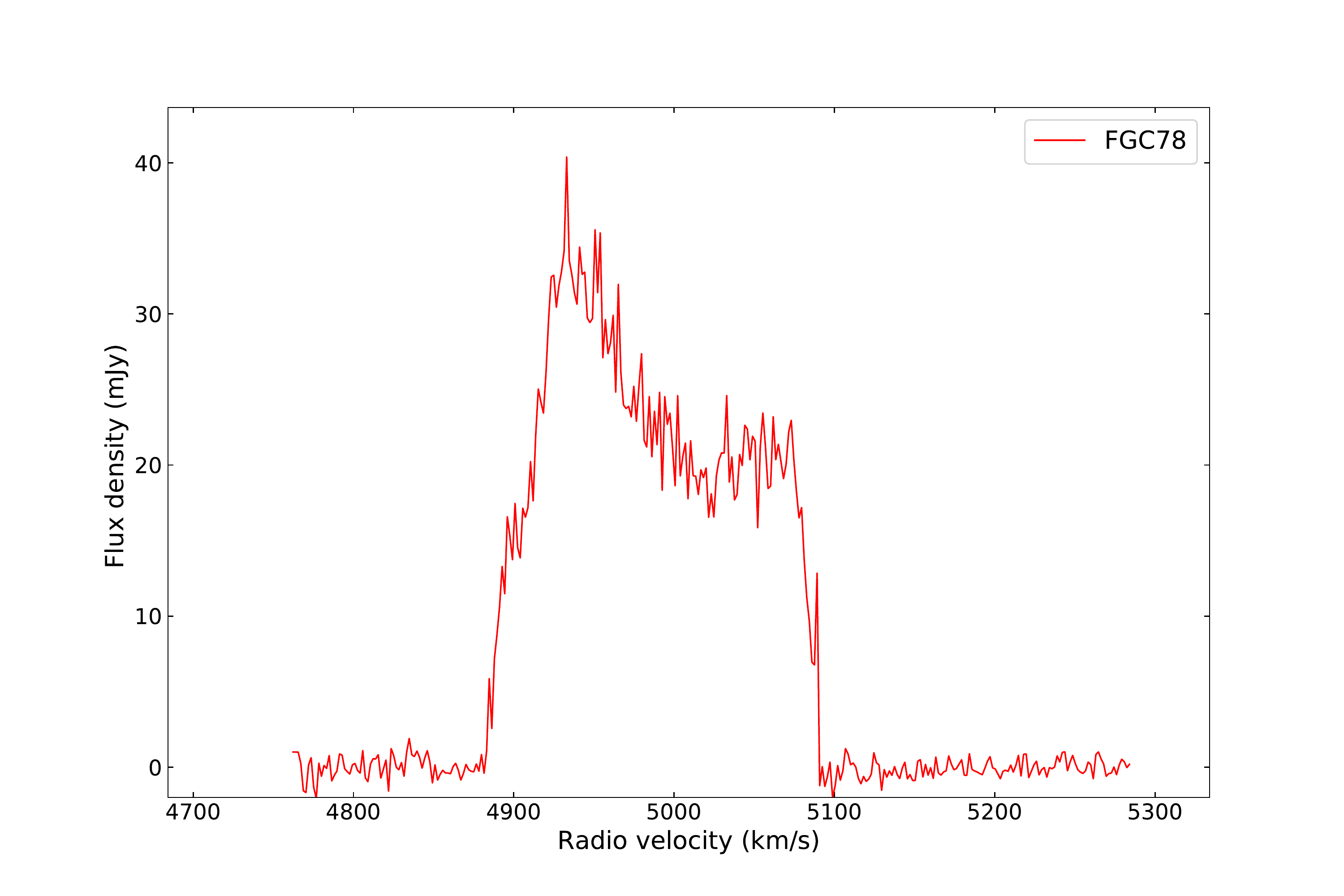}
		\includegraphics[width=0.15\textwidth]{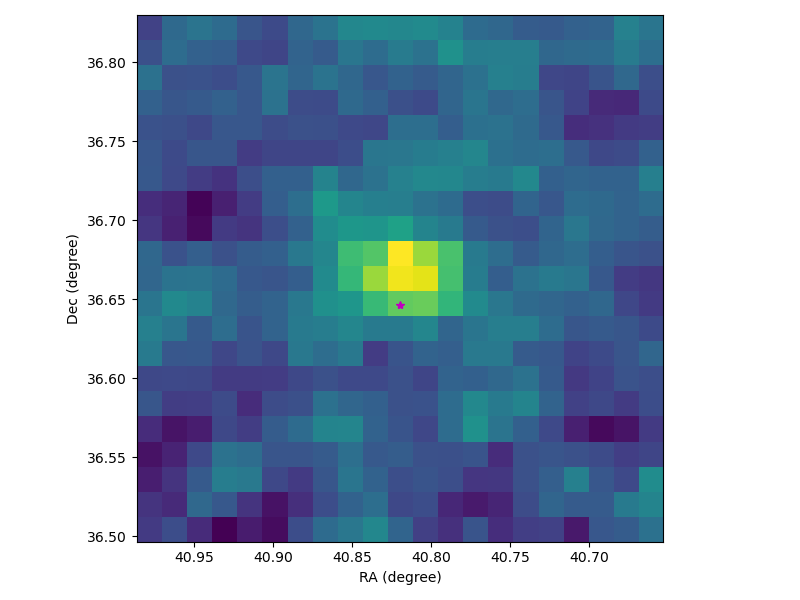}
		\includegraphics[width=0.17\textwidth]{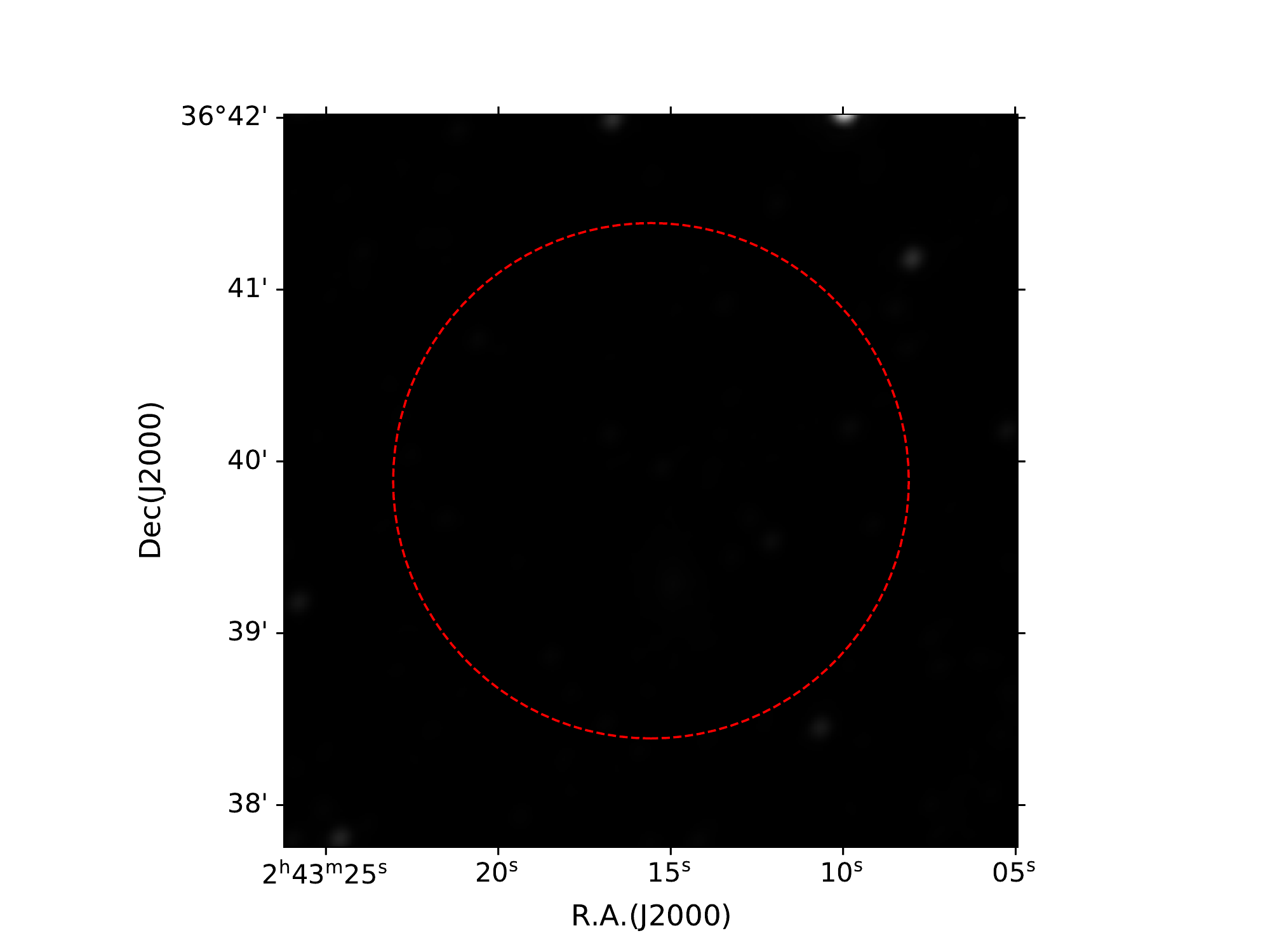}
		\includegraphics[width=0.15\textwidth]{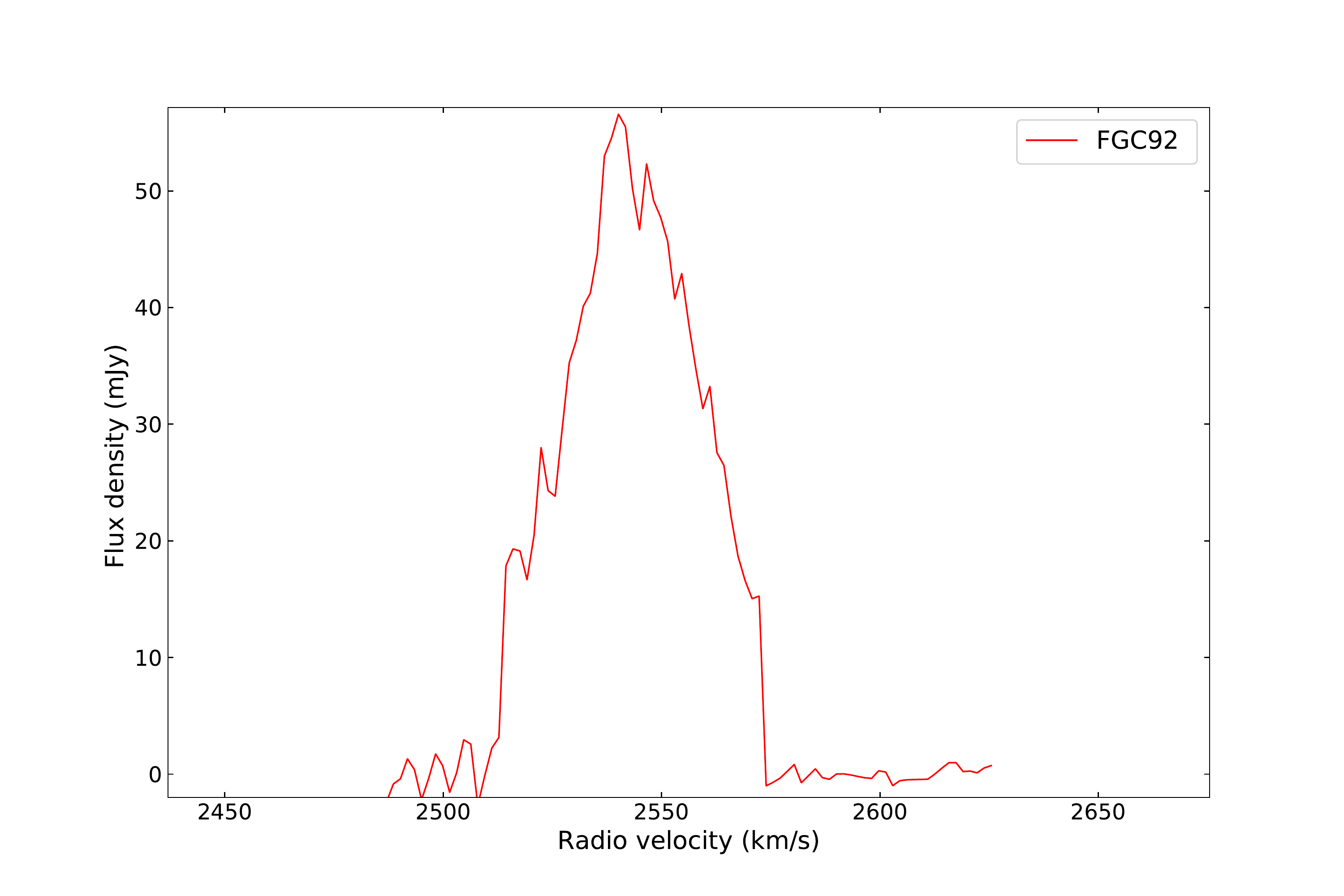}
		\includegraphics[width=0.16\textwidth]{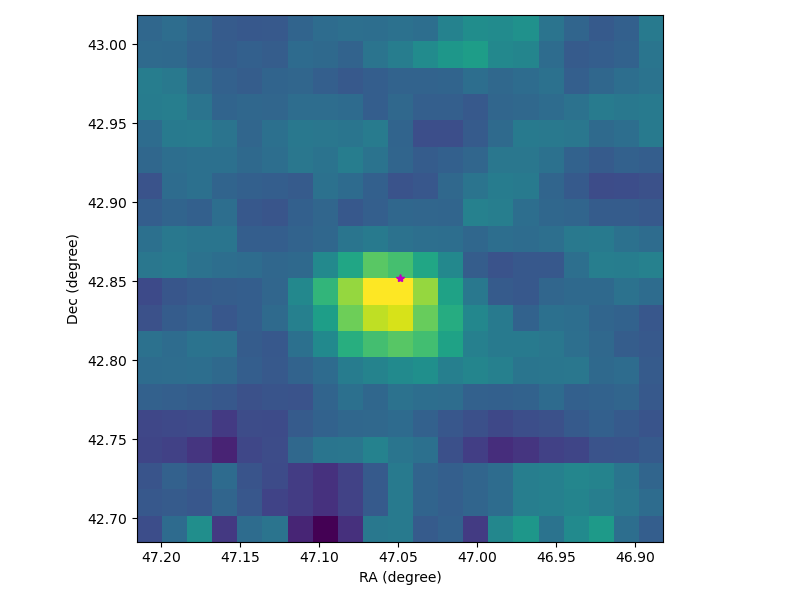}
		\includegraphics[width=0.17\textwidth]{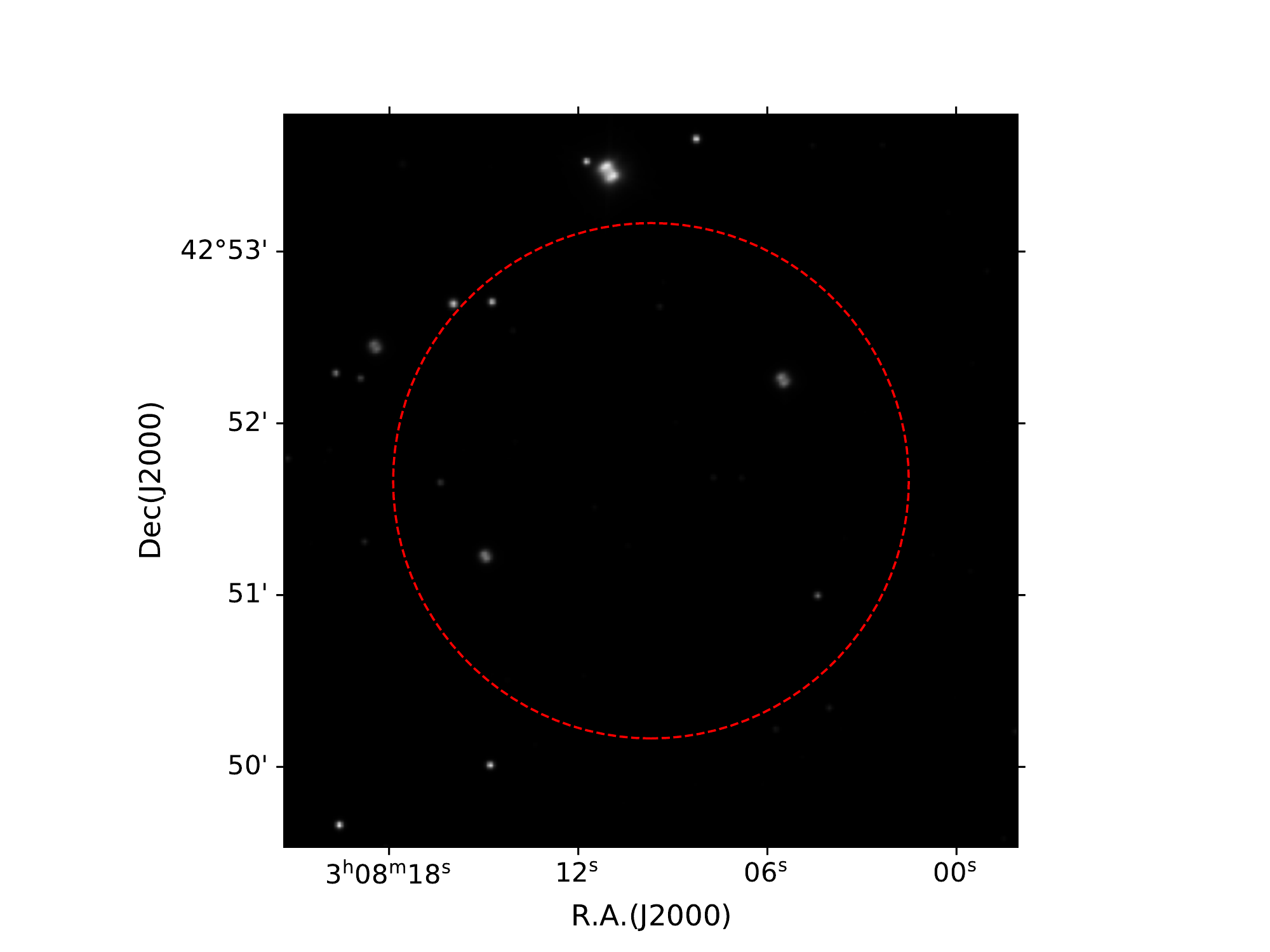}
		\caption{Left and middle plot show  HI source spectrum line, HI intensity mappings,respecitively, and its counterparts image  in unWISE W1/W2 NEO6 band in right panelLeft and middle plot show  HI source spectrum line, HI intensity mappings,respecitively, and its counterparts image  in SDSS grz-band  in right panel.\label{map4}}
	\end{figure*}

\begin{figure*}[ht]
		\includegraphics[width=0.15\textwidth]{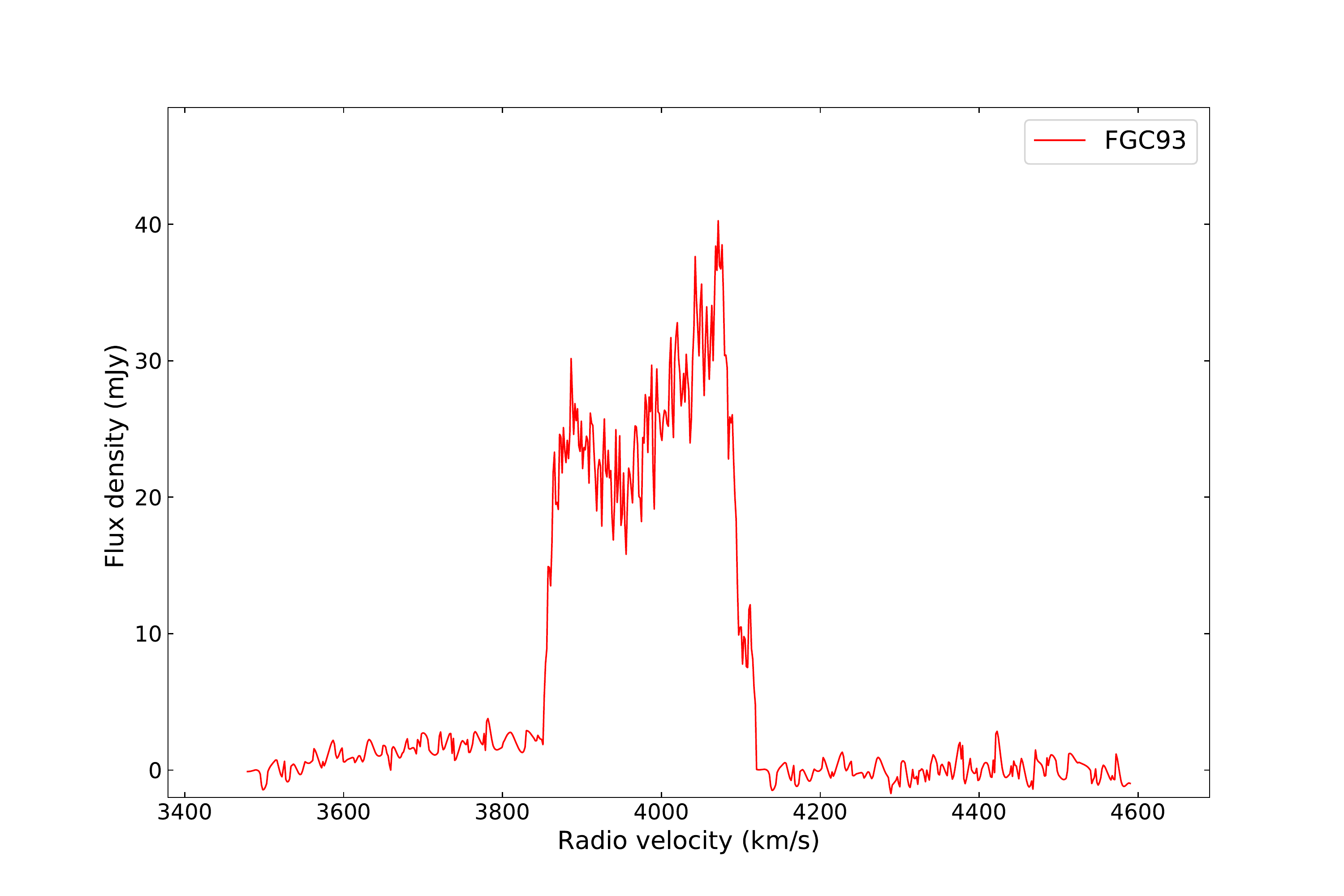}
		\includegraphics[width=0.16\textwidth]{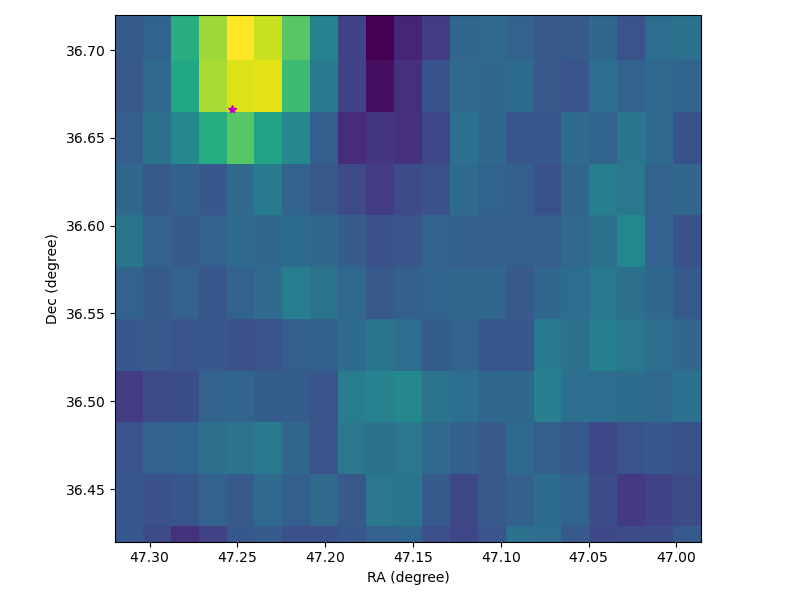}
		\includegraphics[width=0.17\textwidth]{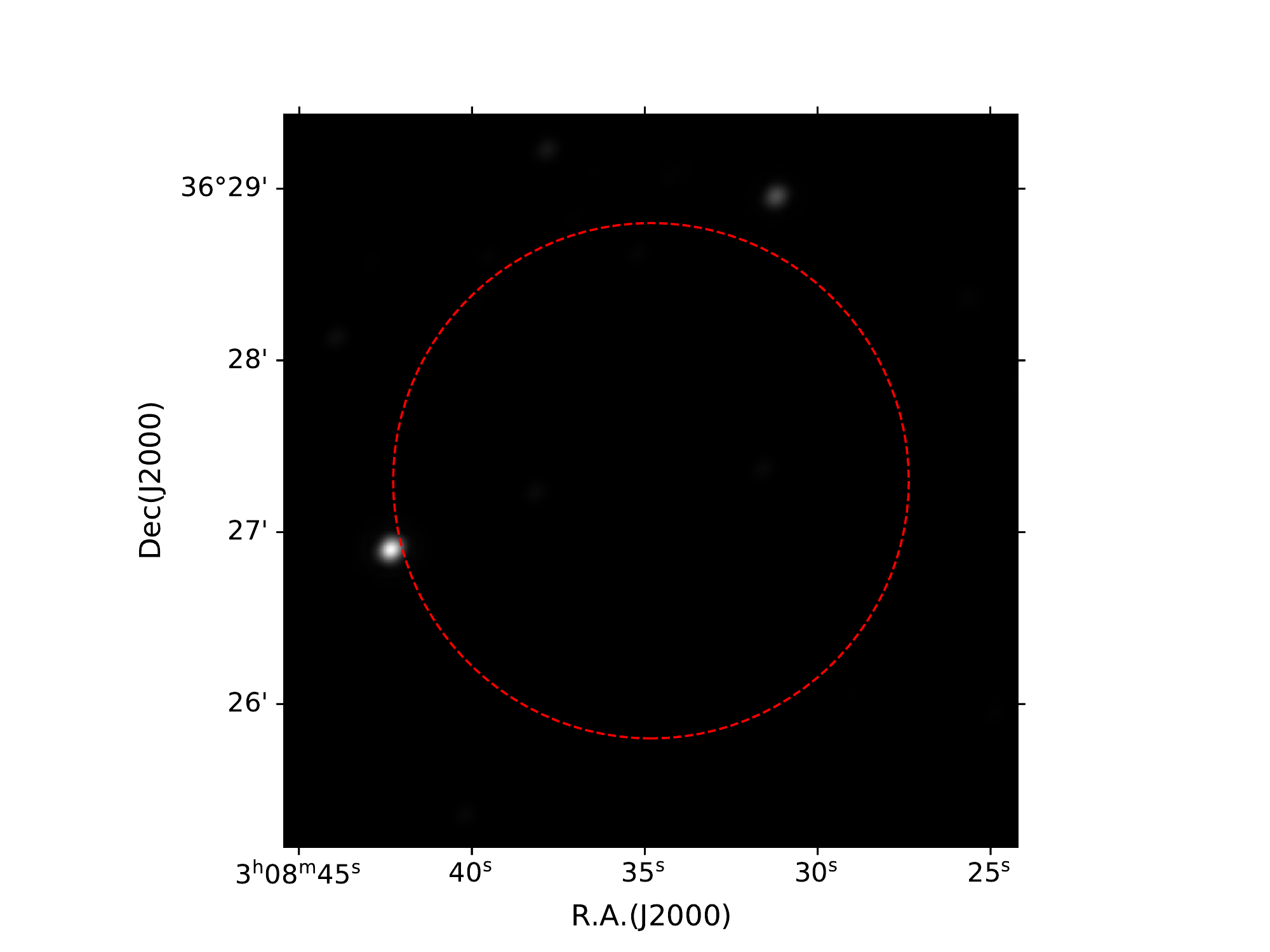}
		\includegraphics[width=0.15\textwidth]{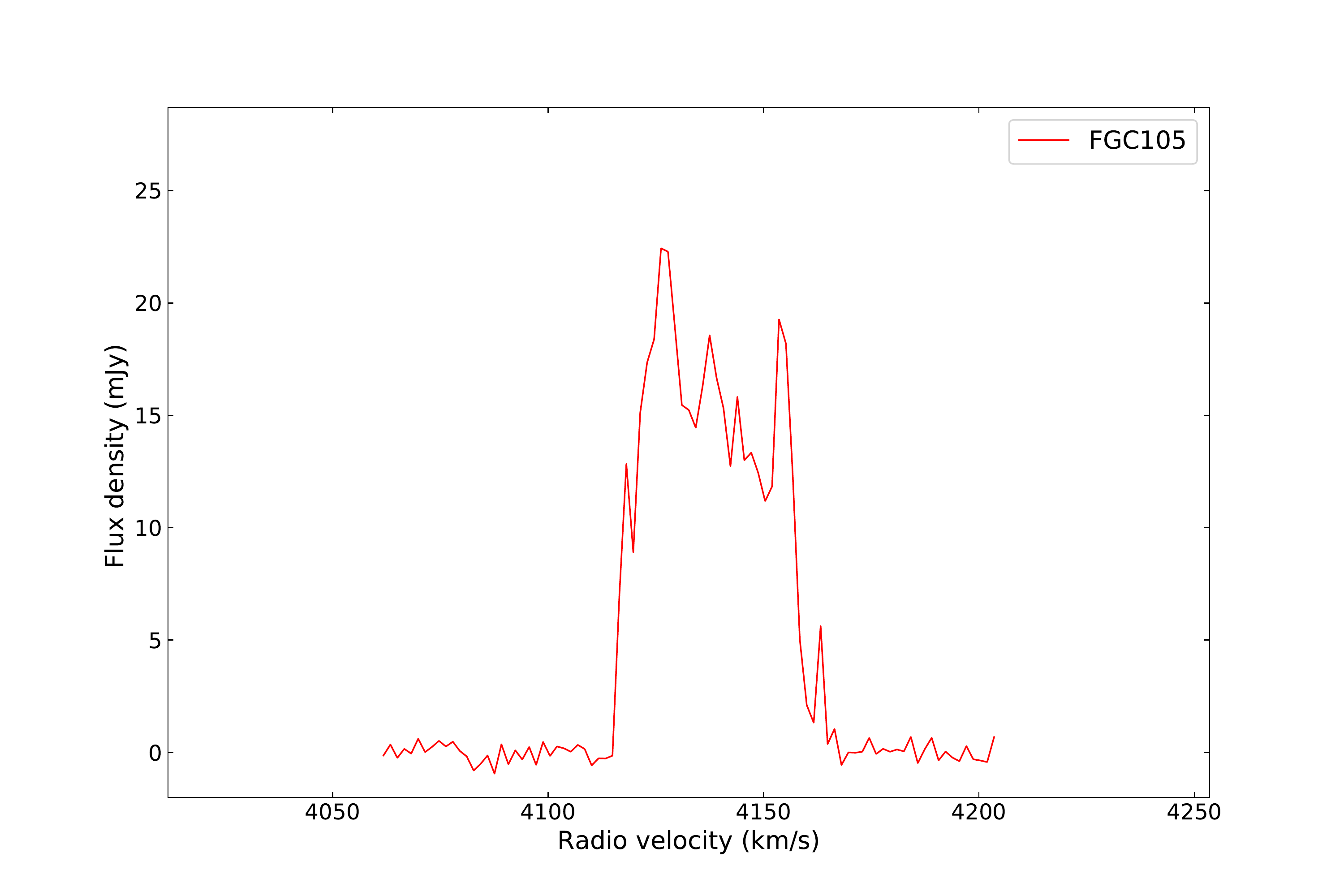}
		\includegraphics[width=0.16\textwidth]{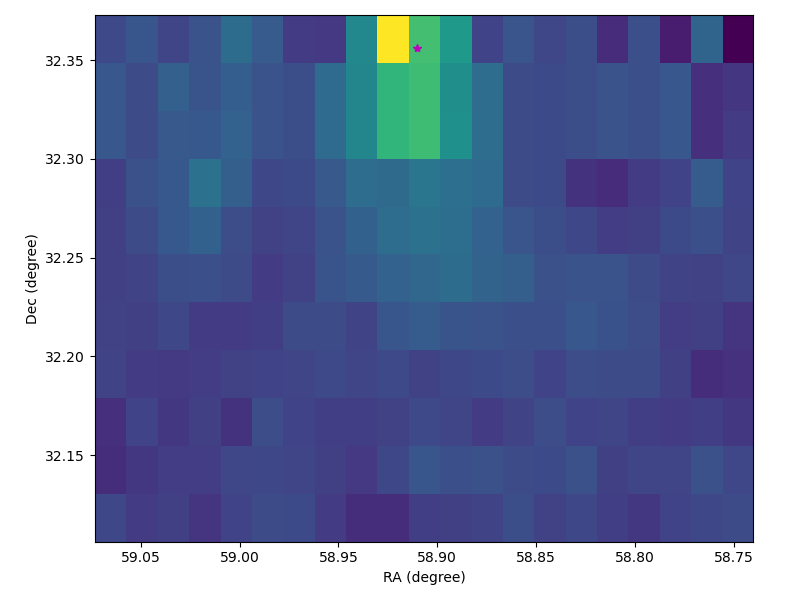}
		\includegraphics[width=0.17\textwidth]{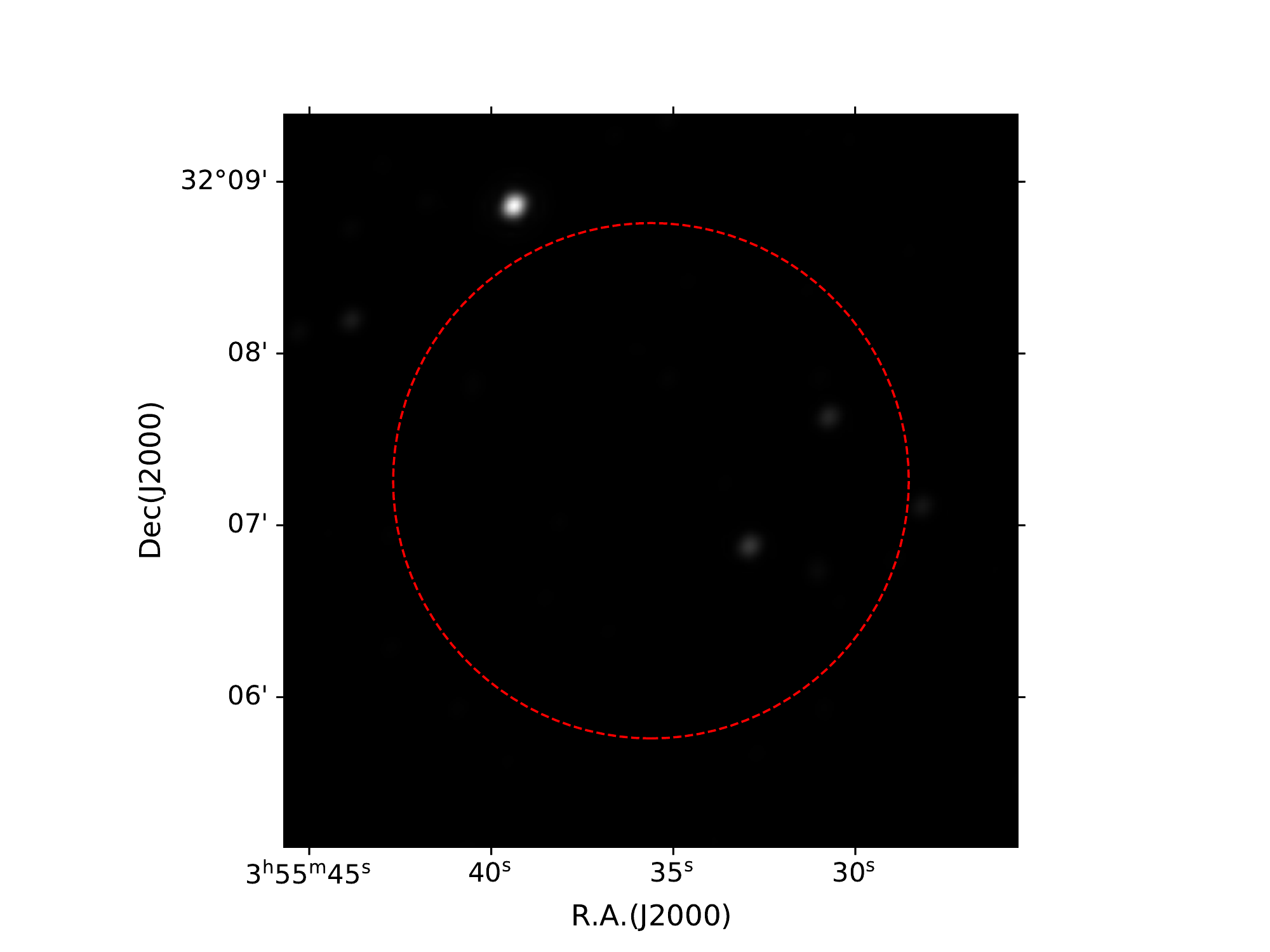}
		\caption{Left and middle plot show  HI source spectrum line, HI intensity mappings,respecitively, and its counterparts image  in unWISE W1/W2 NEO6 band  in right panel,Left and middle plot show  HI source spectrum line, HI intensity mappings,respecitively, and its counterparts image in unWISE W1/W2 NEO6 band  in right panel.\label{map5}}
	\end{figure*}

\begin{figure*}[ht]
		\includegraphics[width=0.15\textwidth]{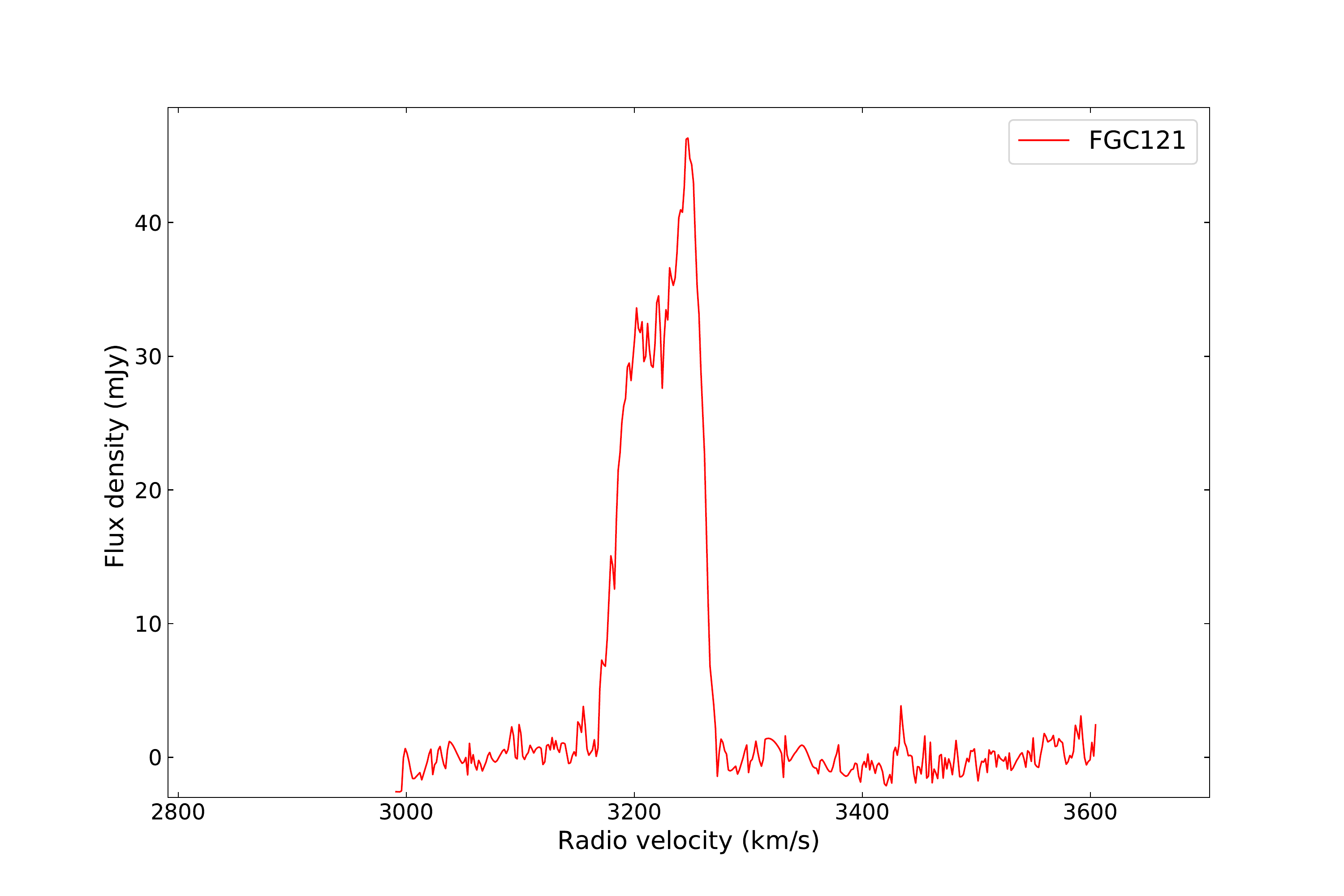}
		\includegraphics[width=0.16\textwidth]{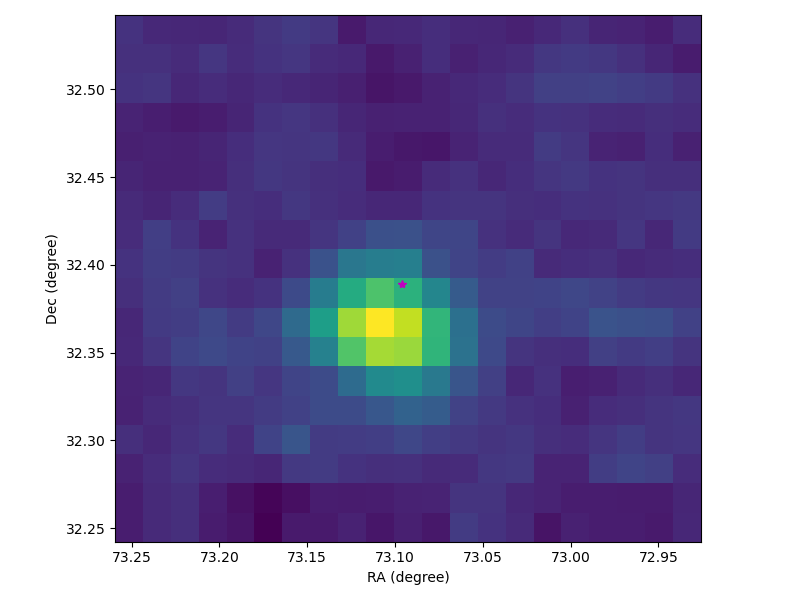}
		\includegraphics[width=0.18\textwidth]{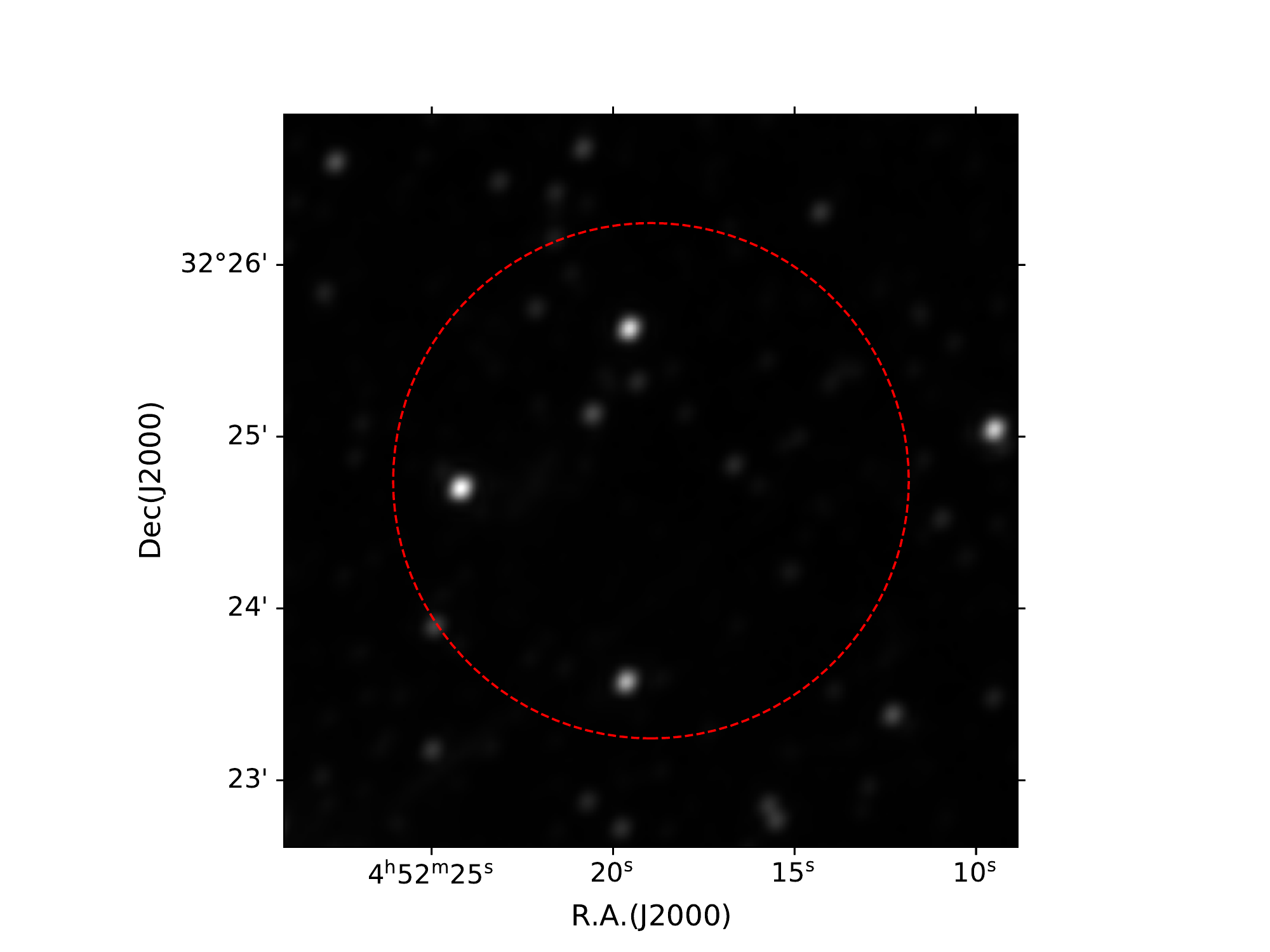}
		\includegraphics[width=0.15\textwidth]{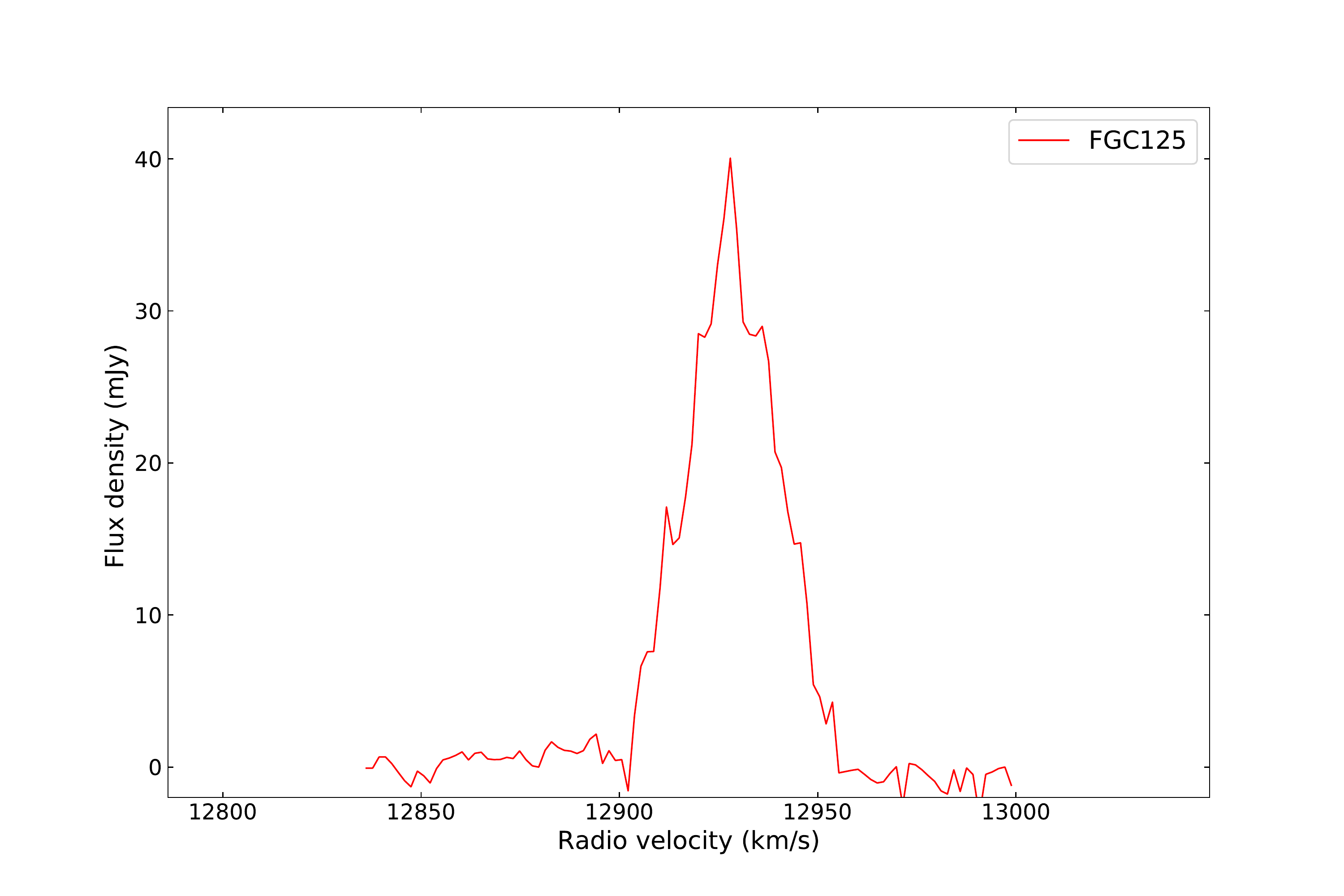}
		\includegraphics[width=0.16\textwidth]{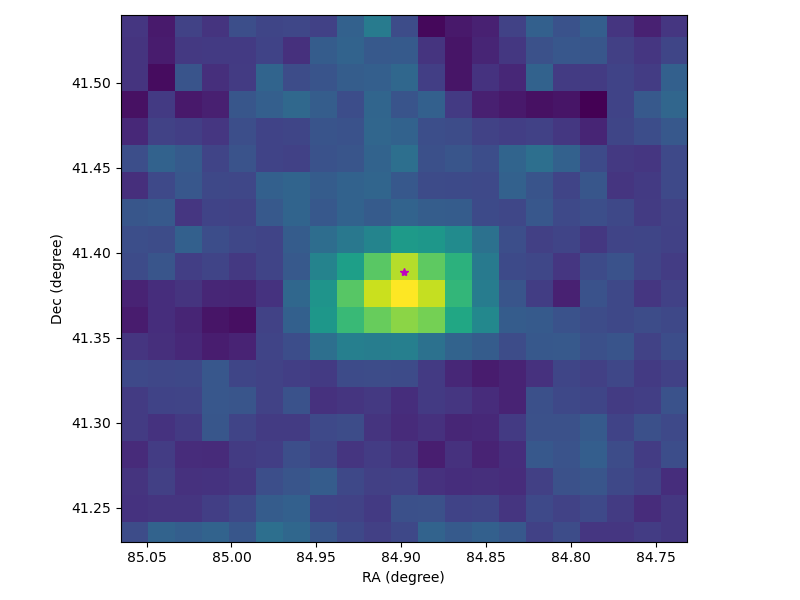}
		\includegraphics[width=0.17\textwidth]{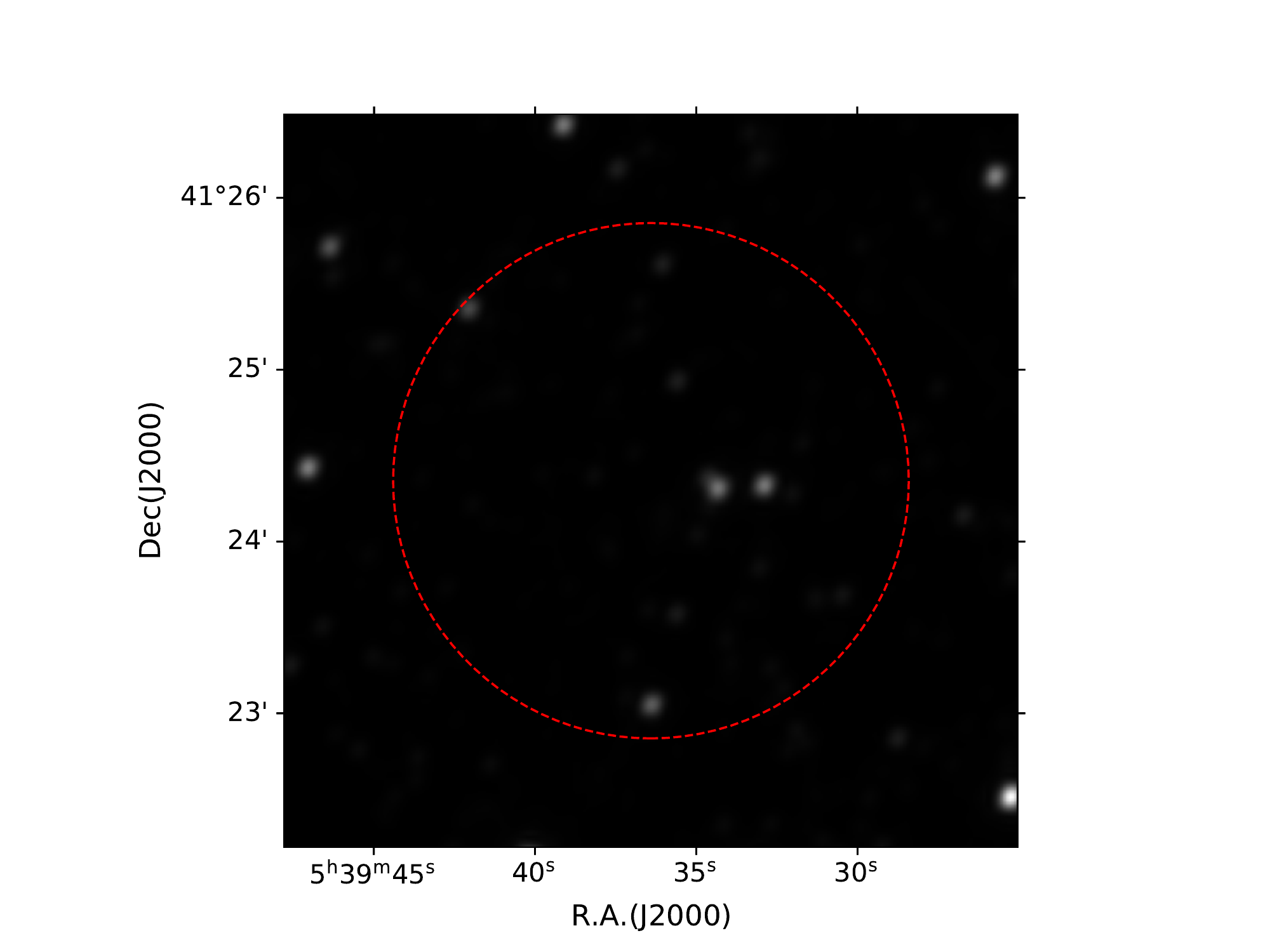}
		\caption{Left and middle plot show  HI source spectrum line, HI intensity mappings,respecitively, and its counterparts image in unWISE W1/W2 NEO6 band  in right panel,Left and middle plot show  HI source spectrum line, HI intensity mappings,respecitively, and its counterparts image in unWISE W1/W2 NEO6 band   in right panel.\label{map6}}
	\end{figure*}

\begin{figure*}[htb]
		\includegraphics[width=0.15\textwidth]{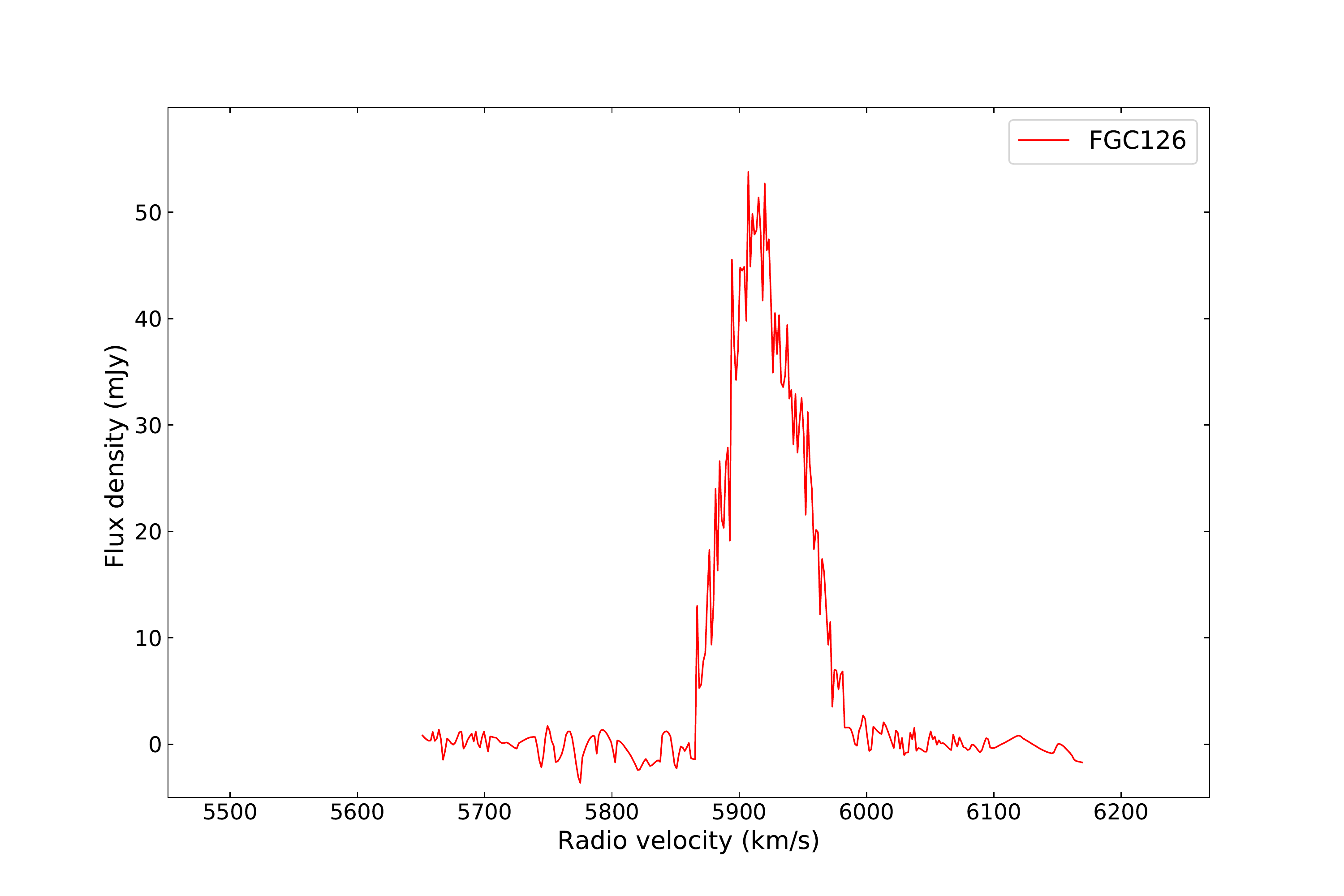}
		\includegraphics[width=0.16\textwidth]{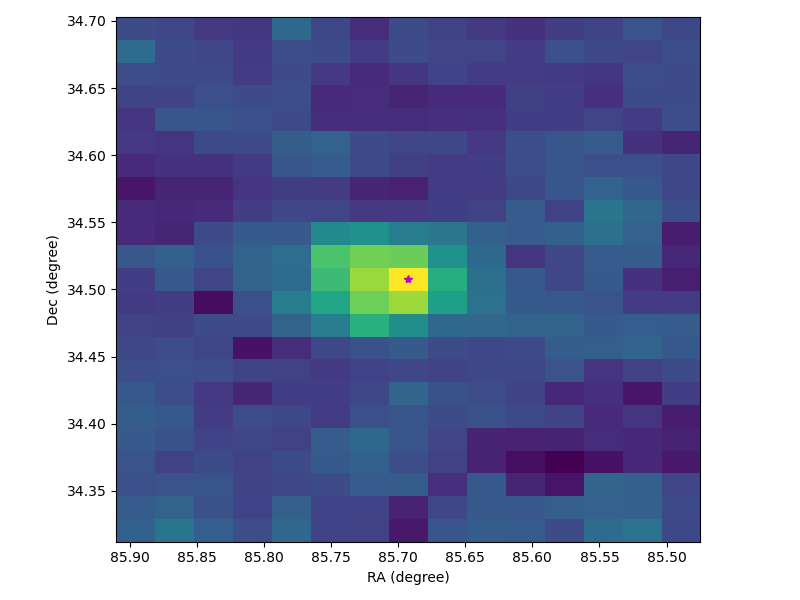}
		\includegraphics[width=0.17\textwidth]{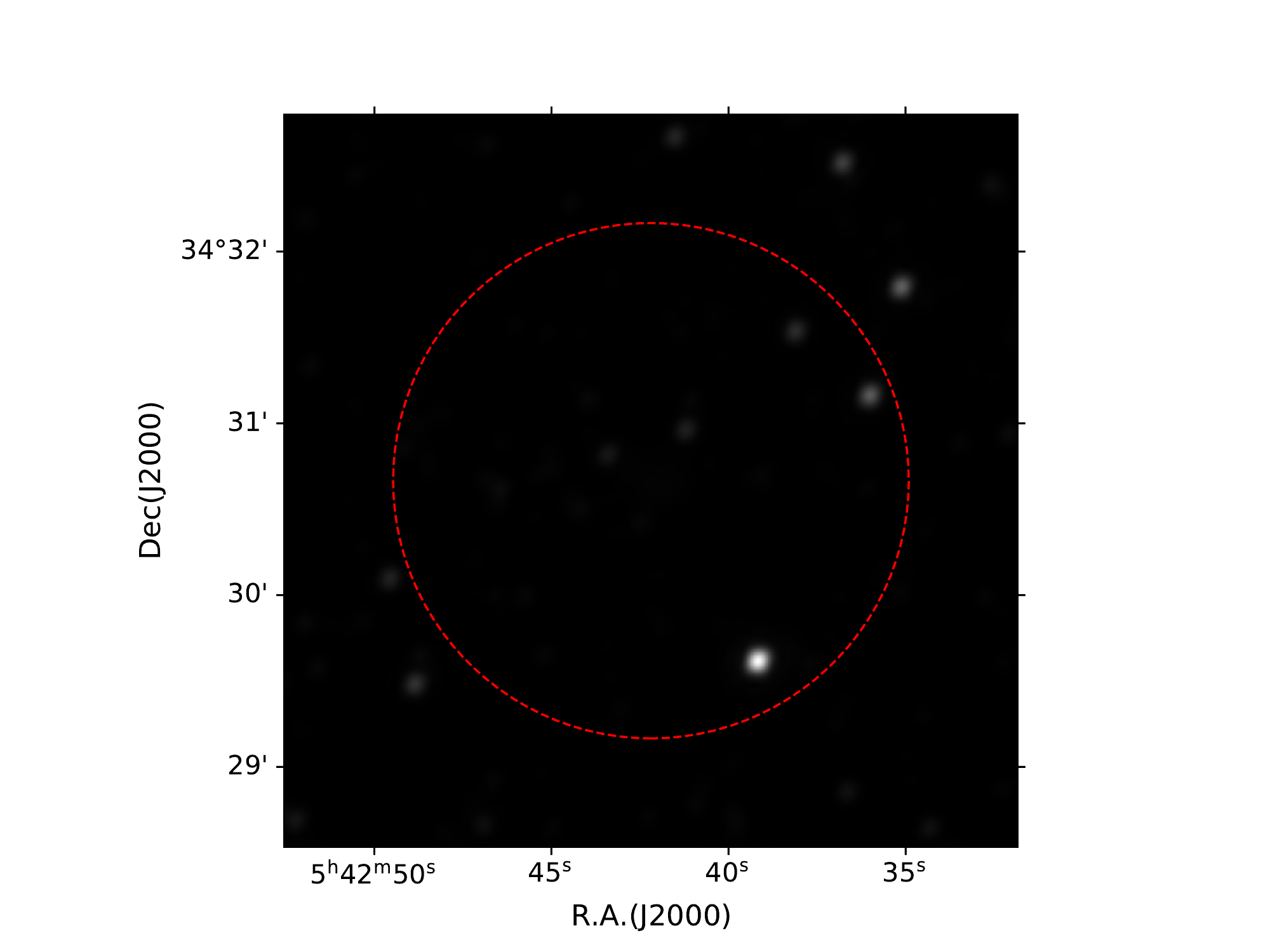}
		\includegraphics[width=0.15\textwidth]{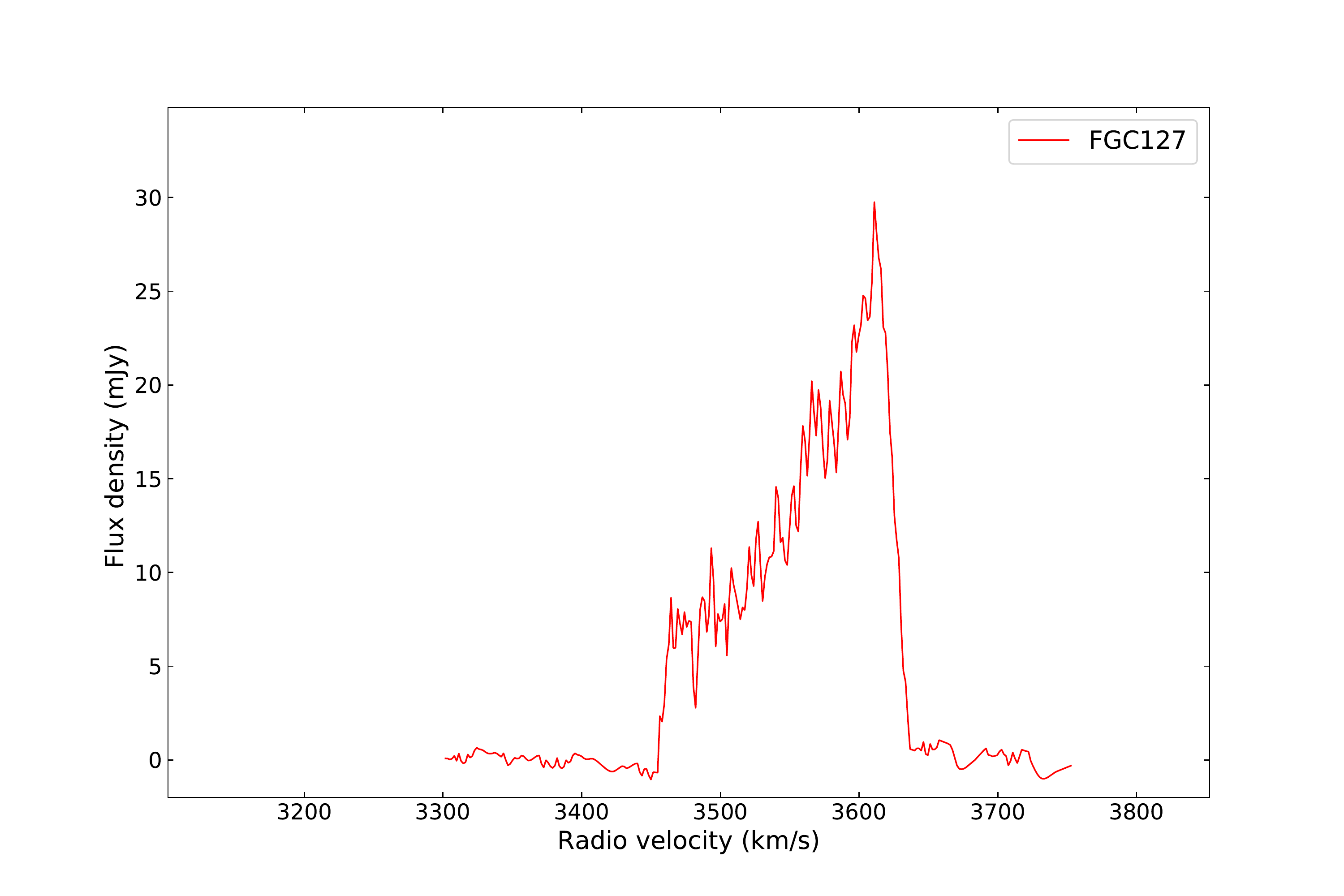}
		\includegraphics[width=0.16\textwidth]{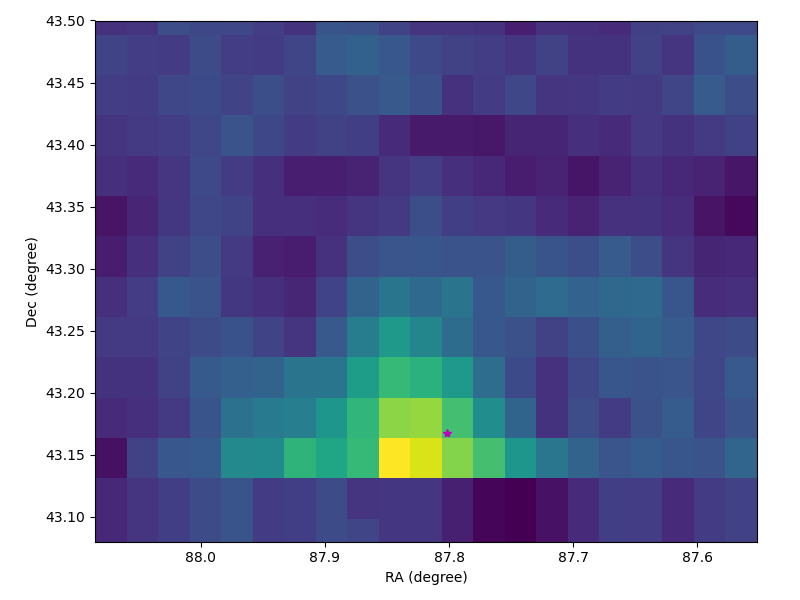}
		\includegraphics[width=0.17\textwidth]{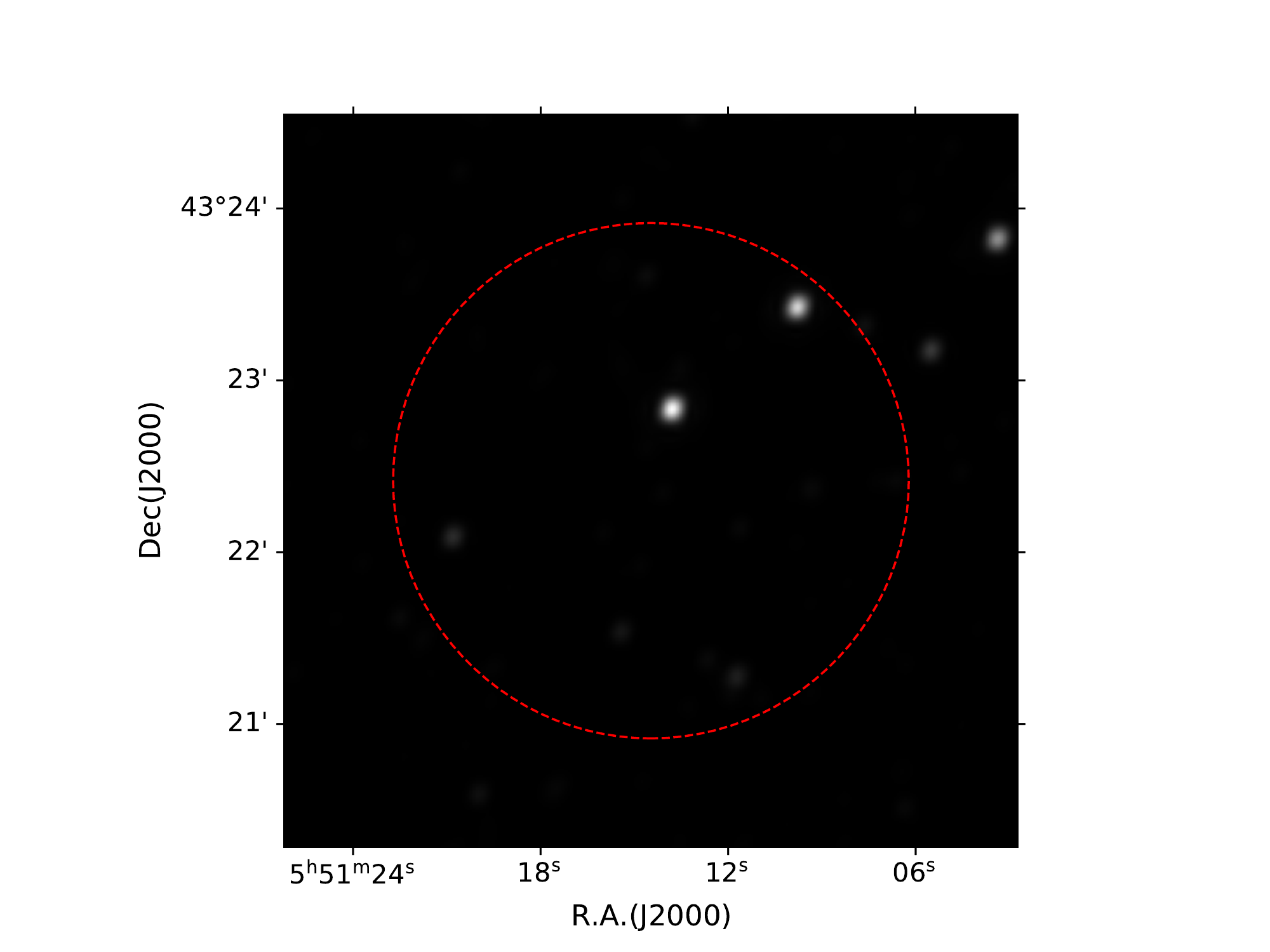}
		\caption{Left and middle plot show  HI source spectrum line, HI intensity mappings,respecitively, and its counterparts image in unWISE W1/W2 NEO6 band   in right panel,Left and middle plot show  HI source spectrum line, HI intensity mappings,respecitively, and its counterparts image in unWISE W1/W2 NEO6 band in right panel.\label{map14}}
	\end{figure*}

\begin{figure*}[ht]
		\includegraphics[width=0.15\textwidth]{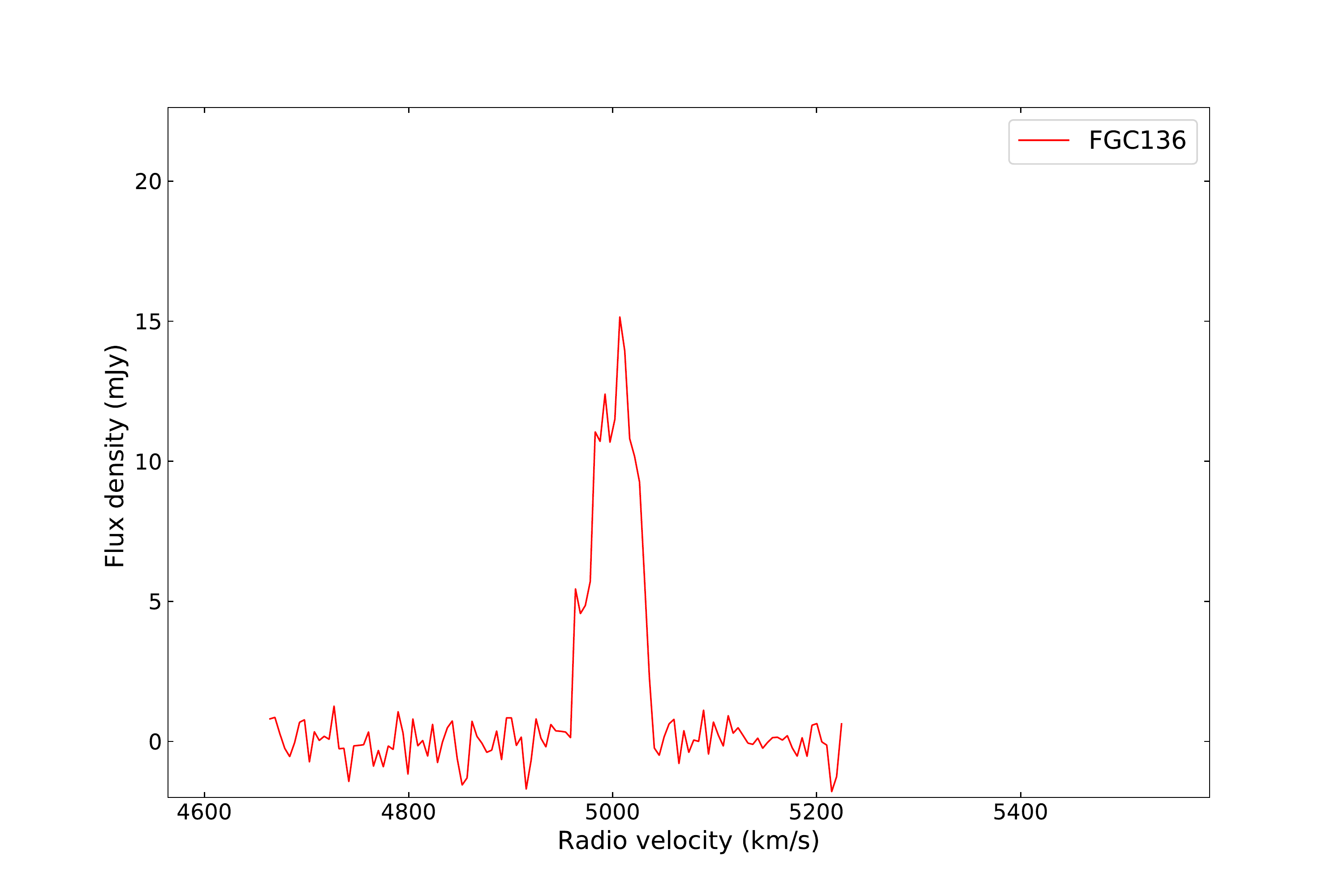}
		\includegraphics[width=0.16\textwidth]{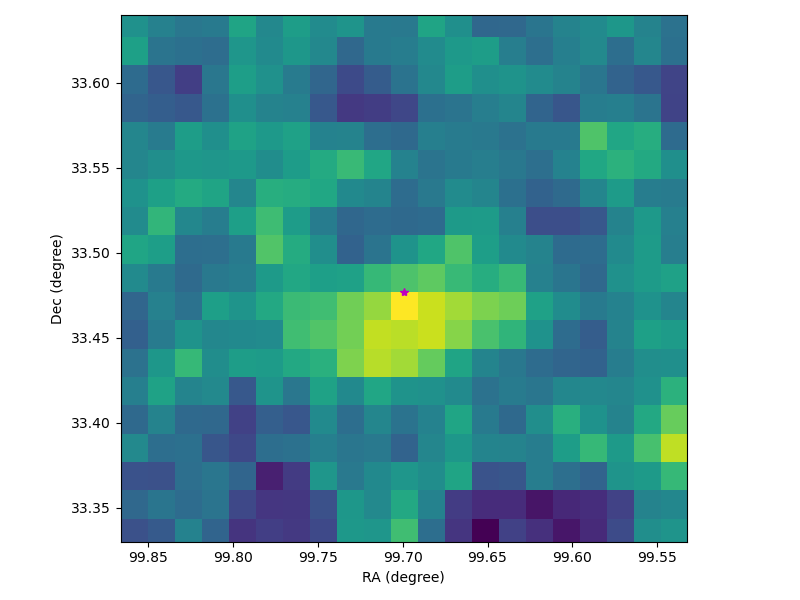}
		\includegraphics[width=0.17\textwidth]{ 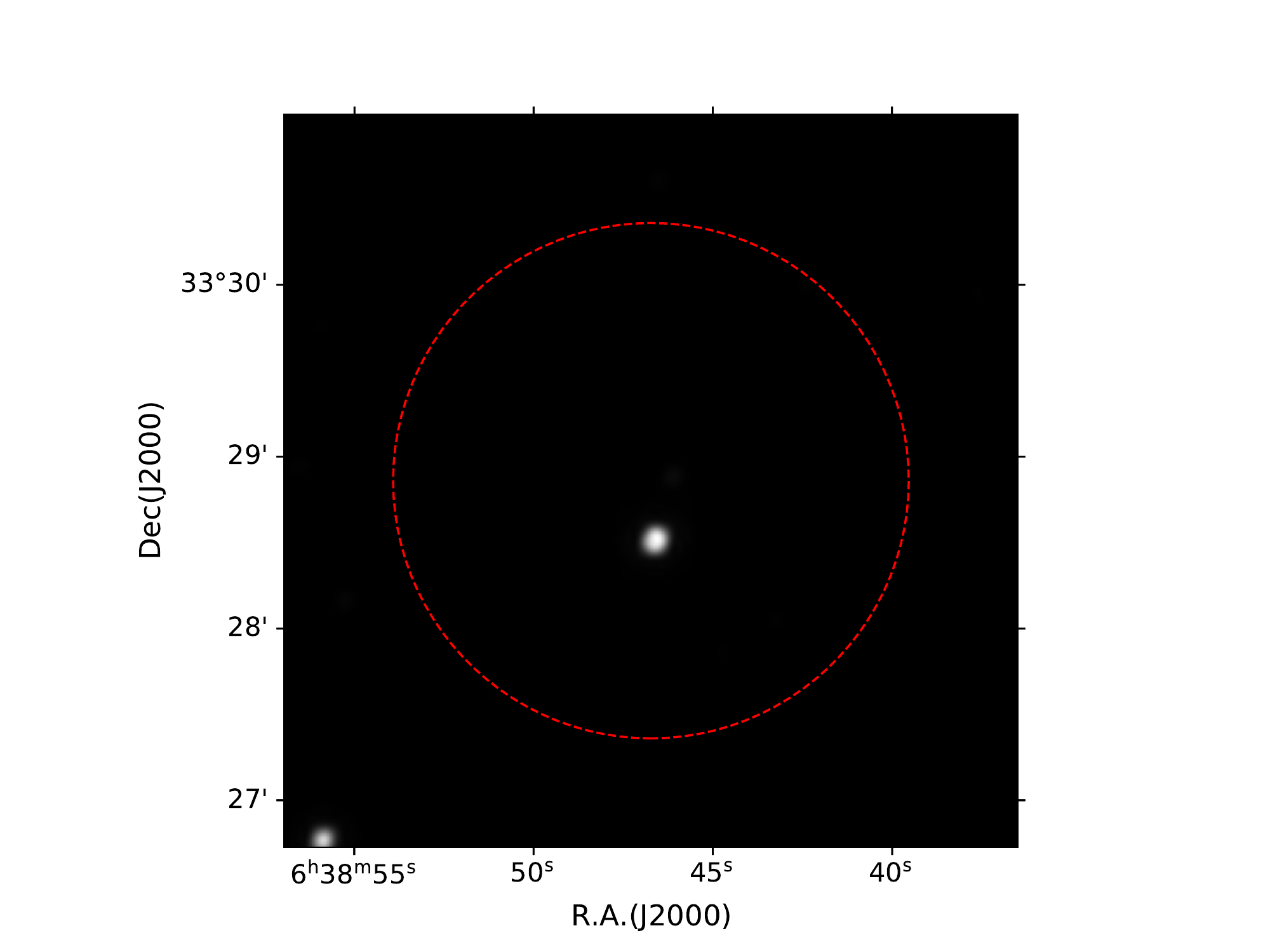}
		\includegraphics[width=0.15\textwidth]{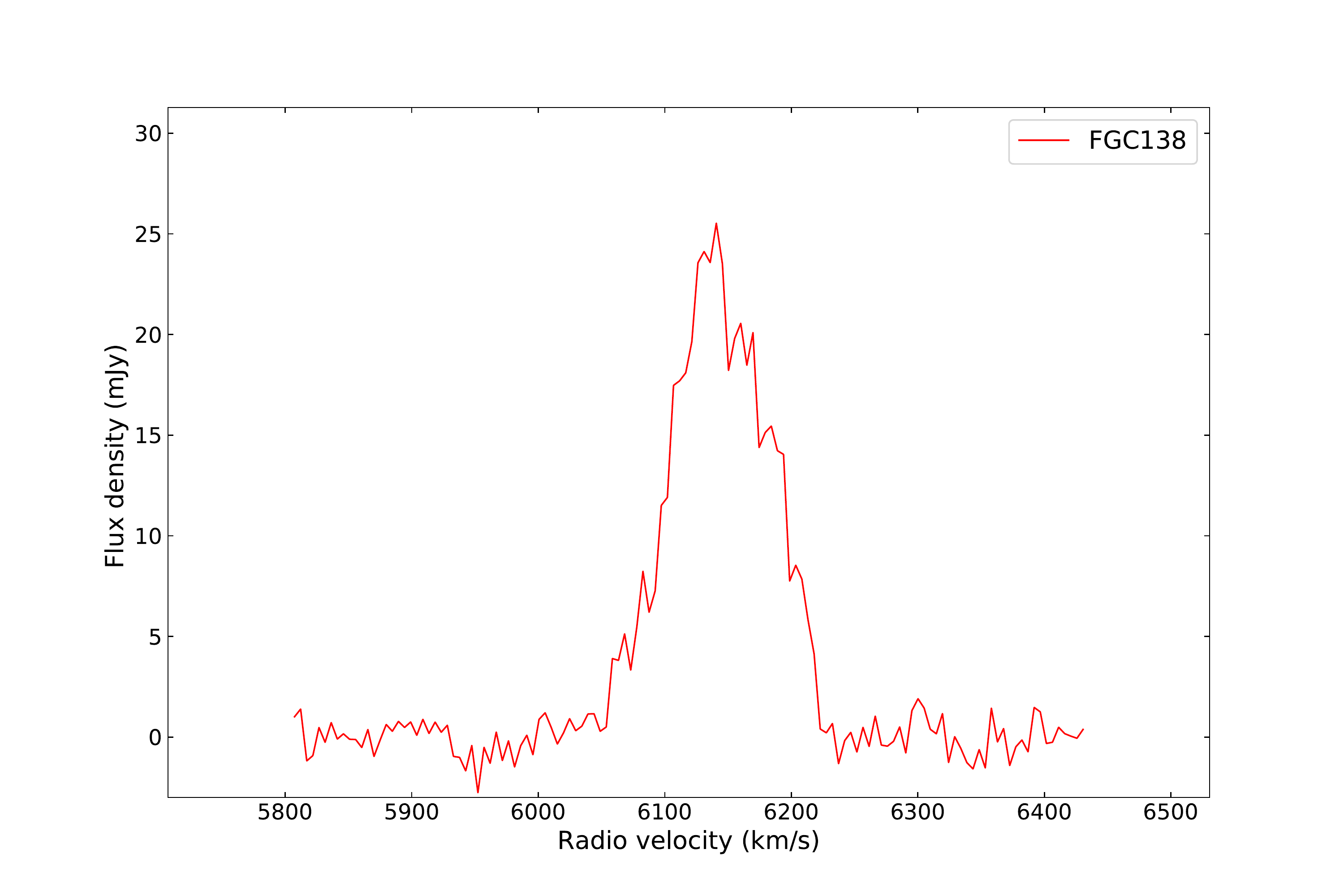}
		\includegraphics[width=0.16\textwidth]{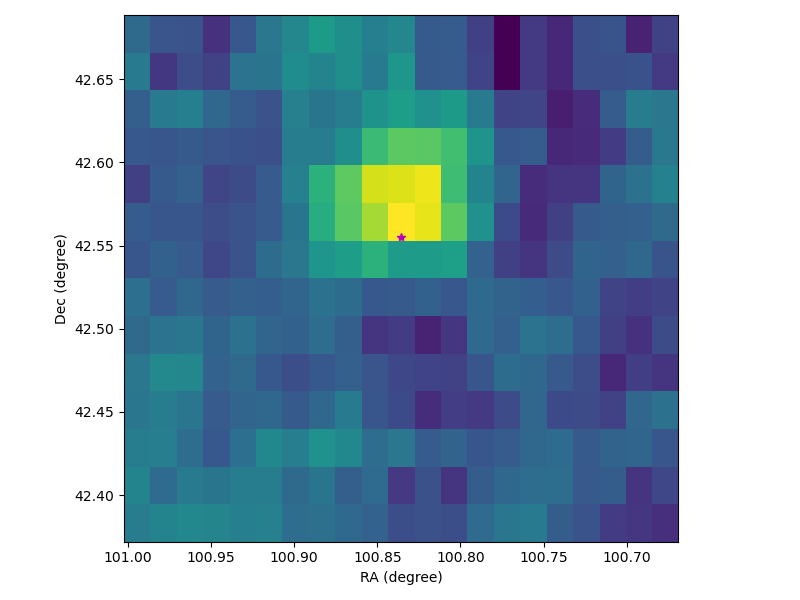}
		\includegraphics[width=0.17\textwidth]{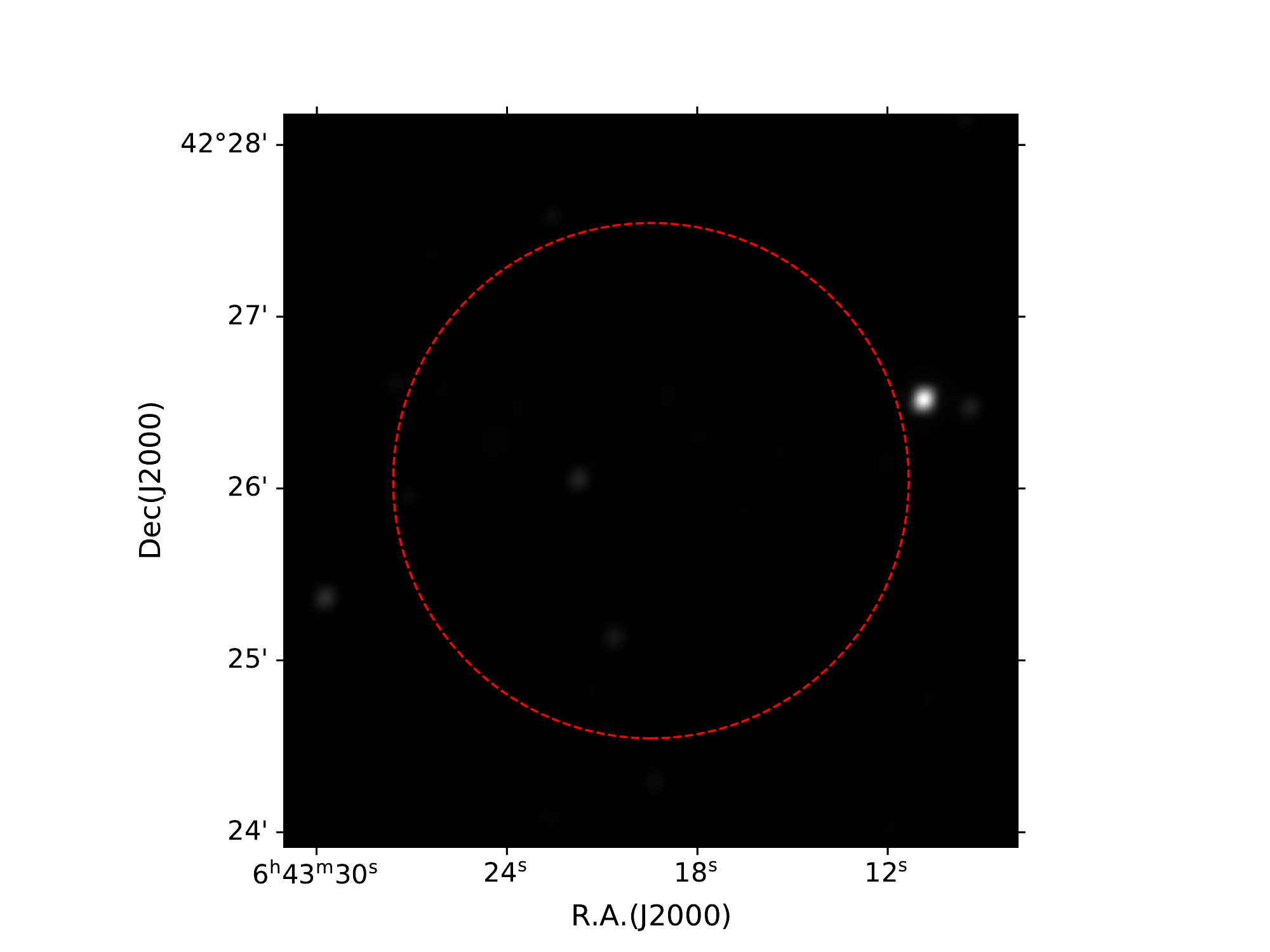}
		\caption{Left and middle plot show  HI source spectrum line, HI intensity mappings,respecitively, and its counterparts image  in unWISE W1/W2 NEO6 band   in right panel,Left and middle plot show  HI source spectrum line, HI intensity mappings,respecitively, and its counterparts image in unWISE W1/W2 NEO6 band in right panel.\label{map7}}
	\end{figure*}

\begin{figure*}[htp]
		\includegraphics[width=0.15\textwidth]{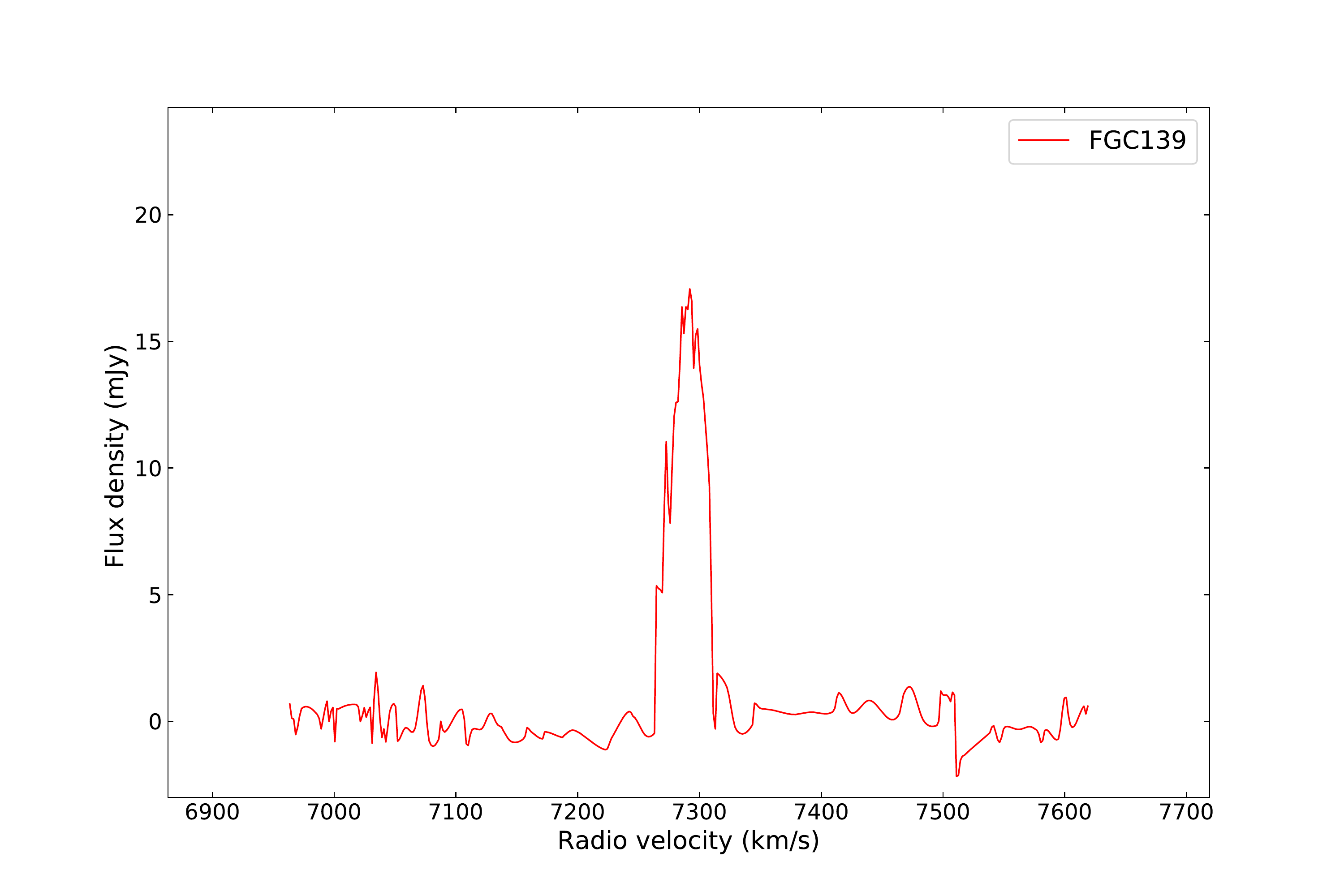}
		\includegraphics[width=0.16\textwidth]{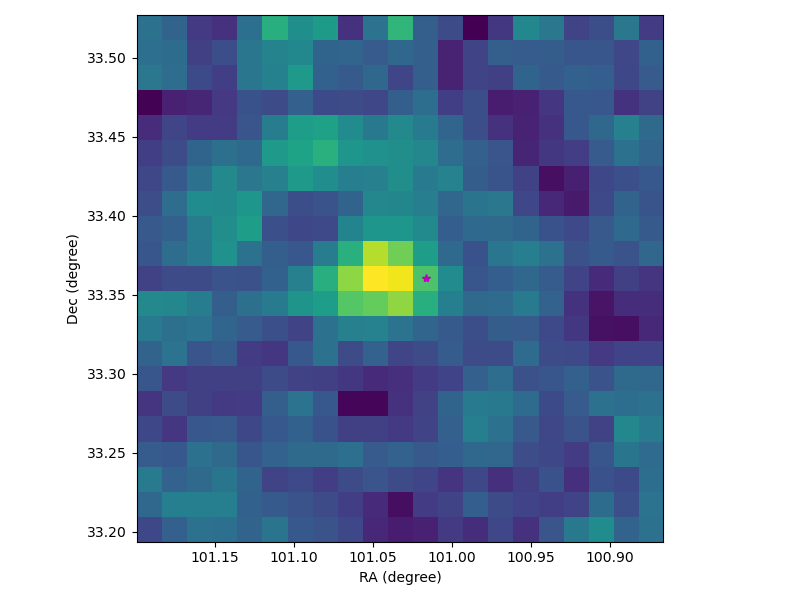}
		\includegraphics[width=0.17\textwidth]{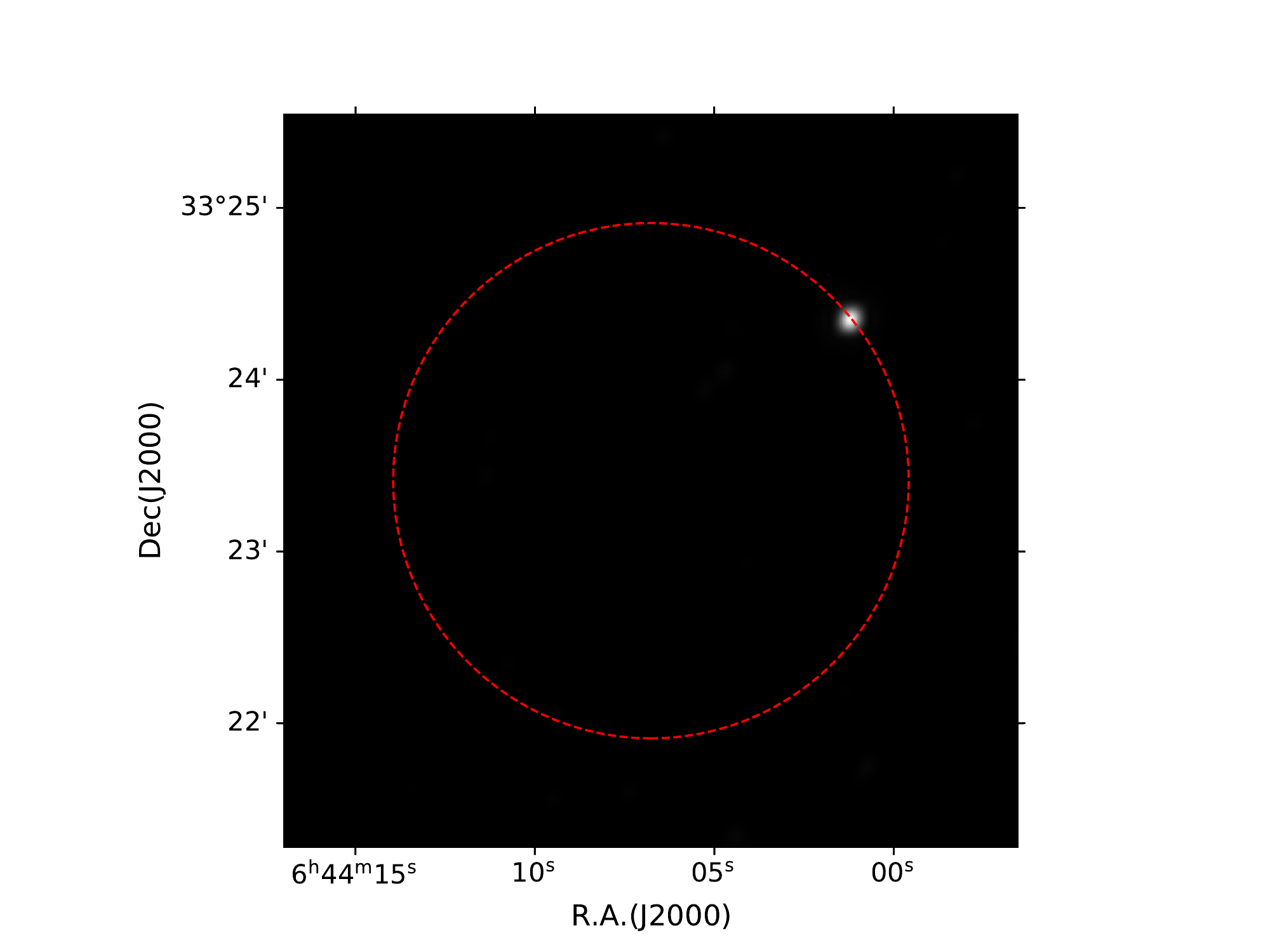}
		\includegraphics[width=0.15\textwidth]{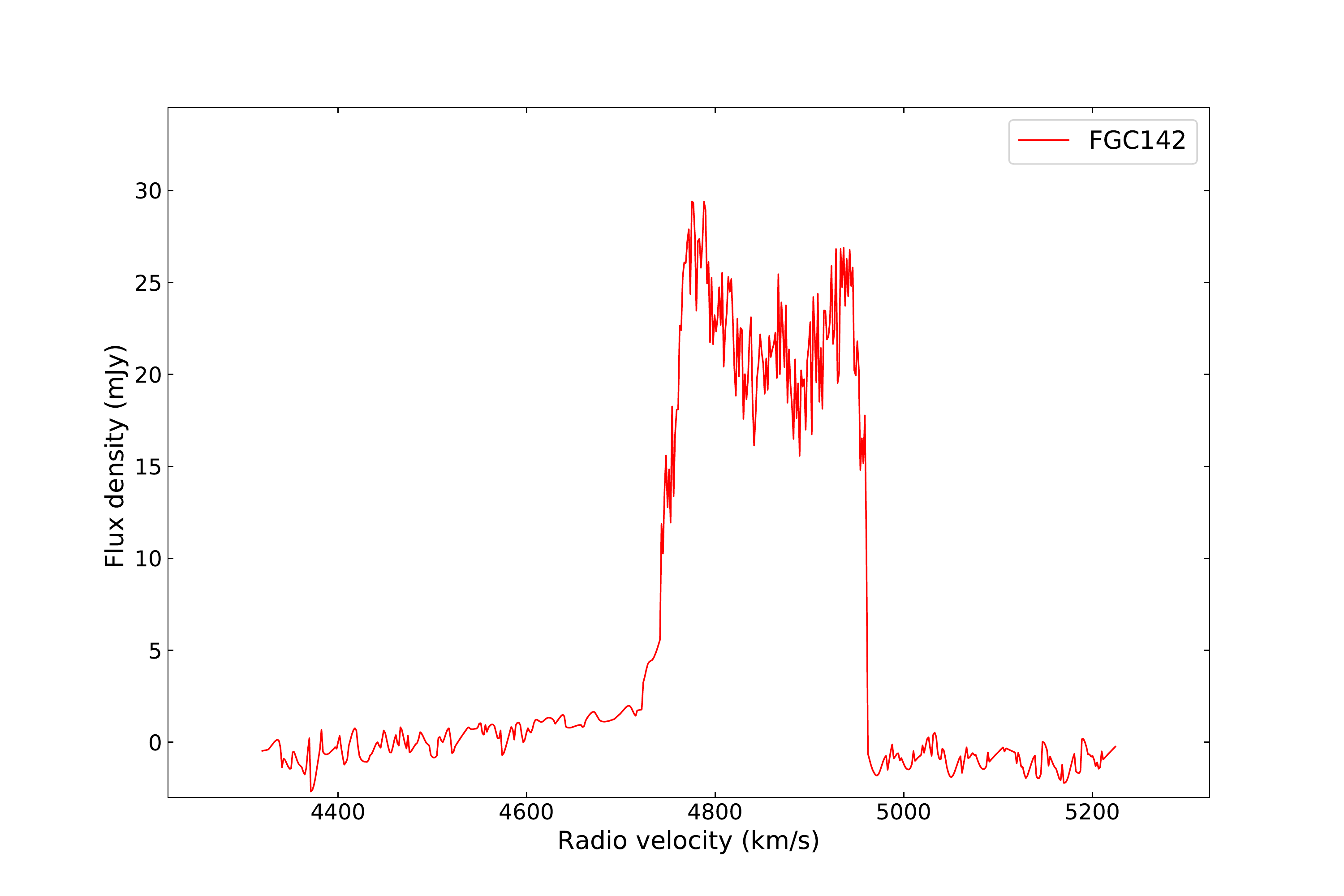}
		\includegraphics[width=0.16\textwidth]{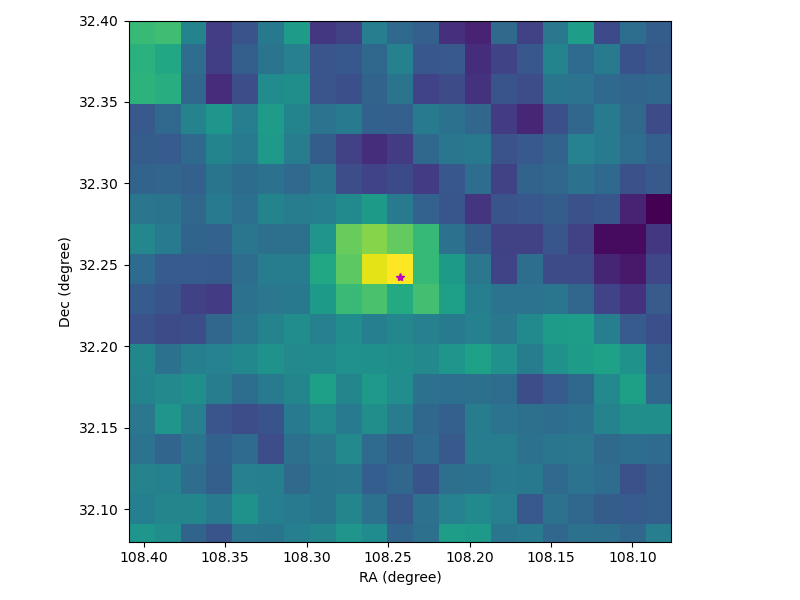}
		\includegraphics[width=0.17\textwidth]{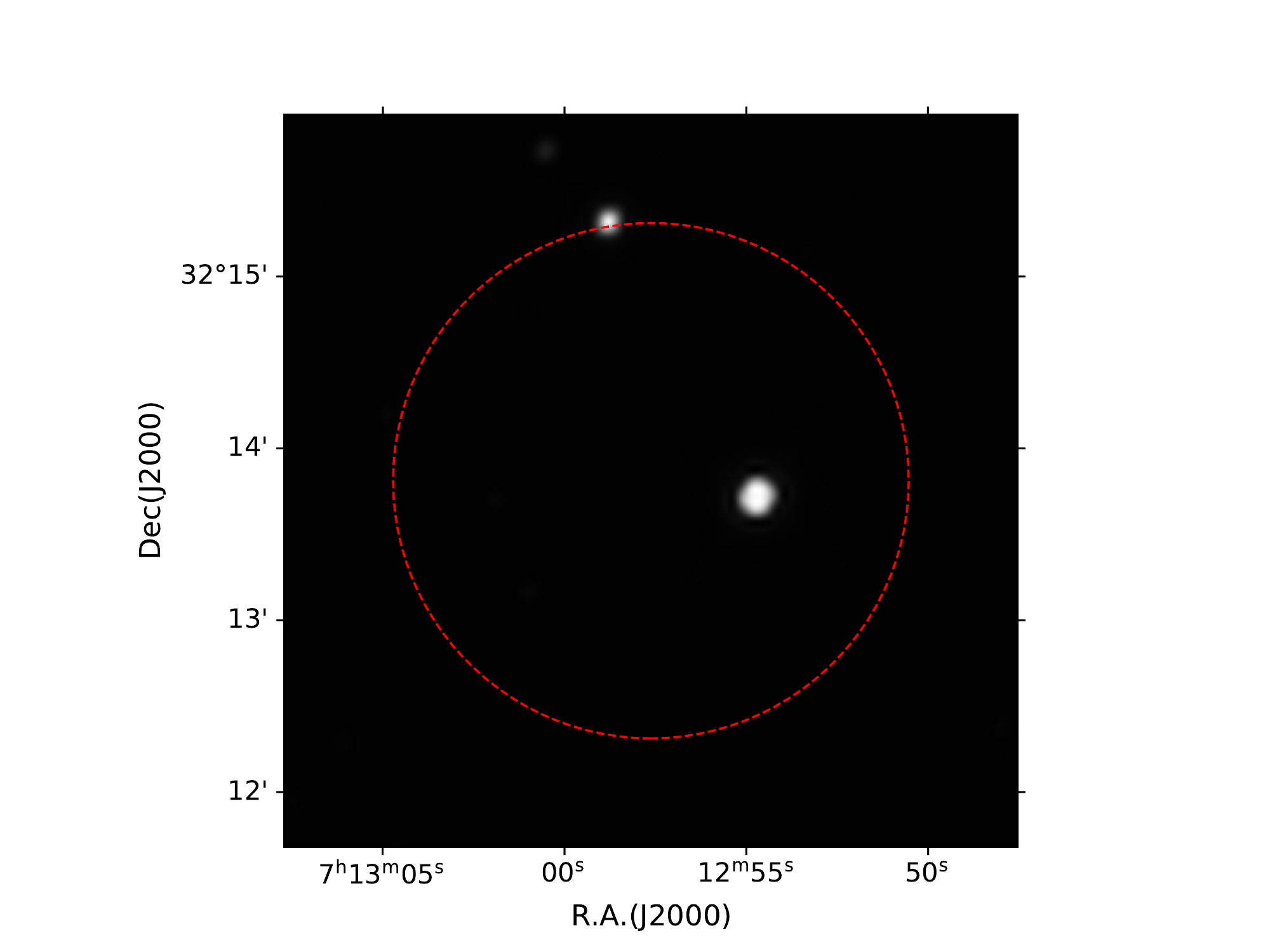}
		\caption{Left and middle plot show  spectrum line for some sources, HI intensity mappings,respecitively, and its counterparts image in unWISE W1/W2 NEO6 band  in right panel, Left and middle plot show  HI source spectrum line, HI intensity mappings,respecitively, and its counterparts image in unWISE W1/W2 NEO6 band   in right panel.\label{map8}}
	\end{figure*}

	Figures \ref{map2} to \ref{map8} present the
	HI spectra and intensity color maps for the 16 
	sources listed in Table \ref{tab:no_optical}, for which no optical counterparts have
	been matched. Four of these sources, e.g. FGC 10,FGC 31, FGC 44, FGC 50, are located in regions not covered by current mainstream optical surveys, while other sources are within the optical survey area but they could be too faint to be detected by the optical surveys.  It is 
	possible that some HI detected sources are indeed optically dark
	galaxies, which have very low optical apparent magnitudes and are not
	detectable by optical telescopes.  In Table \ref{tab:no_optical}, we mark the sources in
	the note column (12), where S stands for the sources that is too faint
	and beyond the SDSS survey detection limit of -22 magntitude in
	grz-band, while W depicts those beyond the WISE survey detection limit
	of -21 magntitude in W1/W2
	band \footnote{https://www.legacysurvey.org/viewer/}. Other parameters
	in this table are identical to that of Table \ref{tab:HI_sources}.
	
	From Table \ref{tab:no_optical} we can see that the HI extragalaxies without optical
	counterparts have an R.A. range from 0 to 110$^{\circ}$ with a
	maximum redshift of 0.024, and their HI masses vary from $10^{8.5}$ to
	$10^{9.7}M_{\odot}$. The total HI flux of these galaxies ranges
	from 0.7 Jy km/s to 5 Jy km/s,and their line widths range
	from 25.9 to 385.4 km/s. Based on the  measurement of the flux and line widths, the gas properties of these optically "dark" galaxies appear to be normal. Future deep optical follow-up observations might be able to reveal some peculiar features in these type of galaxies.

	

	\begin{table*}
		\begin{center}
			\caption{ 16 HI Candidates with no  counterparts\label{tab:no_optical}}
			\resizebox{\textwidth}{!}{
				\begin{tabular}{ccccccccccccc}
					\hline\hline
					FGC ID &$\alpha_{\rm J2000}$(HI) &$\delta_{\rm J2000}$(HI)& cz$_\odot$& W50&$\rm F_{\rm int}$&S/N&$\sigma_{\rm rms}$&Dist&logM$_{\rm HI}$& Code& Note\\
					&      &    & (km s$^{-1}$)  & (km s$^{-1}$) & (Jy km s$^{-1}$) &  &(mJy) & (Mpc) & ($\rm M_{\odot}$)&  & &\\
					(1) &  (2) &  (3) & (4)  & (5) & (6)& (7)            &      (8)  & (9)     &(10) & (11) & (12) \\
					\hline	
					10	&	00:17:15.34	&	42:07:56.85	&	4844.4	&	215.7	&	4.6	&	12	&	2.7	&	69.2	&	9.7	&	1	&	W	\\
					
					31	&	01:14:45.78	&	27:10:52.49	&	3570.7	&	100	&	5	&	11.7	&	1.5	&	51	&	8.8	&	1	&	S	\\
					44	&	01:56:49.57	&	34:08:27.17	&	4739.3	&	385.4	&	2.7	&	21	&	2.1	&	67.7	&	9.5	&	1	& W	\\
					50	&	02:06:31.00	&	43:51:34.95	&	5197.1	&	187.6	&	2.5	&	6.8	&	3	&	74.2	&	9.5	&	1	&	W	\\
					78	&	02:43:16.03	&	36:39:57.10	&	5071.5	&	65.5	&	0.8	&	5.3	&	2.9	&	72.5	&	9	&	2	&W	\\
					92	&	03:08:09.38	&	42:51:36.83	&	2553.2	&	25.9	&	1.1	&	22.6	&	2.4	&	36.5	&	8.5	&	1	&S	\\
					93	&	03:08:35.36	&	36:27:17.10	&	3383.1	&	229	&	4	&	9.8	&	2.9	&	48.3	&	9.4	&	1	&	W	\\
					105 &	03:55:35.52	&   32:08:02.40 &	4135.9	&	40.2	&	0.7	&	10.2	&	2.9	&	59.1	&	8.8	& 1 &		W		\\

					121	&	04:52:18.91	&	32:24:42.31	&	3245.4	&	76.8	&	2.8	&	21.8	&	1.9	&	46.4	&	9.2	&	1	&	W\\
					125	&	05:39:35.00	&	41:24:22.29	&	12936	&	44.4	&	0.7	&	9.3	&	3.2	&	184.8	&	9.8	&	1	&	W	\\
					126	&	05:42:42.34	&	34:30:36.77	&	5918.4	&	74.8	&	3	&	14.7	&	2.9	&	84.6	&	9.7	&	1	&	 W	\\
					127&	05:51:15.75	&	43:22:24.11	&	3615.8	&	186.4	&	1.9	&	9.2	&	2.7	&	51.7	&	9.1	&	1	&	W	\\
					136	&	06:38:46.83	&	33:28:52.43	&	5002.2	&	49	&	0.8	&	5.7	&	2.6	&	71.5	&	9	&	2	&	 W	\\
					138	&	06:43:20.56	&	42:26:04.06	&	6123	&	129.3	&	1.6	&	7.6	&	2.9	&	87.5	&	9.5	&	1	&		 W	\\
					139	&	06:44:6.72	&	33 :21:36.00&7296.9	&	36.1	&	0.4	&	5.8	&	2.6	&	104.2	&	9.1	&	2	&	W	\\
					
					142	&07:12:57.36	&32 :13  48.00	&4781.7	&	189.2	&	1.9	&	12.2	&	2	&	68.3	&	9.3	&	1	&	W	\\
					
					\hline	
			\end{tabular}}
		\end{center}
	\end{table*}

	
	\section{Summary}\label{sums}
	 As first result from the FAST pilot HI survey, a catalog of 544 HI detections at sky region of $+24^{\circ}<\delta<+43^{\circ}$ are discribed in this work. 
		All these sources  as extragalactic objects $cz>100$
		and 527 of these sources can be matched with optical counterparts in
		online data archives.  These detections are searched with a high
		confidence level in the accuracy of position and redshift. In the catalog, we have classified all sources into four categories  based on their S/N and  baseline qualities.
		Among them, 302 sources are also detected by ALFALFA. 
		In regions not affected by RFI and standing waves, the FAST measured HI fluxes and profiles are  consistent with that of ALFALFA. 
	More than $90\%$ of ALFALFA detected sources with peak flux larger
	than 10 mJy were also detected in the FAST pilot HI
	survey. About 10 to 20 percent sources are missed due to RFI
	contamination.  Our pilot
	study found 16 new  detections without optical counterparts in
	redshift range $z < 0.024$. Some of them are quite massive in HI gas,
	but their optical fluxes could be too small to be dectected in optical wavelengths.
	These objects feature a few peculiar physical characteristics that
	deserve more further studies with multi-wavelength follow-up
	observations.
	
	\begin{acknowledgements}
		We acknowledge the supports of the National Key R$\&$D Program of China No. 2017YFA0402600.This work has used the data from the Five-hundred-meter Aperture Spherical radio Telescope (FAST). FAST is a Chinese national mega-science facility, operated by the National Astronomical Observatories of Chinese Academy of Sciences (NAOC).This research has also made use of the NASA/IPAC Extragalactic Database (NED) which is operated by the Jet Propulsion Laboratory, California Institute of Technology, under contract with the National Aeronautics and Space Administration.This research has made use of the VizieR catalogue access tool, CDS,Strasbourg, France (DOI : 10.26093/cds/vizier).
		
	\end{acknowledgements}
	
	\bibliographystyle{raa}
	\bibliography{HI2022}
\end{document}